\documentclass{ws-ijmpd}
\usepackage[super,compress]{cite}
\usepackage[breaklinks]{hyperref}
\hypersetup{colorlinks,urlcolor=black,citecolor=black,linkcolor=black,filecolor=black}
\usepackage{breakurl}

\usepackage{amssymb,amsmath,dsfont}
\usepackage{bm}
\usepackage{subfig}

\newcommand{\beq}{\begin{equation}}
\newcommand{\beqn}{\begin{eqnarray}}
\newcommand{\eeq}{\end{equation}}
\newcommand{\eeqn}{\end{eqnarray}}
\newcommand{\bea}{\begin{eqnarray}}
\newcommand{\eea}{\end{eqnarray}}
\usepackage{comment}
\newcommand{\nc}{\newcommand}
\nc{\ba}{\begin{eqnarray}} \nc{\ea}{\end{eqnarray}}
\nc{\be}{\begin{equation}} \nc{\ee}{\end{equation}}
\newcommand{\id}{\mathbb{I}}

\newcommand\s{\sigma}

\newcommand{\bk}{{\bf{k}}}

\nc{\fb}{\phi_2}
\nc{\ci}{\chi}

\input{epsf.sty}

\begin{document}

%
\catchline{}{}{}{}{}
%

\title{ NONPERTURBATIVE DYNAMICS OF \\REHEATING AFTER INFLATION: A REVIEW}
\author{ MUSTAFA A. AMIN${}^1$\footnote{mustafa.a.amin@gmail.com}, MARK P.~HERTZBERG${}^2$\footnote{mphertz@mit.edu}, \\ DAVID I. KAISER${}^2$\footnote{dikaiser@mit.edu}, JOHANNA KAROUBY${}^2$\footnote{karoubyj@mit.edu}}
\address{$^{1}$Kavli Institute for Cosmology and Institute of Astronomy,\\ University of Cambridge, Madingly Road,Cambridge CB3 0HA, UK\\
$^{2}$Center for Theoretical Physics and Department of Physics,\\
Massachusetts Institute of Technology, 
Cambridge, MA 02139, USA\\
}

\maketitle


\begin{abstract}
Our understanding of the state of the universe between the end of inflation and big bang nucleosynthesis (BBN) is incomplete.  The dynamics at the end of inflation are rich and a potential source of observational signatures. Reheating, the energy transfer between the inflaton and Standard Model fields (possibly through intermediaries) and their subsequent thermalization, can provide clues to how inflation fits in with known high-energy physics. We provide an overview of our current understanding of the nonperturbative, nonlinear dynamics at the end of inflation, some salient features of realistic particle physics models of reheating, and how the universe reaches  a thermal state before BBN. In addition, we review the analytical and numerical tools available in the literature to study preheating and reheating and discuss potential observational signatures from this fascinating era.
\end{abstract}


\ccode{PACS numbers:98.80.Cq,95.30.Cq}

\ccode{Preprint number: MIT-CTP 4560}

\tableofcontents

\section{Introduction}

Our understanding of the early history and evolution of our observable universe is anchored by two major poles: cosmic inflation and big-bang nucleosynthesis. Building on the well-understood framework of quantum field theory in curved spacetime, models of cosmic inflation make specific, quantitative predictions for observable quantities, such as the spatial curvature $\Omega_k$ and the spectral tilt of primordial curvature perturbations, $n_s$. (For reviews of inflation, see Refs. \refcite{Guth:2005zr,Bassett:2005xm,Lyth:2009zz,Baumann:2009ds,Linde:2014nna,Martin:2014vha}.) Likewise, big-bang nucleosynthesis builds on detailed information about nuclear reactions to predict quantities such as light-element abundances.\cite{Steigman:2007xt} Both inflation and nucleosynthesis are tightly constrained by high-precision measurements, and to date both theories match observations to impressive accuracy.

Despite these resounding successes, we still have an inadequate understanding of how to connect these two important eras in cosmic history. If confirmed, the recent announcement by the BICEP collaboration of the detection of primordial gravitational waves implies that inflation occurred at an energy scale of order $E \sim 10^{16}$ GeV, corresponding to a cosmic time $t \sim 10^{-38}$ seconds after the big bang.\footnote{We do not know how old the universe actually is, or even whether this age is finite. We only have a lower bound on how long inflation lasted. When assigning an age to an event in cosmology, we usually assign times using standard big-bang cosmology, ignoring inflation.} Meanwhile, big-bang nucleosynthesis began at an energy scale $E \sim 10^{-3}$ GeV around $t \sim 1$ second.\cite{Steigman:2007xt} Although inflation and nucleosynthesis are each grounded in solid theoretical ideas and are increasingly constrained by robust empirical data, between them stretches a huge range of energy and time scales that are neither well understood nor strongly constrained. The electroweak symmetry-breaking phase transition, for example, which occurred around $E \sim 10^{3}$ GeV at $t \sim 10^{-12}$ seconds, represents an intermediate stage between inflation and primordial nucleosynthesis that is increasingly well-understood from experiments at CERN. That transition, however, was likely many orders of magnitude removed from inflationary energy scales. 

The observational difficulties for constraining this period arise due to two main reasons. (1) For consistency between BBN calculations and measured light-element abundances, the universe has to be thermal (at least in most of the Standard Model species) by the time of BBN. As a result, most of the information about potentially complicated dynamics after inflation and before BBN gets washed away. (2) For simple scenarios, the typical scales over which spatial perturbations are generated during this period are smaller that the Hubble horizon $H^{-1}$ at that time, due to causality considerations. Therefore one does not generically expect effects of these perturbations on scales that affect the cosmic microwave background radiation (CMB) or large-scale-structure observations (though in certain special scenarios, preheating dynamics can generate large fluctuations on superhorizon scales \cite{Taruya:1997iv,Bassett:1999cg,Finelli:2000ya,Tsujikawa:2002nf,Chambers:2007se,Bond:2009xx,Bethke:2013aba,Moghaddam:2014ksa}).

In spite of these difficulties, the period between the end of inflation and BBN is exciting for a number of reasons. On the theoretical/phenomenological side, in order to connect the successes of inflationary cosmology to the standard big-bang scenario, it is crucial to understand how the universe transitioned from the supercooled conditions during inflation to the hot, thermalized, radiation-dominated state required for nucleosynthesis. Presumably this transition occurred as the energy that had driven the exponential expansion of spacetime during inflation was dispersed into more familiar forms of matter. Post-inflation reheating is thus a critical epoch for connecting studies of the early universe with realistic models of high-energy particle physics. If inflation put the ``bang" in the big bang, it is post-inflation reheating that populated our universe with matter more like the kind we see around us today, and of which we are made. 

Dynamics at the end of inflation also impact how we connect inflationary predictions to empirical observations of temperature anisotropies in the CMB. Relating those two epochs depends upon understanding the expansion history of the universe between inflation and later times; and the expansion history, in turn, depends on whether the post-inflation transition to a hot, thermalized state occurred quickly or slowly. Uncertainty in the energy scale and duration of post-inflation reheating accounts for why inflationary predictions for spectral observables are typically evaluated at $N_* = 55 \pm 5$, where $N_*$ is the number of e-folds before the end of inflation when cosmologically relevant perturbations first crossed outside the Hubble radius.\cite{Dodelson:2003vq,Liddle:2003as,Tegmark:2004qd} In simple models, observable quantities such as the spectral index, $n_s$, and the tensor-to-scalar ratio, $r$, vary inversely with $N_*$, and hence the residual uncertainties from reheating will become important as more and more spectral observables are measured to percent-level accuracy or higher.\cite{Adshead:2010mc,Dodelson:2014exa,Dai:2014jja,Creminelli:2014fca} As we discuss below, the energy scale of inflation also has implications for potential observational consequences that could derive from the reheating process itself.

Physicists began to study post-inflation reheating soon after the introduction of the first inflationary models. The original calculations focused on perturbative decays of individual inflaton particles into other forms of matter.\cite{Albrecht:1982mp,Dolgov:1982th,Abbott:1982hn} About a decade later, several groups recognized that the energy transfer at the end of inflation could involve highly nonperturbative resonances, as the inflaton field oscillated around the minimum of its potential.\cite{Dolgov:1989us,Traschen:1990sw,Kofman:1994rk,Shtanov:1994ce,Boyanovsky:1995ud,Yoshimura:1995gc,Kaiser:1995fb,Kofman:1997yn} Such a ``preheating" phase, governed by parametric resonance, may be studied in a linearized approximation, to first order in field fluctuations.  Yet the resonant, exponential growth of fluctuations means that linearized analyses can only remain self-consistent for a limited duration, before fully nonlinear effects must be considered. Moreover, perturbative decays of the inflaton still play a critical role in the reheating process: at late stages, perturbative decays help to complete the transfer of energy from the inflaton and avoid an extended phase of matter-dominated evolution.

\begin{figure}[h!]
\centering
\includegraphics[width=5in]{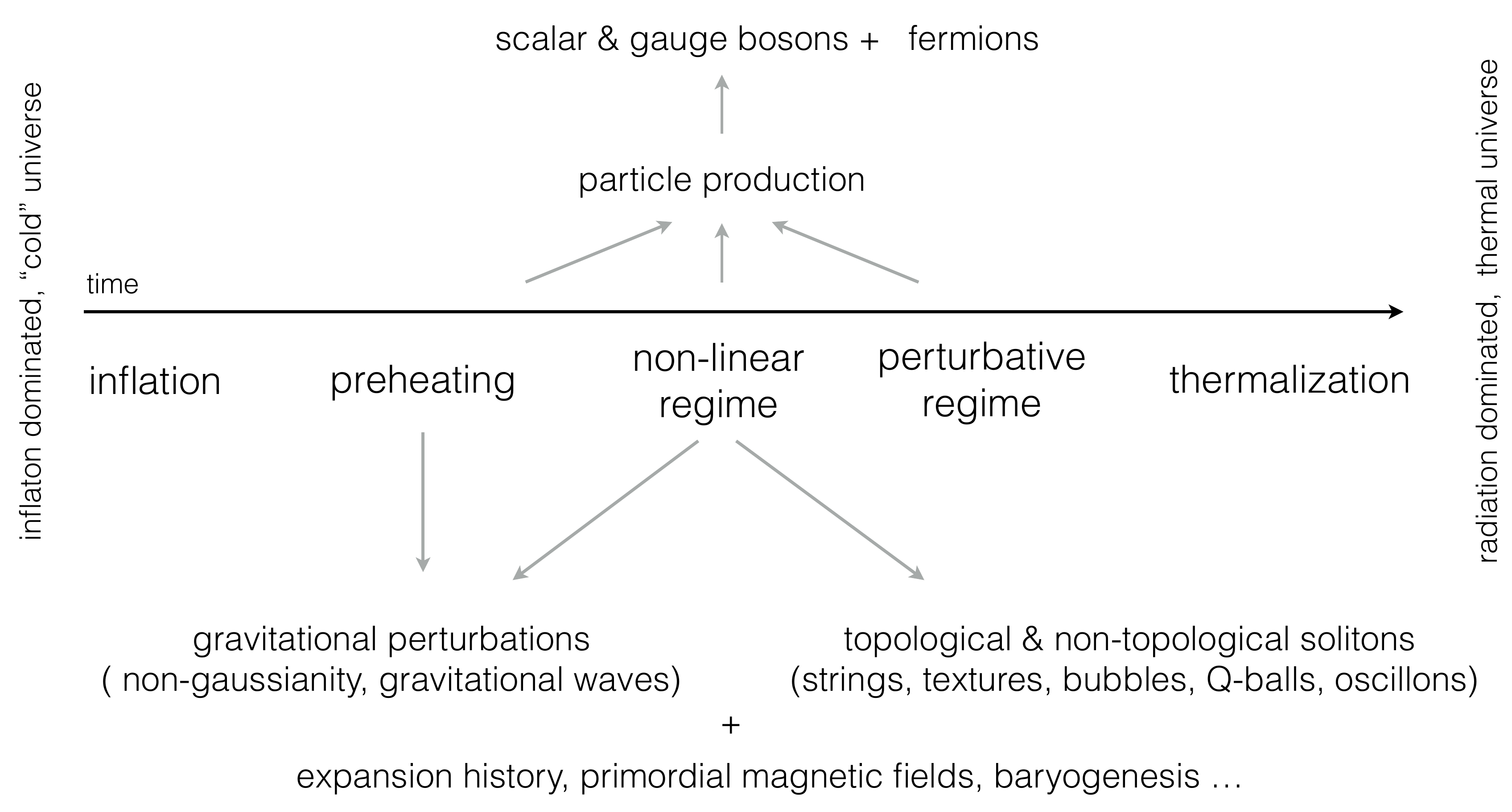} 
\caption{\label{fig1scheme} \small Post-inflation reheating consists of several distinct stages. At the end of inflation, parametric resonance (``preheating") often dominates the early transfer of energy from the inflaton into other types of particles. This highly nonperturbative process can modify the predictions for observational signatures from inflation or create its own unique signals. As preheating enters a fully nonlinear regime, nontrivial field configurations such as solitons and oscillons may be created. Late-stage reheating typically involves perturbative decays of the inflaton to complete the energy transfer and avoid a matter-dominated phase of expansion, before the universe achieves thermal equilibrium in a radiation-dominated phase at some temperature $T_{\rm reh}$.}
\end{figure}

Much of the exciting recent work in post-inflation dynamics has therefore focused on three main challenges: (i) Accounting for the production of ordinary matter after inflation, which requires studying how inflation and reheating can occur within realistic models of high-energy particle physics. (ii) Delineating and understanding several distinct stages in the reheating process, from nonlinear fragmentation of the fields soon after inflation to the stage that leads to thermalization of the universe in a radiation-dominated phase at some reheat temperature, $T_{\rm reh}$. (iii) Understanding observational constraints on and observational implications of post-inflation dynamics. We highlight different stages of reheating and their distinct phenomenological features in Fig. \ref{fig1scheme}.

In Section 2, we consider the initial conditions for studying dynamics near the end of inflation. Section 3 provides an introduction to parametric resonance and preheating. Section 4 focuses on nonlinear effects and Section 5 on thermalization. Section 6 surveys recent efforts to embed inflation and reheating in realistic models of particle physics, and Section 7 discusses some possible observational consequences of the post-inflation epoch. Section 8 turns to discussion and prospects for further research.

A number of excellent previous reviews on reheating may be found in the literature (see, for example, Refs. \refcite{Bassett:2005xm,Boyanovsky:1996sv,Allahverdi:2010xz,Frolov:2010sz}, and Section 5.5 of Ref. \refcite{Mukhanov:2005sc}). Indeed, the literature on post-inflation reheating has grown rapidly over the past two decades; at the time of writing, some of the seminal early papers that helped launch the new understanding of preheating (e.g., Refs. \refcite{Kofman:1994rk,Kofman:1997yn}) have each been cited nearly 900 times in the SLAC-SPIRES database. Rather than aim at encyclopedic coverage of such a vast literature, we have instead tended to cite original work, recent work, and a limited number of references in between, with the goal of providing the reader enough guidance to dig into the subject further. Our aim is to provide a reasonably self-contained, pedagogical treatment of the subject, especially the linearized analysis. Because realistic models of high-energy physics generically include multiple scalar fields at inflationary energy scales, we devote particular emphasis to multifield dynamics.

Throughout this review we use the following notation and conventions. We work in $3 + 1$ spacetime dimensions, with metric signature $(-, +, +,+)$. Greek letters label spacetime indices, $\mu, \nu = 0, 1, 2, 3$; lower-case Latin letters label spatial indices, $i, j = 1, 2, 3$; and upper-case Latin letters label fields, $I, J = 1, 2, 3, ..., N$ for models with $N$ scalar fields. We adopt ``natural" units in which $\hbar = c = 1$, and we use the reduced Planck mass, $M_{\rm pl} \equiv 1 / \sqrt{ 8 \pi G} \simeq 2.43 \times 10^{18}$ GeV.


\section{Degrees of Freedom and Initial Conditions}

Models of inflation with $N$ scalar fields may be constructed from an action of the form
\beq
S = \int d^4 x \sqrt{-g} \left[ \frac{M_{\rm pl}^2}{2} R - \frac{1}{2} {\cal G}_{IJ} g^{\mu\nu} \partial_\mu \phi^I \partial_\nu \phi^J - V (\phi^I ) \right] ,
\label{action}
\eeq
where ${\cal G}_{IJ} (\phi^K)$ is a metric on the field-space manifold.\footnote{Models that incorporate nonminimal couplings between the fields $\phi^I$ and the Ricci spacetime curvature scalar, $R$, may be put in this form following a conformal transformation.\cite{Kaiser:2010ps} }. In order to understand post-inflation dynamics, we must consider the conditions at the moment when the universe stopped inflating. We take the end of inflation to be when $\epsilon (t_{\rm end}) = 1$, where $\epsilon$ is the standard slow-roll parameter, 
\beq
\epsilon \equiv - \frac{\dot{H}}{H^2} .
\label{epsilon}
\eeq
Here $H \equiv \dot{a} /a$ is the Hubble parameter, $a (t)$ is the scale factor of the Friedmann-Robertson-Walker (FRW) line element, and overdots denote derivatives with respect to cosmic time, $t$. The condition $\epsilon = 1$ is equivalent to $\ddot{a} = 0$. Around $t_{\rm end}$, the inflaton field(s) typically begin oscillating around the minimum of the potential, $V (\phi^I)$.

We are interested in the behavior of quantum fluctuations during this era. We expand both the scalar fields and the spacetime metric to first order around their background values,
\beq 
\phi^I (x^\mu) = \varphi^I (t) + \delta \phi^I (x^\mu) , \>\> g_{\mu\nu} (x^\lambda) = \bar{g}_{\mu\nu} (x^\lambda) + h_{\mu\nu} (x^\lambda) .
\label{phivarhpi}
\eeq
The scalar degrees of freedom of the perturbed line-element may be written
\beq
ds^2 = - (1 + 2A) dt^2 + 2a (\partial_i B ) dt dx^i + a^2 \left[ (1 - 2 \psi) \delta_{ij} + 2 \partial_i \partial_j E \right] dx^i dx^j .
\label{ds}
\eeq
We expand around a spatially flat FRW metric because we are interested in conditions near the end of inflation. 

From the (scalar) metric perturbations, one may construct the well-known gauge-invariant Bardeen potentials, $\Phi \equiv A - \partial_t [ a^2 ( \dot{E} - B / a )]$ and $\Psi \equiv \psi + a^2 H (\dot{E} - B/a)$.\cite{Bardeen:1980kt,Bassett:2005xm} For models with an action of the form in Eq. (\ref{action}), the anisotropic stress vanishes to first order in the perturbations: $\delta T^i_{\> j} = 0$ for $i \neq j$, where $\delta T_{\mu\nu} \equiv T_{\mu\nu} (\phi^I, g_{\mu\nu} ) - T_{\mu\nu} (\varphi^I, \bar{g}_{\mu\nu})$. We will work in Newtonian gauge, with $E=B=0$. In absence of the anisotropic stress, the $i\ne j$ Einstein's field equation yields $\Phi=\Psi$. The fluctuations in a model governed by an action of the form in Eq. (\ref{action}) thus include $N + 1$ scalar degrees of freedom ($\delta \phi^I$ and $\Psi$), though the other Einstein's field equations provide an additional constraint, leaving $N$ physical degrees of freedom among the scalar fluctuations.

To background order, the equations of motion that follow from the action in Eq. (\ref{action}) may be written
\beq
{\cal D}_t \dot{\varphi}^I + 3 H \dot{\varphi}^I + {\cal G}^{IK} V_{, K} = 0
\label{phieom}
\eeq
and 
\beq
H^2 = \frac{1}{ 3 M_{\rm pl}^2 } \left[ \frac{1}{2} {\cal G}_{IJ} \dot{\varphi}^I \dot{\varphi}^J + V (\varphi^I) \right] , \>\> \dot{H} = - \frac{1}{ 2 M_{\rm pl}^2} {\cal G}_{IJ} \dot{\varphi}^I \dot{\varphi}^J ,
\label{Heq}
\eeq
where $V_{, K} \equiv \partial V / \partial \phi^K$, and the quantities ${\cal G}_{IJ}$, $V$, and their derivatives are evaluated at background order, as functions of $\varphi^I$. Here we have introduced a (covariant) directional derivative, ${\cal D}_t A^I = \dot{A}^I + \Gamma^I_{\> LJ} \dot{\varphi}^L A^J$, appropriate to the nontrivial field-space metric.\footnote{For any vector in the field-space manifold, $A^I$, we may introduce a covariant derivative, ${\cal D}_J A^I = \partial_J A^I + \Gamma^I_{\> JK} A^K$, where the Christoffel symbol for the field-space is defined in the usual way, $\Gamma^I_{\> LJ} \equiv {1 \over 2} {\cal G}^{IK} [ \partial_L {\cal G}_{KJ} + \partial_J {\cal G}_{LK} - \partial_K {\cal G}_{LJ} ].$\cite{Sasaki:1995aw,GrootNibbelink:2001qt,Langlois:2008mn,Weinberg:2008zzc,Gong:2011uw,Peterson:2011yt,Elliston:2012ab,Kaiser:2012ak}}

To first order, the fluctuations obey
\beq
{\cal D}_t^2 \delta \phi^I + 3 H {\cal D}_t \delta \phi^I + \left[ -\frac{1}{a^2} \delta^I_{\> J} \nabla^2 + {\cal M}^I_{\> J} \right] \delta \phi^J = - 2 {\cal G}^{IK} V_{, K} \Psi + 4 \dot{\varphi}^I \dot{\Psi} ,
\label{deltaphieom}
\eeq
where
\beq
{\cal M}^I_{\> J} \equiv {\cal G}^{IK} \left( {\cal D}_J {\cal D}_K V \right) - {\cal R}^I_{\> LMJ} \dot{\varphi}^L \dot{\varphi}^M ,
\label{MIJ}
\eeq
and ${\cal R}_{IJKL}$ is the Riemann tensor constructed from ${\cal G}_{IJ}$ and its first two derivatives, and again we evaluate $V$, ${\cal G}_{IJ}$, and their derivatives to background order. The metric perturbation $\Psi$ satisfies
\beq
\begin{split}
\dot{\Psi} + H \Psi &= \frac{1}{ 2 M_{\rm pl}^2 } {\cal G}_{IJ} \dot{\varphi}^I \delta \phi^J , \\
\left( \dot{H} - \frac{1}{a^2} \nabla^2 \right) \Psi &=  \frac{1}{ 2M_{\rm pl}^2} {\cal G}_{IJ} \left[  \left( {\cal D}_t \dot{\varphi}^I \right) \delta \phi^J -  \dot{\varphi}^I \left( {\cal D}_t \delta \phi^J \right) \right] .
\end{split}
\label{Psieom}
\eeq
Upon using Eq. (\ref{Psieom}) and performing a Fourier transform,\footnote{Our convention for Fourier transforms is $F (x^\mu) = \int d^3 k \> F_\bk (t) e^{i {\bf k} \cdot {\bf x} }$.} we may cast Eq. (\ref{deltaphieom}) in the suggestive form\cite{Lozanov:2014zfa}
\beq
\mathds{L}_k (t) \cdot \delta \vec{\phi}_\bk (t) = \left[ \id \partial_t^2 - \Pi (k, t) \partial_t - {\cal F} (k, t) \right] \cdot \delta \vec{\phi}_\bk (t) = 0.
\label{deltaphieomL}
\eeq
Here $\mathds{L}_k$ is a linear differential operator, second-order in $t$, which depends on $k$ and $t$. The identity matrix is represented by $\id$. Because $\delta \vec{\phi}_\bk = [ \delta \phi^1_\bk, \delta \phi^2_\bk , ... , \delta \phi^N_\bk ]^T$ is an $N$-component vector, the operator $\mathds{L}_k$ is an $N \times N$ matrix. By writing the operator $\mathds{L}_k$ in terms of $\partial_t$ rather than ${\cal D}_t$, the explicit expressions for the matrices $\Pi_{IJ}$ and ${\cal F}_{IJ}$ no longer appear manifestly covariant with respect to ${\cal G}_{IJ}$:
\beq
\Pi^I_{\> J} (k, t)  = - 3 H \delta^I_{\> J} - 2 \Gamma^I_{\> LJ} \dot{\varphi}^L + \frac{1}{M_{\rm pl}^2} \left[ \frac{ {\cal G}^{IK} V_{, K} + 2 H \dot{\varphi}^I }{ \dot{H} + (k^2 / a^2 ) } \right] {\cal G}_{KJ} \dot{\varphi}^K
\label{PiIJ}
\eeq
and
\beq
\begin{split}
{\cal F}^I_{\> J} (k, t) &= - \frac{k^2}{a^2} \delta^I_{\> J} - {\cal N}^I_{\> J} + \frac{2}{ M_{\rm pl}^2 } {\cal G}_{KJ} \dot{\varphi}^I \dot{\varphi}^K \\
&\quad - \frac{1}{ M_{\rm pl}^2}  \left[ \frac{ {\cal G}^{IK} V_{, K} + 2 H \dot{\varphi}^I }{ \dot{H} + (k^2 / a^2 ) } \right] \left[ {\cal G}_{KJ} \ddot{\varphi}^K + \left( {\cal G}_{KJ} \Gamma^K_{\> BC} - {\cal G}_{KC} \Gamma^K_{\> JB} \right) \dot{\varphi}^B \dot{\varphi}^C \right] ,
\end{split}
\label{FIJ}
\eeq
where we have defined
\beq
{\cal N}^I_{\> J} \equiv  {\cal G}^{IK} V_{, KJ} - {\cal G}^{IA} {\cal G}^{KB} \left( \partial_J {\cal G}_{AB} \right) V_{, K} + \left( \partial_J \Gamma^I_{\> LM} \right) \dot{\varphi}^L \dot{\varphi}^M .
\label{NIJ}
\eeq
The terms in $\Pi_{IJ}$ and ${\cal F}_{IJ}$ proportional to $M_{\rm pl}^{-2}$ arise from the metric perturbations on the righthand side of Eq. (\ref{deltaphieom}). 

In general, neither $\Pi_{I J}$ nor ${\cal F}_{I J}$ will be diagonal in multifield models. Nonvanishing cross-terms link the behavior of $\delta \phi^J$ and $\delta \phi^K$, even at linear order.\footnote{The fluctuations $\delta \phi^I$ are gauge-dependent. One may choose instead to work in terms of $N$ gauge-invariant fluctuations, by generalizing the Mukhanov-Sasaki variable, $Q^I \equiv \delta \phi^I + ( \dot{\varphi}^I / H ) \psi$.\cite{Mukhanov:1985rz,Sasaki:1986hm,Bassett:1999mt,Bassett:2005xm} Then one may write a source-free equation of motion for the $N$-component vector $Q^I$, in similar form to Eq. (\ref{deltaphieomL}), again with nonvanishing cross-terms.\cite{Kaiser:2012ak} In such a basis, the gauge-invariant comoving curvature perturbation, ${\cal R}_c$, may be constructed as a linear combination of the $Q^I$. In our present formulation, ${\cal R}_c$ is given by a simple linear combination of $\Psi$ and $\dot{\Psi}$. } The fact that the equations of motion for the fluctuations become coupled has important physical implications. In particular, the nonvanishing cross-terms in $\Pi_{IJ}$ and ${\cal F}_{IJ}$ can drive complex dynamics among the coupled fields --- a qualitatively different evolution of the fluctuations than in single-field models, or in multifield models in which one neglects the cross-terms within the operator $\mathds{L}_k (t)$.\cite{Podolsky:2002qv,Jin:2004bf,Bond:2009xx}

Eq. (\ref{deltaphieomL}) is a set of $N$ linear second-order, ordinary differential equations; any such system will have $2N$ linearly independent solutions. For real-valued scalar fields, we may therefore write the solutions as (see, for example, Refs. \refcite{Salopek:1988qh,Weinberg:2008zzc,Lozanov:2014zfa,Price:2014xpa}):
\beq
\delta \vec{\phi}_\bk (t) = \sum_{n = 1}^N \vec{u}_n (k, t) a_{ {\bf k} n}  +  \vec{u}^*_n (k, t ) a^*_{- {\bf k} n} ,
\label{deltavecphik}
\eeq
which in component form is given by
\beq
\delta \phi^J_\bk (t) = \sum_{n = 1}^N u^J_n (k, t) a_{ {\bf k} n}  +  u^{J*}_n (k, t ) a^*_{- {\bf k} n} .
\eeq
For each $n$, $\mathds{L}_k (t) \cdot \vec{u}_n (k, t) = 0$. For a two-field model involving fields $\phi$ and $\chi$, for example, we may label $a_{ {\bf k} n} \rightarrow b_{\bf k}$ for $n = 1$ and $c_{\bf k}$ for $n = 2$, and likewise label the components of $\vec{u}_{1,2}$ as $\vec{u}_1 =  [v, z]^T$ and $\vec{u}_2 =  [w, y]^T$. Then Eq. (\ref{deltavecphik}) becomes
\beq
\begin{split}
\delta \phi_{\bf k} (t) &= v (k, t) b_{\bf k} + w (k, t) c_{\bf k} + v^* (k, t) b^*_{- \bf k} + w^* (k, t) c^*_{- \bf k} , \\
\delta \chi_{\bf k} (t) &= y (k, t) c_{\bf k} + z (k, t) b_{\bf k} + y^* (k, t) c^*_{- \bf k} + z^* (k, t) b^*_{- \bf k} .
\end{split}
\label{deltaphitwofield}
\eeq
Eq. (\ref{deltaphieomL}) and Eq. (\ref{deltavecphik}) couple the functions $v$ and $z$ together, while $w$ and $y$ become coupled. Because $\Pi_{IJ}$ and ${\cal F}_{IJ}$ depend on time, any rotation of the basis that sets $w(k, t_0) = z (k, t_0) = 0$ at some moment $t_0$ will evolve over time, so that the general time-dependent solutions for $\delta \phi_{\bf k} (t)$ and $\delta \chi_{\bf k} (t)$ will include a mixture of functions as in Eq. (\ref{deltaphitwofield}).\footnote{An analogy may be found with neutrino oscillations: in general the flavor-eigenstates and the mass-eigenstates are distinct, which gives rise to the phenomenon of neutrino-flavor oscillations.\cite{Camilleri:2008zz} }

We may quantize the fluctuations by promoting $a_{ {\bf k} n}$ and $a_{ { \bf k} n}^*$ to operators, $a_{ {\bf k} n} \rightarrow \hat{a}_{ {\bf k} n}$ and $a_{ {\bf k} n}^* \rightarrow \hat{a}^\dagger_{ {\bf k} n}$, which obey the usual commutation relations,
\beq
\left[ \hat{a}_{ {\bf k} n} , \hat{a}_{ {\bf q} m} \right] = 0 , \> \> \left[ \hat{a}_{ {\bf k} n} , \hat{a}^\dagger_{ {\bf q} m} \right] = \delta^{(3)} ( {\bf k} - {\bf q} ) \delta_{nm} .
\label{commutators}
\eeq
Correlation functions among the fields are then given by
\beq
\langle 0 \vert \delta \hat{\phi}^I_{ \bf k} (t) \delta \hat{\phi}^{J \dagger}_{ \bf q} (t) \vert 0 \rangle = \delta^{(3)} ( {\bf k} - {\bf q} ) P^{IJ} (k, t),
\label{correlation}
\eeq
where
\beq
P^{IJ} (k, t) = \sum_{n = 1}^N u_n^I (k, t) u_n^{J*} (k, t) .
\label{powerspectra}
\eeq
We emphasize that cross correlations need not vanish, and in fact can become significant, especially on superhorizon scales --- a point that can have important implications for the initial conditions at the start of post-inflation dynamics.\cite{Lozanov:2014zfa}

During inflation, one typically considers fluctuations around the Bunch-Davies vacuum state.\cite{Birrell:1982ix,Weinberg:2008zzc} When modes of comoving wavenumber $k$ are sufficiently deep inside the Hubble radius during inflation, with $k \gg aH$, the mode functions evolve as\cite{Weinberg:2008zzc}
\beq
u_n^I (k, t) \rightarrow  \frac{e_n^I (t)}{ (2 \pi)^{3/2} a(t) \sqrt{2k} } \exp \left[ - i k \int_{t_{\rm in}}^t \frac{d t' }{a (t') } \right] ,
\label{BunchDaviesmode}
\eeq
where $t_{\rm in}$ is some initial time, early in inflation. Here $e_n^I (t)$ is a vielbein of the field-space metric, defined via ${\cal G}^{IJ} (\varphi^K (t)) = \sum_n e_n^I (t) e_n^J (t)$.\cite{GrootNibbelink:2001qt,Langlois:2008mn,Weinberg:2008zzc} One may then evolve the fluctuations forward in time, to the end of inflation. When performing lattice simulations for dynamics near the end of inflation, one typically sets initial conditions by treating the power spectra for the quantum fluctuations (and their velocities), $P^{IJ} (k, t_{\rm end})$, as a probability distribution for an ensemble of {\it classical} (stochastic, Gaussian-distributed) values for the fluctuations.\cite{Polarski:1995jg} 

\begin{figure}[t]
\centering
\includegraphics[width=3in]{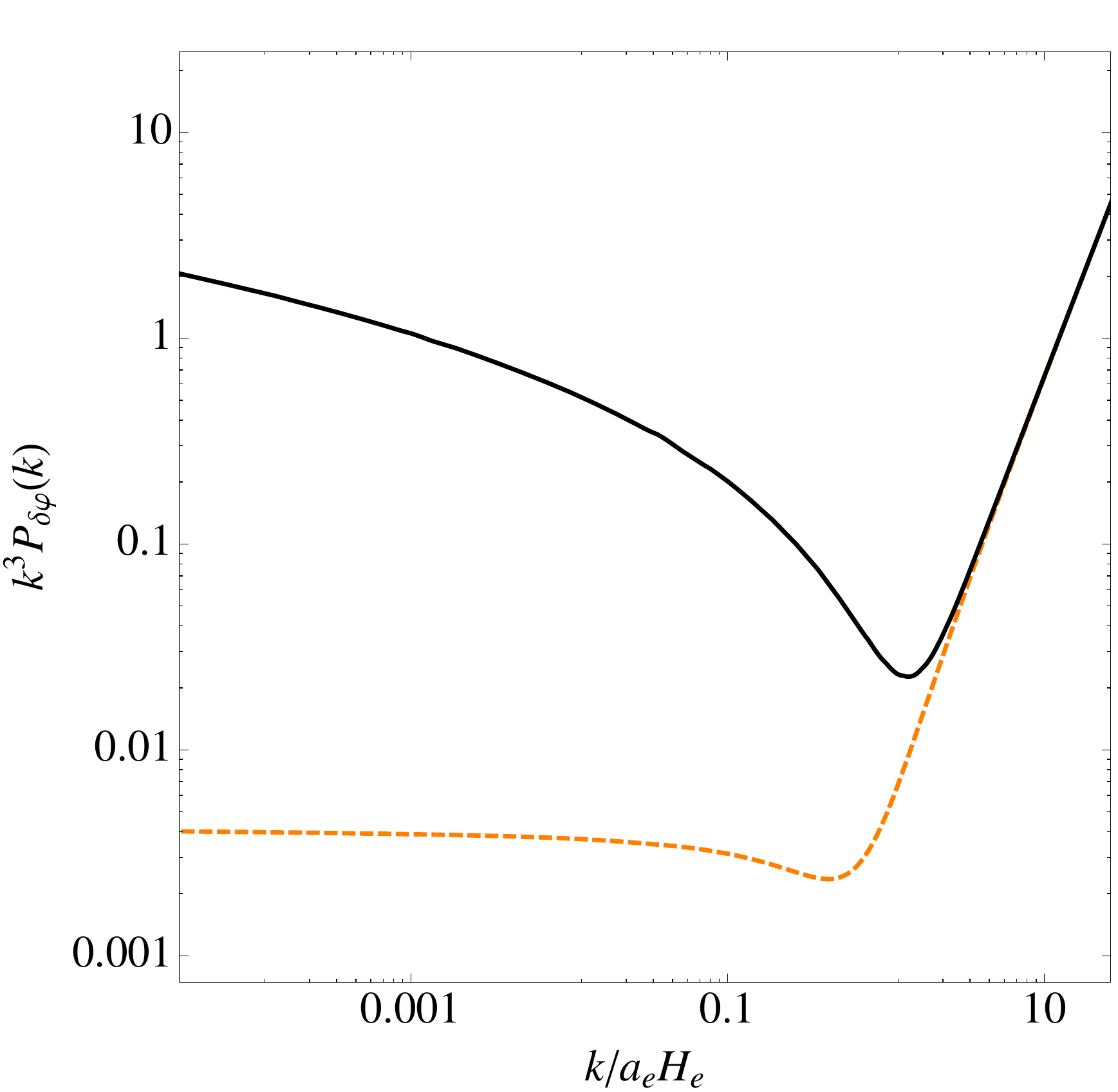}
\caption{\label{powerspectrumfig} \small Power spectrum of the field fluctuations (in arbitrary units)  versus wavenumber at the end of inflation, when $\epsilon = 1$ . The black solid line shows the spectrum for the field fluctuations for the canonical single-field inflation model with $V (\phi) = {1 \over 2} m_{\phi}^2 \phi^2$ in the Newtonian gauge. The dashed orange line shows the spectrum for the fluctuations $\delta \phi$ when metric perturbations $\Psi$ are ignored. The two spectra coincide for wavelengths shorter than the Hubble scale at the end of inflation, $k > a_e H_e$, but differ substantially for modes with $k < a_e H_e$, where $a_e H_e = a (t_{\rm end}) H (t_{\rm end})$.}
\end{figure}

It is important to include the contributions arising from metric perturbations --- proportional to $M_{\rm pl}^{-2}$ in Eqs. (\ref{PiIJ}) and (\ref{FIJ}) --- when evolving the system governed by Eq. (\ref{deltaphieomL}) up to the end inflation. Although such terms are typically slow-roll suppressed during inflation, they can become significant near the end of inflation (when the slow-roll parameter $\epsilon \rightarrow 1$). The discrepancies between power spectra at $t_{\rm end}$ calculated with or without the inclusion of the metric perturbations become especially pronounced for wavelengths longer than the Hubble radius at the end of inflation, $k < a (t_{\rm end}) H (t_{\rm end} )$. See Fig. \ref{powerspectrumfig}. 

An additional subtlety arises when considering fluctuations near the end of inflation. In general, the Hamiltonian for the system depends explicitly on time (given the expansion of the universe, $\dot{a} \neq 0$), and hence so do the eigenvectors. A state that minimizes the energy of the system at one time (the ``vacuum") will not, in general, remain the state of lowest energy at later times: even a noninteracting scalar field can undergo particle production strictly from the change of the FRW scale factor \cite{Birrell:1982ix,Mukhanov:2007}. Early lattice simulations of post-inflation reheating therefore tended to measure changes with respect to the state that instantaneously minimizes the energy of the system at the end of inflation.\cite{Khlebnikov:1996mc,Khlebnikov:1996wr,Khlebnikov:1996zt,Prokopec:1996rr} Just like the effects of metric perturbations, changes in $P^{IJ} (k, t_{\rm end})$ arising from the choice of these two different vacuum states can be significant on superhorizon lengthscales, but rapidly vanish for $k > a (t_{\rm end} ) H (t_{\rm end} )$.

\section{Inflaton Energy Transfer}
\label{PreheatingSection}

At the end of inflation, the inflaton field(s) must decay into other particles, eventually yielding the particle content of the Standard Model. Those decay products need to thermalize at some equilibrium temperature, $T_{\rm reh}$, before the onset of big-bang nucleosynthesis. 

Reheating was originally studied as a perturbative process, in which individual quanta of the inflaton field decay independently of each other. Later studies emphasized the importance of collective, nonperturbative resonances in the initial transfer of energy from the inflaton to other species of matter.

\subsection{Perturbative decay}
\label{PerturbativeSubsection}

A simple model of perturbative decay involves a three-point interaction of the form $\sigma \phi \chi^2$, where $\phi$ is the inflaton field, $\chi$ is some (scalar) decay product, and $\sigma$ is a coupling constant with dimensions of mass. For an interaction of this type, with $\phi \rightarrow \chi \chi$ decays, the decay rate may be calculated in the usual way.\cite{PeskinSchroeder} To tree-level order, one finds for the decay rate
\beq
\Gamma= \frac{\s^2}{8 \pi m_\phi} ,
\label{Gammaperturb}
\eeq
where $m_\phi$ is the mass of the inflaton field. The effect of these decays on the evolution of $\phi$ may be approximated by including an extra friction term, $\Gamma \dot{\phi}$, in the equation of motion for $\phi$,\cite{Kofman:1997yn,Mazumdar:2013gya} though the full effects of dissipation on the inflaton field can, in general, become rather complicated.\cite{Gleiser:1993ea,Boyanovsky:1994me,Boyanovsky:1996xx,Boyanovsky:1996rw}

Early estimates of the reheat temperature arose from equating the Hubble parameter to the inflaton decay rate, $H \sim \Gamma$.\cite{Albrecht:1982mp,Dolgov:1982th,Abbott:1982hn,Kofman:1997yn} Assuming that the decay products are light compared to $H$, they should behave like radiation, and hence one may set
\beq
\rho = \frac{\pi^2}{30} g_* T^4 = 3 M_{\rm pl}^2 H^2 ,
\label{rhoradiation}
\eeq
where $g_* \sim {\cal O} (10^2)$ is the number of relativistic degrees of freedom one would expect for Standard Model species at high energies. Setting $H \sim \Gamma$ yields
\beq
T_{\rm reh} \sim \left( \frac{ 90 }{ g_* \pi^2 } \right)^{1/4} \> \sqrt{ M_{\rm pl} \> \Gamma} .
\label{Trehesetimate}
\eeq
This simple estimate may be modified by other interaction terms such as $g^2 \phi^2 \chi^2$, which allows for $\phi \phi \rightarrow \chi\chi$ scattering; if the inflaton decays primarily into fermions instead of scalar bosons;\cite{Berges:2002wr, Boyanovsky:1995ema,Mazumdar:2013gya} or if one takes into account the back-reaction of the bath of decay products on the inflaton dynamics.\cite{Drewes:2013iaa,Drewes:2014pfa} Nonetheless, trilinear couplings of the form that lead to Eq. (\ref{Gammaperturb}) are important to include so that the transfer of energy from the inflaton field may eventually become complete. Otherwise the universe could end up cold, empty, and unsuitable for life.\cite{Kofman:1997yn,Podolsky:2005bw,Mazumdar:2013gya}. 

\subsection{Preheating and parametric resonance}
\label{ParametricResonanceSection}
At the end of inflation, the homogeneous field(s) start sloshing about the minimum of their potential. Perturbative calculations of reheating neglect the fact that the oscillations of the inflaton(s) at the end of inflation can be large and coherent. Such oscillations drive {\it parametric resonances}, which are much more efficient than single-body decays at transferring energy from homogeneous inflaton fields to their own perturbations and to other coupled fields. \cite{Dolgov:1989us,Traschen:1990sw,Kofman:1994rk,Shtanov:1994ce,Boyanovsky:1995ud,Yoshimura:1995gc,Kaiser:1995fb,Kofman:1997yn,Boyanovsky:1996sv,Bassett:2005xm,Allahverdi:2010xz,Frolov:2010sz, Amin:2010xe, Amin:2010dc, Amin:2011hj} Such resonances require a nonperturbative description, to which we now turn. 
This phase precedes the final thermalization and is referred to as ``preheating".

The rapid growth of small fluctuations in a background of oscillating homogeneous fields may be captured by {\it Floquet analysis}, which applies to linear equations of motion with periodic coefficients. We note that strict periodicity of the background fields is rare at the end of inflation due to two reasons. First, in an FRW universe, expansion causes field amplitudes to decay. If the oscillation time-scales are fast compared to the typical expansion time-scale, one may ignore expansion as a first approximation, though this is not always self-consistent. Second, even without expansion in a general multifield case, periodic motion at the background level only occurs for special trajectories (e.g., effectively one-dimensional, oscillatory motion in field space and Lissajous curves). Nevertheless, a Floquet analysis is an important first step in determining whether significant instabilities exist in the evolution of perturbations, as well as the length scales associated with such instabilities. For {\it quasi-periodic} motion, a somewhat modified Floquet analysis can be applied, see Refs. \refcite{Braden:2010wd,Zanchin:1997gf,Zanchin:1998fj}.

To avoid loss of periodicity, as a first approximation, we set $M_{\rm pl}\rightarrow \infty$ while keeping the fields' energy density finite. This corresponds to a rigid spacetime: from Eq. (\ref{Heq}), we find $H (t) \rightarrow 0$ (and hence $a(t) \rightarrow {\rm constant}$) in this limit, and the metric perturbations likewise remain negligible.\footnote{The gauge-invariant Bardeen potential obeys a generalized Poisson equation, $(1/a^2) \nabla^2 \Psi = \delta \rho_m / (2 M_{\rm pl}^2)$, where $\delta \rho_m \equiv \delta \rho - 3 H \delta q$ is the comoving density perturbation.\cite{Bassett:2005xm} For finite $\delta \rho_m$, we therefore find $\Psi \rightarrow 0$ in the limit $M_{\rm pl} \rightarrow \infty$.} We will therefore ignore gravity in this section. After calculating the rate-of-growth of fluctuations in a Minkowski spacetime, we may compare it to the expansion rate and see if significant growth is possible when we relax the restriction $M_{\rm pl} \rightarrow \infty$. 

Without gravity, Eqs. (\ref{deltaphieomL}), (\ref{PiIJ}), and (\ref{NIJ}) for the field fluctuations become 
\beq
\mathds{L}_k \cdot \delta \vec{\phi}_\bk (t) = \left[ \id \partial_t^2 - \Pi (k,t) \partial_t - {\cal F} (k, t) \right] \cdot \delta \vec{\phi}_\bk (t) = 0,
\label{eq:2EOM}
\eeq
where
\beq
\begin{aligned}
\Pi^I_{\> J} (k,t)  &= - 2 \Gamma^I_{\> LJ} \dot{\varphi}^L, \\
{\cal F}^I_{\> J} (k, t) &= - \frac{k^2}{a^2} \delta^I_{\> J} - \mathcal{N}^I_J.
\end{aligned}
\eeq
In an attempt to understand the instabilities of the fluctuations, we will formally derive a complete basis of linearly independent solutions of the above equations. We will not need to worry about the quantum operator nature of $\delta\vec{\phi}_\bk$ at this stage, since the operator aspects will be contained in the coefficients multiplying this basis of solutions. For a more detailed quantum field theory treatment of the parametric resonance regime see Ref. \refcite{Berges:2002cz}.

Before moving to a general Floquet analysis of this equation, we consider a simple example. Let us assume that inflation is driven by a single field $\phi$, but is coupled to a subdominant $\chi$ field (with $\langle \chi\rangle \approx 0$). For concreteness, we consider the potential 
\beq
V (\phi, \chi) = {1 \over 2} m_\phi^2 \phi^2 + {1 \over 2} m_\chi^2 \chi^2 + {1 \over 2} g^2 \phi^2 \chi^2,
\label{phichiPot}
\eeq
with a trivial field-space metric $\mathcal{G}_{IJ}=\delta_{IJ}$ where $I,J=\{\phi,\chi\}$. Note that for this simple case $\Pi^I_{\> J} = 0$ and $\mathcal{N}^I_{\> J} = \partial^I\partial_K V$. In this case the linearized equations of motion for the fluctuations $\delta\vec{\phi}_k=(\delta\phi_\bk,\delta\chi_\bk)^T$ are 
\beq
\begin{aligned}
\delta\ddot{\phi}_\bk+(k^2+m_{\phi}^2)\delta\phi_\bk=0,\\
\delta\ddot{\chi}_\bk+(k^2+m_{\chi}^2+g^2\varphi^2)\delta\chi_\bk=0.
\label{eomchik}
\end{aligned}
\eeq
Note that the equations are decoupled because we have assumed $\langle \chi \rangle = 0$. The equation for $\delta\phi_\bk$ is that of a simple harmonic oscillator, with solutions $\delta\phi_\bk\sim e^{i\sqrt{k^2+m_\phi^2}t}$. The equation for $\delta\chi$, however, has a time-dependent frequency
\beq
\omega_k^2 (t) = k^2 + m_\chi^2 + g^2 \varphi^2 (t).
\label{omegakMinkowski}
\eeq
Because the background field satisfies $\ddot{\varphi} + V_{, \phi} \simeq 0 $, the inflaton solution $\varphi(t) =\Phi \sin (m_\phi t)$. As a result, $\omega_k^2(t)$ is periodic. Eq. (\ref{eomchik}) for $\delta \chi_{\bf k}$ is therefore an example of Hill's equation\cite{HillBook,dif}, which is typically written in the form
\beq
\frac{d^2 y_\bk }{ dz^2}  + \left[ A_k + q F (z) \right] y_\bk (z) = 0 ,
\label{Hillseq}
\eeq
where $z$ is a dimensionless time-like variable and $F(z)$ is some periodic function in $z$ with unit amplitude. For our model, we may take $z = m_\phi t$ and find $A_k = ( k^2 + m_\chi^2 + {1 \over 2} g^2 \Phi^2 ) / m_\phi^2$ and $q = g^2 \Phi^2 / (2 m_\phi^2)$. In the special case in which $F (z)$ is harmonic (and not just periodic), Hill's equation is known as the Mathieu equation. 

\begin{figure}[t!]
\centering{
\includegraphics[width=3.0in]{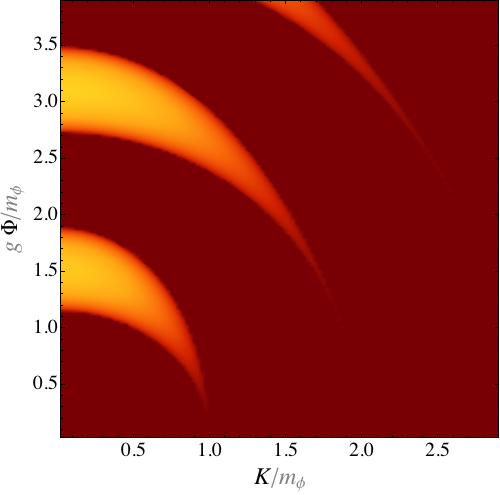}
\includegraphics[width=0.83in]{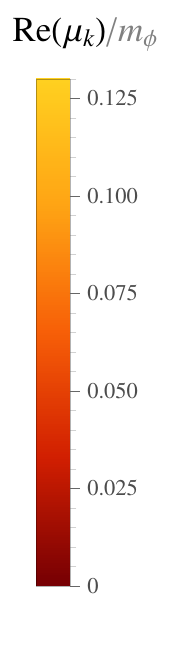}}
\caption{\small A plot of the instability band structure for the model $V (\phi, \chi) = {1\over 2} m_\phi^2 \phi^2 + {1 \over 2} m_\chi^2 \chi^2 + {1 \over 2} g^2 \phi^2 \chi^2$. The color represents the real part of the Floquet exponent rescaled by the mass of the inflaton, $\Re(\mu_k) / m_{\phi}$. The two other axes represent the scaled amplitude of the inflaton field, $g \Phi / m_\phi$, and the rescaled wavenumber $K / m_\phi$, where $K =\sqrt{k^2+m_\chi^2}$. }
\label{Floquetchart}
\end{figure}

Properties of Hill's equation have been studied extensively.\cite{HillBook,dif}. In particular, Floquet's theorem (discussed in detail below) states that solutions to Eq. (\ref{Hillseq}) are of the form
\beq
y_k (z) = e^{\tilde{\mu}_k z} g_1 (z) + e^{- \tilde{\mu}_k z} g_2 (z) ,
\label{Floquet}
\eeq
where $g_1 (z)$ and $g_2 (z)$ are periodic functions and $\tilde{\mu}_k$ is a complex number known as the ``Floquet exponent" (or ``characteristic exponent"). The exponent $\tilde{\mu}_k$ depends on wavenumber $k$ as well as other parameters such as the coupling $g^2$ and the inflaton amplitude $\Phi$. For wavenumbers $k$ such that $\Re [ \tilde{\mu}_k ] > 0$, the corresponding modes $\delta \chi_\bk (t)$ grow exponentially, whereas for $\tilde{\mu}_k$ pure imaginary, the modes are stable and no parametric resonance occurs. In general, the system exhibits a band structure, revealing boundaries between regions of stability and instability as functions of $A_k$ and $q$. Instead of plotting the instability bands as a function of $A_k$ and $q$, in Fig. \ref{Floquetchart} we plot the bands as a function of the amplitude of oscillations $\Phi$ and rescaled wavenumber $K \equiv \sqrt{k^2 + m_\chi^2}$. Note that $\tilde{\mu}_k\equiv \mu_k/m_{\phi}$. 

The exponential instabilities correspond to the breakdown of the WKB approximation. In particular, the time dependence of $\omega_k (t)$ means that the system repeatedly violates the adiabatic condition:
\beq
\left| \frac{\dot{\omega}_k(t)}{\omega_k^2(t)} \right| \ll 1\qquad\qquad \textrm{adiabatic condition}
\eeq
 with $| {\dot{\omega}_k(t)}/{\omega_k^2(t)}| > 1$ around the times when $\varphi (t)$ passes through zero. Rather than evolve as $\delta \chi_\bk \sim e^{\pm i \int \omega_k dt}$, modes grow as $\delta \chi_\bk \sim e^{\mu_k t}$, with $\Re [ \mu_k ] > 0$.

Couplings beyond ${1 \over 2} g^2 \phi^2 \chi^2$ can also yield efficient resonances. For example, trilinear couplings, such as either $\phi^2 \chi$ or $\phi \chi^2$, can drive parametric resonances.\cite{Bassett:1999ta,Dufaux:2006ee,DeMelo:2001nr} Meanwhile, tachyonic instabilities can develop in models with spontaneous symmetry breaking: in the phase with negative mass-squared, modes with wavenumbers that satisfy $k^2 < \vert m^2 \vert$ will grow as $\delta \chi_\bk (t) \sim e^{\sqrt{ \vert m^2 \vert - k^2} \> t }$. Models in this class include those with couplings of the form $g^2 \phi^2 \chi^2$, with $g^2 < 0$.\cite{Greene:1997ge,Dufaux:2006ee,tach1,Bassett:1999ta,Arrizabalaga:2004iw}

Physically, the exponential amplification of modes that lie within a resonance band corresponds to rapid particle production. The energy per mode and the number density of particles per mode may be written\cite{Kofman:1997yn}
\beq
\begin{split}
E_\bk &= \frac{1}{2} \vert \delta \dot{\chi}_\bk \vert^2 + \frac{1}{2} \omega_k^2 \vert \delta \chi_\bk \vert^2 , \\
n_\bk &= \frac{1}{ 2 \omega_k} \left[ \vert \delta \dot{\chi}_\bk \vert^2 + \omega^2_k \vert \delta \chi_\bk \vert^2 \right] - \frac{1}{2} ,
\end{split}
\label{Eknk}
\eeq
and hence for modes that fall within some resonance band $\Delta k$, one finds $n_k \sim e^{2 \mu_k t}$. This rapid, early burst of particle production is known as ``preheating." Given band structure as in Fig. \ref{Floquetchart}, the resulting spectrum from the preheating phase is highly nonthermal.

Apart from some special cases, in general it is difficult to derive closed-form expressions for the Floquet exponents.\cite{Boyanovsky:1996sq,Boyanovsky:1996sv,Greene:1997fu,Kaiser:1995fb,Kaiser:1997mp,Kaiser:1997hg,Kofman:1997yn,Lachapelle:2008sy} For a single degree of freedom, a simple algorithm for calculating Floquet exponents numerically may be found, for example, in Refs. \refcite{Frolov:2010sz,Karouby:2011xs,Hertzberg:2014iza}, and in the appendix to Ref. \refcite{Amin:2011hu}. However, there does not appear to exist in the literature a general, pedagogical treatment for the case of multiple, coupled scalar fields. In what follows we therefore provide a very general framework for calculating Floquet solutions and exponents, which is applicable to multifield scenarios with and without canonical kinetic terms.\footnote{The following discussion is based on notes prepared by one of us (MA), in part, for the undergraduate students of the ``Density Perturbation Group" at MIT in 2010-2011. A simple numerical code for calculating the exponents and generating Floquet instability charts is available on request: mustafa.a.amin@gmail.com. MA would like to acknowledge many fruitful interactions with Leo Stein geared towards the development of a more general numerical code.} To state the algorithm in a general form, we first establish useful notation and preliminaries.

The second-order equation of motion in Eq. (\ref{eq:2EOM}) may be cast as a first-order system as follows:
\beq
\begin{aligned}
&\delta\pi_\bk^I\equiv\delta \dot{\phi}_\bk^I,\\
&\delta\dot{\pi}^I_\bk=\Pi^I_{\> J} \delta\pi_\bk^J+\mathcal{F}^I_{\> J}\delta\phi_\bk^J.
\end{aligned}
\eeq
In matrix form, this first-order system of linear equations may be written as
\beq
\partial_t{x}(t)=U(t)x(t),
\eeq
where 
\[
x(t)=[\delta\phi_\bk^1,\hdots,\delta\phi_\bk^N,\delta\pi_\bk^1,\hdots,\delta\pi_\bk^N]^T,
\]
and
\beq
U(t)
=
\left( {\begin{array}{c|c}
0& \id\\ \hline
\mathcal{F}& \Pi
 \end{array} } \right)
=
\left( {\begin{array}{cccccccc}
0& 0&\hdots& 0&1& 0&\hdots& 0\\
0& 0&\hdots& 0&0& 1&\hdots& 0\\
\vdots& \vdots&\vdots&\vdots& \vdots& \vdots&\vdots& \vdots\\
0& 0&\hdots& 0&0& 0&\hdots& 1\\
\mathcal{F}^1_1& \mathcal{F}^1_2&\hdots& \mathcal{F}^1_N&\Pi^1_1& \Pi^1_2&\hdots& \Pi^1_N\\
\mathcal{F}^2_1& \mathcal{F}^2_2&\hdots& \mathcal{F}^2_N&\Pi^2_1& \Pi^2_2&\hdots& \Pi^2_N\\
\vdots& \vdots&\vdots&\vdots& \vdots& \vdots&\vdots& \vdots\\
\mathcal{F}^N_1& \mathcal{F}^N_2&\hdots& \mathcal{F}^N_N&\Pi^N_1& \Pi^N_2&\hdots& \Pi^N_N\\
 \end{array} } \right).
 \eeq
Note that $U(t)$ and hence the solutions $x (t)$ depend on the wavenumber $k$. We suppress that dependence in what follows to reduce clutter. 

Before we state Floquet's theorem, it is useful to recall the idea of the ``fundamental matrix" of solutions. The fundamental matrix $\mathcal{O}(t,t_0)$ is defined as
\beq
\begin{aligned}
\partial_t\mathcal{O}(t,t_0)&=U(t)\mathcal{O}(t,t_0),\\
\mathcal{O}(t_0,t_0)&=\id ,
\end{aligned}
\eeq
where $\id$ is the $N\times N$ identity matrix. 
 Explicitly, $\mathcal{O}(t,t_0)$ consists of $2N$ columns representing $2N$ independent solutions. The fundamental matrix solution evolves the initial conditions $x(t_0)$ in time: 
 \beqn
 x(t)=\mathcal{O}(t,t_0)x(t_0).
 \eeqn
The theorem and the proof below are based in part on Ref. \refcite{dif}.
\subsubsection{Floquet theorem}
\label{FloquetTheorem}
Consider the linear system
\beqn
\partial_t x(t)=U(t)x(t),
\eeqn
where $x$ is a column vector and $U$ is a real, $2N\times 2N$ matrix satisfying $U(t+T)=U(t)$ for all $t$.
Floquet's theorem states that the fundamental solution can be expressed as
 \beqn
\mathcal{O}(t,t_0)=P(t,t_0)\exp[(t-t_0)\Lambda(t_0)],
\eeqn
where $\Lambda(t_0)$ is defined via $\mathcal{O}(t_0+T,t_0)=\exp[T \Lambda(t_0)]$ and $P(t+T,t_0)=P(t,t_0)$.\\ \\
{\bf Proof: } Any invertible matrix can be represented as the exponential of some other matrix (not necessarily real). Hence it is possible to define a matrix $\Lambda(t_0)$ such that $\mathcal{O}(t_0+T,t_0)=\exp[T \Lambda(t_0)].$

Suppose the fundamental solution takes the form
\beqn
\mathcal{O}(t,t_0)=P(t,t_0)\exp[(t-t_0)\Lambda(t_0)],
\eeqn
where the form of $P$ is to be determined. Then
\beqn
P(t,t_0)=\mathcal{O}(t,t_0)\exp[-(t-t_0)\Lambda(t_0)].
\eeqn
For $t\rightarrow t+T$ we have
\beqn
\begin{aligned}
P(t+T,t_0)
&=\mathcal{O}(t+T,t_0)\exp[-(t+T-t_0)\Lambda(t_0)],\\
&=\mathcal{O}(t+T,t_0)\mathcal{O}^{-1}(t_0+T,t_0)\exp[-(t-t_0)\Lambda(t_0)],\\
&=\mathcal{O}(t+T, t_0+T)\exp[-(t-t_0)\Lambda(t_0)],\\
&=\mathcal{O}(t, t_0)\exp[-(t-t_0)\Lambda(t_0)],\\
&=P(t, t_0).
\end{aligned}
\eeqn
In the second line we used $\mathcal{O}^{-1}(t_0+T,t_0)=\exp[-T\Lambda(t_0)]$ whereas in the third we used $\mathcal{O}(t_1,t_2)\mathcal{O}(t_2,t_3)=\mathcal{O}(t_1,t_3)$ and $\mathcal{O}^{-1}(t_1,t_2)=\mathcal{O}(t_2,t_1)$.
The fourth line follows from the fact the $U(t+T)=U(t)$ since that implies that $\mathcal{O}(t+T,t_0+T)$ and $\mathcal{O}(t,t_0)$ both satisfy the same differential equation. 

Thus, $\mathcal{O}(t,t_0)=P(t,t_0)\exp[(t-t_0)\Lambda(t_0)]$ is a solution with $\Lambda(t_0)$ defined via $\mathcal{O}(t_0+T,t_0)=\exp[T \Lambda(t_0)]$ and $P(t+T,t_0)=P(t,t_0)$. This completes the proof.

The eigenvalues  $\mu^1_k,\mu^2_k,\hdots,\mu^{2N}_k$ of $\Lambda(t_0)$ are known as Floquet exponents. As we will see below, we have unstable, growing solutions iff for some $s=1,2\hdots 2N$, $\Re[\mu^s_k]>0$. We have used $k$ in the subscript as a reminder that the Floquet exponents are in general functions of the wavenumber $k$. 

\subsubsection{Floquet solutions}
\label{FloquetSolutions}

For simplicity we will assume that $\Lambda(t_0)$ has $2N$ distinct eigenvectors $\{e_1(t_0),e_2(t_0),\hdots,e_{2N}(t_0)\}$ corresponding to the (not necessarily distinct) eigenvalues $\mu^1_k,\mu^2_k, \hdots,\mu^{2N}_k$.

An arbitrary initial condition $x(t_0)$ can be written in terms of this eigenbasis as $x(t_0)=\sum_{s=1}^{2N} c_s e_s(t_0)$.
The general solution $x(t)$ is then given by 
\beqn
\begin{aligned}
x(t)
=\mathcal{O}(t,t_0)x(t_0)
&=\sum_{s=1}^{2N}c_sP(t,t_0)\exp[(t-t_0)\Lambda(t_0)]e_s(t_0)\\
&= \sum_{s=1}^{2N}c_sP(t,t_0)e_s(t_0)e^{\mu_k^s(t-t_0)},\\
\end{aligned}
\eeqn
where in the last step we used $\Lambda(t_0)e_s(t_0)=\mu_k^s e_s(t_0)$. A solution in {\it Floquet form} is
\beqn
\begin{aligned}
x(t)
&=\sum\limits_{s=1}^{n}c_s\mathcal{P}_s(t,t_0)e^{\mu_k^s (t-t_0)},\quad \textrm{with}\quad
\mathcal{P}_s(t,t_0)&=P(t,t_0)e_s(t_0).
\end{aligned}
\eeqn
Note that $\mathcal{P}_s(t,t_0)$ is a column vector with period $T$ for each $s$.  From the above form of the solution we can now see that we get exponentially growing solutions iff at least one of the eigenvalues $\mu_k^s$ of $\Lambda(t_0)$ satisfies $\Re[\mu_k^s]>0$. The coefficients $c_s$ contain all the necessary information about the operator-valued coefficients for the quantum problem. 

Note that we can construct the entire solution, including $P(t,t_0)$, $e_s(t_0)$, and $\mu_k^s$, from $\mathcal{O}(t,t_0)$ evaluated on a single period $0\le t -t_0\le T$. To see this, recall that $\mu_k^s$ and $e_s(t_0)$ are the eigenvalues and eigenvectors of $\Lambda(t_0)= T^{-1}\ln \mathcal{O}(t_0+T,t_0)$. Using $O(t,t_0)$ for $0\le t -t_0\le T$, we can evaluate $P(t,t_0)=\mathcal{O}(t,t_0)\exp\left[-(t-t_0)\Lambda(t_0)\right]$. Since $P(t,t_0)=P(t+T,t_0)$, we have $P(t,t_0)$, and hence $\mathcal{P}_s(t,t_0)$, for all time. 

 \subsubsection{Calculating Floquet exponents: A simple algorithm}
 \label{sec:algo}
 Based on the above analysis, we describe a simple algorithm to determine the Floquet  exponents. Of particular importance is whether there exist exponentially growing Floquet solutions.
\begin{enumerate}
\item Find the period $T$ of the system from $U(t)$.
\item Solve $\partial_t \mathcal{O}(t,t_0)=U(t)\mathcal{O}(t,t_0)$ from $t_0$ to $t_0+T$ to obtain $\mathcal{O}(t_0+T,t_0)$.
\item Diagonalize $\mathcal{O}(t_0+T,t_0)$ to obtain the (in general complex) eigenvalues $o^s_k=|o_k^s| e^{i\theta_k^s}$. Since $O(t_0+T,t_0)=\exp[T\Lambda(t_0)]$, the Floquet exponents are given by
\beq
\mu_k^s=\frac{1}{T}[\ln |o^s_k|+i\theta^s_k].
\eeq
\item We have exponentially growing solutions if for any $s$,
\beq
\Re[{\mu_k^s}]=\frac{1}{T}\ln |o^s_k|>0.
\eeq
\end{enumerate}

\subsection{Worked examples}
We will apply the above algorithm to calculate the Floquet exponents for two simple examples.

\subsubsection{Self-resonance}
\label{Self-resonance}
For the case in which we have a single inflaton without couplings to other fields, the equation of motion of the inflaton fluctuations (with $M_{\rm pl}\rightarrow \infty$) is given by
\beq
\delta \ddot{\phi}_\bk+\left[k^2+V_{,\phi\phi} (\varphi)\right]\delta\phi_\bk=0.
\label{deltaphisinglefield}
\eeq
For $V(\phi)= {1 \over 2} m_\phi^2\phi^2$, Eq. (\ref{deltaphisinglefield}) is that of a simple harmonic oscillator with a {\it time-independent} frequency, $\omega_k^2=k^2+m_\phi^2$. However, if the potential has nonlinearities, and if the field is oscillating, the frequency becomes periodic and time-dependent, a scenario in which fluctuations can grow exponentially via parametric resonance. This is often called ``self-resonance." Such a phenomenon is important for all inflationary models in which the self-coupling is significant compared to the coupling to other fields (see for example, Refs. \refcite{Khlopov:1985,Amin:2011hj}). For long wavelengths, the Floquet exponent can be derived analytically as $\mu_k^{\pm} = \pm \, i\, c_s\, k$, where $c_s$ is a sound speed associated with the time averaged pressure of the background\cite{Johnson:2008se,Hertzberg:2014iza}. For arbitrary wavelengths, the Floquet exponent requires a full numerical analysis, as we now describe.

To calculate the growth rate of the instabilities, we first get the equations of motion in first-order matrix form. Given $\delta\pi_\bk=\delta\dot{\phi}_\bk$, we have $x(t)=\left[\delta\phi_\bk,\delta\pi_\bk\right]^T$, and Eq. (\ref{deltaphisinglefield}) becomes $\partial_t x(t)=U(t)x(t)$ where
\beq
U(t)=\left(\begin{array}{cc}
0 & 1\\
-k^2-V_{, \phi \phi} (\varphi) & 0
\end{array}  \right).
\eeq
We now follow the steps described in the algorithm above to find the Floquet exponents for this single-field scenario. Note that this can be adapted to the first example discussed in this section (see Eq. \eqref{phichiPot}) with the replacement $V_{, \phi \phi} (\varphi)\rightarrow g^2\varphi^2$.
\begin{enumerate}
\item First we need the period $T$ of $U$. The period of $U(t)$ will depend on the initial amplitude of the homogeneous field $\varphi(t_0)$ (assuming $\partial_t\varphi(t_0)=0)$ and is given by
\beq
T(\varphi_{\rm max})=2\int_{\varphi_{\rm min}}^{\varphi_{\rm max}}\frac{d\varphi}{\sqrt{2V(\varphi_{\rm max})-2V(\varphi)}}.
\eeq
Usually we will end up specifying either $\varphi(t_0)=\varphi_{\rm max}$ or $\varphi_{\rm min}$. The other can be found by solving $V(\varphi_{\rm min})=V(\varphi_{\rm max})$. For a symmetric potential with $V(\varphi)=V(-\varphi)$, we have $\varphi_{\rm max}=-\varphi_{\rm min}=\varphi(t_0)$. One can also find the period of $U(t)$ by solving the equation of motion of the background field, $\ddot{\varphi}+V_{, \phi} (\varphi)=0$.
\item Next we need to solve $\partial_t \mathcal{O}(t,t_0)=U(t)\mathcal{O}(t,t_0)$ from $t_0$ to $t_0+T$ to obtain $\mathcal{O}(t_0+T,t_0)$. Explicitly we wish to obtain
\beq
\mathcal{O}(t_0+T,t_0)=\left(\begin{array}{cc}
\delta\phi^{(1)}_k(t_0+T) & \delta\phi^{(2)}_k(t_0+T) \\
\delta\pi^{(1)}_k(t_0+T) & \delta\pi^{(2)}_k(t_0+T) 
\end{array}  \right),
\eeq
where the initial conditions $\mathcal{O}(t_0,t_0)=\id$. Note that the superscripts represents two sets of solutions, not different fields. This is of course equivalent to solving Eq. (\ref{deltaphisinglefield}) for the two set of initial conditions, $\{\delta\phi^{(1)}_k(t_0)=1, \delta\dot{\phi}^{(1)}_k(t_0)=0\}$, and $\{\delta\phi^{(2)}_k(t_0)=0, \delta\dot{\phi}^{(2)}_k(t_0)=1\}$, from $t_0$ to $t_0+T$. We have suppressed the dependence of $T$ on $\varphi(t_0)$ to reduce clutter.
\item Now we need to find the eigenvalues of $\mathcal{O}(t_0+T,t_0)$. Explicitly, these are
\beqn
o^{\pm}_k=\frac{\delta\phi^{(1)}_k +\delta\pi^{(2)}_k}{2}\pm\frac{\sqrt{\left\{\delta\phi^{(1)}_k -\delta\pi^{(2)}_k\right\}^2+4\delta\phi_k^{(2)}\delta\pi_k^{(1)}}}{2},
\eeqn
where all the quantities are evaluated at $t_0+T$. 
\item The real parts of the Floquet exponents are given by 
\beq
\Re[\mu^{\pm}_k]=\frac{1}{T}\ln |o^\pm_k|.
\eeq
If $\Re[\mu^{\pm}_k]> 0$, then we have exponential growing solutions. For the case above, it is easy to check that $\mu^+_k+\mu^-_k=0$. Hence, if one of the solutions is growing, the other is always decaying. In Fig. \ref{FloquetMonodromy} we plot the Floquet bands for the monodromy model\cite{Silverstein:2008sg,McAllister:2014mpa,Amin:2011hj}, with the potential $V(\phi)=m^2M^2\left[\sqrt{1+(\phi/M)^2}-1\right]$ where $M$ is the scale where the potential changes from quadratic to linear. 

\begin{figure}[t!]
\centering{
\includegraphics[width=5.20in]{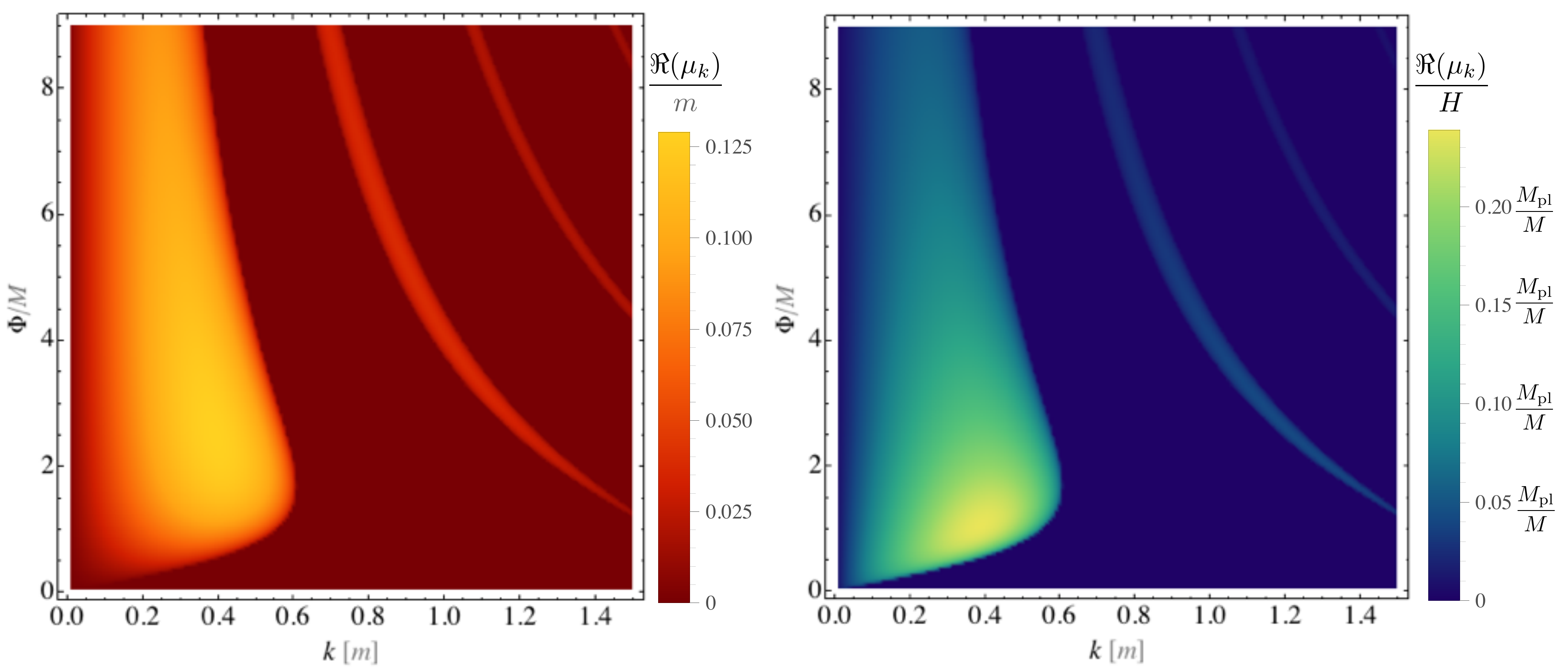}
\caption{\small Left: A plot of the instability band structure for the field fluctuations $\delta\phi_\bk$ in the case of self-resonance in the model $V(\phi)=m^2M^2\left[\sqrt{1+(\phi/M)^2}-1\right]$ (see Ref.\cite{Amin:2011hj}). $\Phi$ is the amplitude of oscillation of the background  inflaton field and $k$ is the wavenumber of the fluctuation. The color represents the real part of the Floquet exponent rescaled by the mass $m$: $\Re(\mu_k) / m$. Right: Same as the left plot, but now the Floquet exponent is scaled  by the instantaneous Hubble parameter $\Re(\mu_k)/H$, where $H$ is calculated using $H^2=V(\Phi) / (3 M_{\rm pl}^2)$. In an expanding universe, the exponential growth is counteracted by expansion. Heuristically,  $\Re(\mu_k)/H\gg 1$ is a strong indicator of rapid growth of fluctuations. For this model, such growth is possible for $M \ll M_{\rm pl}$. } 
\label{FloquetMonodromy} }
\end{figure}

\end{enumerate}

\subsubsection{$O(N)$ symmetric potential}
\label{ParallelPerpendicular}

Multifield models can often lead to background motion that is not periodic. However, there exist multifield scenarios in which periodicity at the end of inflation is guaranteed. This occurs for example, when the Lagrangian carries an $O(N)$ symmetry among the fields as examined by Ref. \refcite{Hertzberg:2014iza}. In this case, inflation generally redshifts away any angular motion in field space, leaving only radial motion as the attractor solution.\cite{Greenwood:2012aj}  Let us consider a scenario with a trivial field space metric $\mathcal{G}_{IJ}=\delta_{IJ}$ and an $O(N)$ symmetric potential, $V(\vec{\phi})=V(|\vec{\phi}|)$, with a minimum at $\vec{\phi}=\vec{0}$. We take the radial motion, without loss of generality, to be along the $\varphi^N$ direction, which we label $\sigma$. In this case, the field perturbations parallel to the direction of motion of the homogeneous field satisfy the following equation:
\beq
\delta\ddot{\sigma}_\bk+[k^2+V''(\sigma)]\delta\sigma_\bk=0.
\eeq
On the other hand, the field perturbations perpendicular to the direction of motion satisfy
\beq
\delta\ddot{s}^I_\bk+\left[ k^2+\frac{V'(\sigma)}{\sigma} \right]\delta s^I_\bk=0 ,
\eeq
where $I\ne N$. Note that the parallel and perpendicular components decouple in this scenario, because inflation drives the background motion to be radial. (In models that break the symmetry, more complicated evolution among the $\varphi^I$, including trajectories that turn in field space, will generically couple the $\delta \sigma$ and $\delta s^I$ perturbations.\cite{Gordon:2000hv,Bassett:2005xm,Langlois:2008mn,Peterson:2011yt,Kaiser:2012ak})

For each of these uncoupled components, we may now apply our algorithm to calculate the Floquet exponents, with the important feature that the perpendicular components have a different ``auxiliary potential".\cite{Hertzberg:2014iza} The periodic $U$ matrices for these two scenarios are given by
\beq
U_{\sigma}(t)=\left(\begin{array}{cc}
0 & 1\\
-k^2-V''(\sigma) & 0
\end{array}  \right), \qquad U_{s}(t)=\left(\begin{array}{cc}
0 & 1\\
-k^2-\frac{V'(\sigma)}{\sigma} & 0
\end{array}  \right).
\eeq
As discussed in Refs. \refcite{Hertzberg:2014iza} and \refcite{Lozanov:2014zfa}, in some cases this leads to complementarity between the most unstable, low-momentum modes: either the components parallel to the direction of motion are significantly resonant or the components perpendicular to the direction of motion, but not both. By performing a long wavelength fluid analysis in Ref. \refcite{Hertzberg:2014iza}, this complementarity and stability structure is derived from the pressure that governs the adiabatic mode $\delta\sigma$, and a type of ``auxiliary pressure" that governs the isocurvature modes $\delta s^I$. Alternatively, this can be understood in terms of the attraction/repulsion of the underlying many particle quantum mechanics of bosons\cite{Hertzberg:2014jza}.

We emphasize that although the examples in Sections \ref{Self-resonance} and \ref{ParallelPerpendicular} involve effectively single-field examples, the method described in Sections \ref{FloquetTheorem} - \ref{sec:algo} is valid quite generally, and applies to multifield models with non-canonical kinetic terms as long as the background motion of the fields is periodic. 

Let us now re-introduce the effects of expansion by relaxing the assumption that $M_{\rm pl}\rightarrow \infty$. In an expanding universe, the growth of perturbations in counteracted by the expansion. Parametric resonance results in significant growth only if the growth rate of fluctuations is much larger than the expansion rate, as in Fig. \ref{FloquetMonodromy},  
\begin{equation}
\frac{\Re(\mu_k)}{H}\gg 1 ,
\label{mukH}
\end{equation}
for a sufficiently long time. One should imagine passing though Floquet bands as the wavenumbers as well as the field amplitudes redshift. If the  condition of Eq. (\ref{mukH}) is satisfied for a sufficiently long time, the perturbations eventually grow large enough to enter the fully nonlinear regime.  Mode-mode coupling and other forms of nonlinear interactions begin to dominate, which can transfer power between modes of different wavenumbers.\cite{Khlebnikov:1996mc,Khlebnikov:1996wr,Khlebnikov:1996zt,Prokopec:1996rr,Kofman:1997yn,Bassett:1999mt,Frolov:2010sz,Levasseur:2010rk}.  To address the behavior of fields after such mode coupling begins, we need to turn to a nonlinear analysis and numerical simulations, and will be discussed in Section \ref{NonlinearSection}. 


\section{Nonlinear Effects}
\label{NonlinearSection}

As emphasized in Section \ref{PreheatingSection}, in a large class of models the homogenous oscillations of the inflaton(s) lead to rapid growth of spatially varying perturbations via parametric or tachyonic resonance. However, such growth cannot proceed forever. It is eventually shut off due to backreaction of perturbations on the homogeneous fields. Such backreaction leads to a fragmentation of the homogeneous inflaton(s), and the subsequent evolution of the combined inflaton-daughter fields system is dynamically rich and a potential source of observational signatures. In this Section we focus on the period after the initial burst of particle production but before thermalization.

\subsection{Numerical simulations}

The dynamically rich behavior of the inflaton and daughter fields during the nonlinear phase makes numerical simulations invaluable. Typically it is necessary to solve the coupled system of fields, including gravity, on a lattice, subject to 
\begin{equation}
\begin{aligned}
\nabla^\mu\nabla_\mu\phi^I + \Gamma^I_{\> LJ} \partial_\mu \phi^L \partial^\mu \phi^J &= {\cal G}^{IK} V_{, K} (\phi^J),\\
G^{\mu}_{\> \nu}(g_{\mu\nu})&= \frac{1}{M_{\rm pl}^2} T^{\mu}_{\> \nu} (\phi^J, g_{\mu\nu} ),
\end{aligned}
\end{equation}
where $\nabla_\mu$ is a covariant spacetime derivative. Some fields $\phi^J$ may be important during inflation, and some may act as daughter fields into which the inflaton transfers its energy after inflation. There can be non-canonical kinetic terms even beyond the ansatz ${\cal G}_{IJ} \neq \delta_{IJ}$, and the daughter fields do not have to be scalars (unlike the expressions above).

A number of publicly available computer programs already exist for evolving (mostly) scalar fields on a lattice in an expanding universe. A limited number of them include the calculation of metric perturbations, and even fewer include the backreaction of the metric perturbations on the field dynamics. We list a few of them below. Each comes with its pros and cons, and the choice depends on the user's familiarity with the programming language used as well as the nature of the problem at hand.
\begin{itemize}
\item Lattice Easy (Ref. \refcite{Felder:2000hq}) is perhaps the most widely used and has detailed documentation. It is a finite-difference code, with a possibility of running over multiple machines. 
\item Defrost (Ref. \refcite{Frolov:2008hy}) is a finite-difference code. It has sophisticated templates for spatial derivatives and has excellent energy conservation.
\item PSpectre (Ref. \refcite{Easther:2010qz}) is a pseudo-spectral code, unlike the previously mentioned finite-difference codes.
\item HLattice (Ref. \refcite{Huang:2011gf}) includes metric perturbations and their backreaction on the scalar fields.
\item GABE (Ref. \refcite{Child:2013ria}) can simulate fields with noncanonical kinetic terms.
\item CudaEasy (Ref. \refcite{Sainio:2009hm}) is a GPU accelerated lattice code for cosmological scalar field evolution. It has the potential to significantly reduce the time required for detailed, long-time simulations.
\item PyCool (Ref. \refcite{Sainio:2012mw}) is a Python based, GPU accelerated lattice code with symplectic integrators.
\end{itemize}

We briefly consider the zoo of possible phenomena that can occur during the nonlinear phase, and their implications. 

\subsection{Nonlinear dynamics in single-field models}

In the simplest case, consider an effectively single-field model in which the inflaton is very weakly coupled to other fields, such that other fields may be neglected during the period of interest. Even for a single massive inflaton with no self-couplings, with $V (\phi) = {1 \over 2} m_\phi^2 \phi^2$, homogenous oscillations of the inflaton do not remain stable indefinitely. Gravitational interactions eventually lead to the formation of nonlinear structures, akin to the gravitational instability in pressureless matter in the late universe. This fragmentation has been explored in detail.\cite{Jedamzik:2010dq,Jedamzik:2010hq,Easther:2010mr}

 \begin{figure}[t!] 
   \centering
   \includegraphics[width=2in]{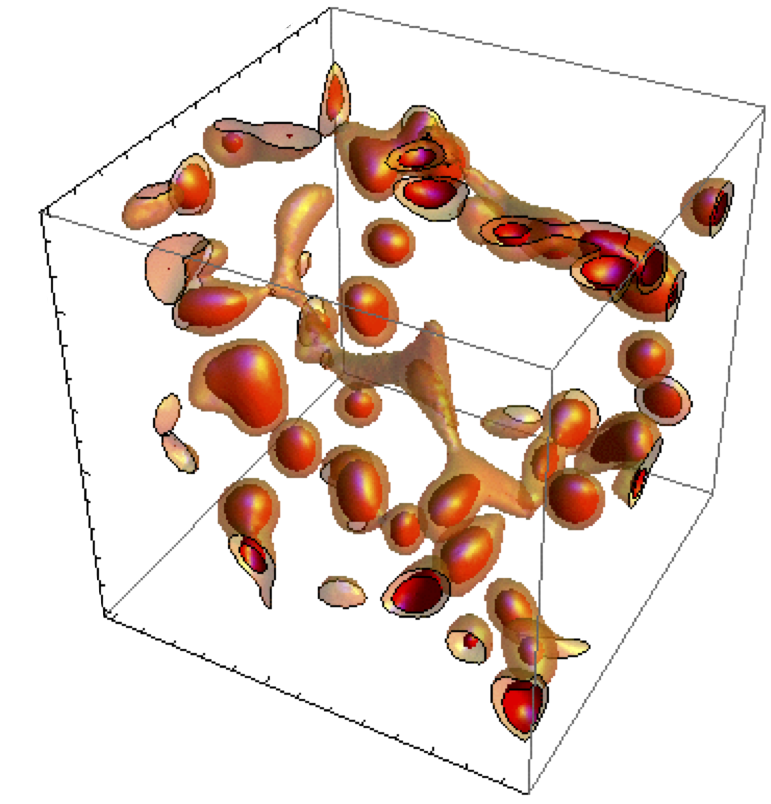}
    \includegraphics[width=2in]{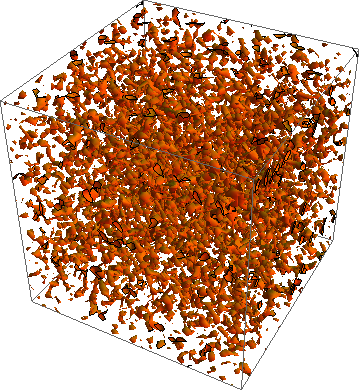}
   \caption{Left: Self interactions of the inflaton can lead to fragmentation and soliton formation in the inflaton field at the end of inflation. The plot shows soliton (oscillon) formation after inflation where the inflaton potential flattens away from the minimum\cite{Amin:2011hj}. Right: Fragmentation in a model where the inflaton is governed by a quadratic potential, and  is coupled to a daughter field through a quartic interaction term $g^2\phi^2\chi^2$ [figure on the right, courtesy K. Lozanov]. The surfaces are iso-density surfaces (several times the average density). In both cases the size of the box is smaller than $H^{-1}$ at that time.}
   \label{fig:oscillons}
\end{figure}

If the inflaton potential includes self-interactions,
\begin{equation}
V(\phi) = \frac{1}{2} m_\phi^2\phi^2 + \frac{\lambda_3}{3} \phi^3 + \frac{ \lambda_4}{4} \phi^4+\hdots,
\end{equation}
the oscillating inflaton can fragment on a much faster timescale compared to the gravitational one. Such self-interactions are present in all but the simplest models, and depending on their form they can lead to complex, nonlinear phenomena. 

In a class of models in which the potential opens up away from the minimum, such fragmentation can lead to the formation of soliton-like configurations known as ``oscillons."\cite{Bogolyubsky:1976yu, Gleiser:1993pt, CopGle95, Amin:2010dc, Amin:2011hj, Salmi:2012ta} (See Fig. \ref{fig:oscillons}.) Oscillons can dominate the energy density of the universe for a large class of of observationally consistent models (for example, see Ref. \refcite{Amin:2011hj}). Oscillons eventually decay away\cite{Hertzberg:2010yz,Salmi:2012ta}, leading to a radiation-dominated universe. In models in which the scalar field is complex, one can also get nontopological solitons called Q-balls. \cite{Coleman:1985ki,Lee:1991ax}. Oscillons and Q-balls could play an important role in baryogenesis (see, for example, Refs. \refcite{Kasuya:2001hg,Lozanov:2014zfa}), generate high-frequency gravitational waves \cite{2010PhRvD..81h3503C,Zhou:2013tsa}, change the expansion history or delay thermalization. Along with self-interactions, non-canonical kinetic terms can also lead to  nontrivial dynamics during this phase. \cite{Child:2013ria,Amin:2013ika}

\subsection{Nonlinear dynamics in multifield models}
In models in which the inflaton's couplings to other fields dominate the inflaton's self-couplings, the nonlinear evolution of the system often leads to the formation of temporary bubble-wall-like structures which collide and fragment further.\cite{Felder:2006cc,GarciaBellido:2002aj} In most cases the initial structures have coherence on large spatial scales (still smaller than the horizon at that time), and subsequent fragmentation and evolution tends to transfer momentum to higher and higher momenta \cite{Greene:1997fu,GarciaBellido:2002aj}. In certain cases, multifield models can also lead to the formation of defects and solitons. \cite{Copeland:2002ku,Gleiser:2010,Gleiser:2011xj,Gleiser:2014ipa}
 
\begin{figure}[t!] 
   \centering
 \includegraphics[width=2in]{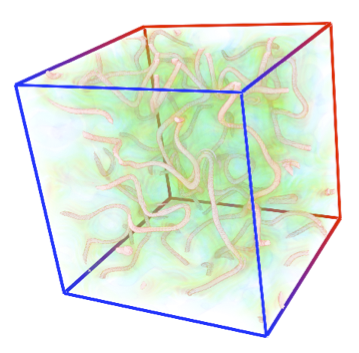}
  \caption{When gauge fields are coupled to the inflaton, defect-like configurations (strings) tend to form at the end of inflation. Shown here is the magnetic field density associated with Abelian gauge fields soon after the end of inflation. (From Ref. \cite{Dufaux:2010cf}.) }
   \label{fig:strings}
\end{figure}

Models that include more than just scalar fields can also lead to new phenomenology at the end of inflation. A number of authors have considered lattice simulations of reheating involving Abelian gauge fields\cite{Mazumdar:2008up,Dufaux:2010cf,Deskins:2013dwa} and non-Abelian gauge fields\cite{GarciaBellido:2003wd} (the latter in a non-expanding background). In such models, gauge fields can lead to the formation of defects such as cosmic strings at the end of inflation.\cite{Dufaux:2010cf} (See Fig. \ref{fig:strings}.) The coupling of the inflaton to gauge fields can also generate magnetic fields after inflation (as discussed below in Section \ref{MagneticSubsection}). Finally, gauge fields may also speed up the transition to a radiation-dominated universe\cite{Deskins:2013dwa}.

\section{Thermalization}
\label{ThermalizationSection}
In the last major stage of reheating, the universe achieves a radiation-dominated state in thermal equilibrium at some reheat temperature $T_{\rm reh}$. The process is governed by out-of-equilibrium quantum field theory.\cite{Boyanovsky:1995ema,Boyanovsky:1996rw,Boyanovsky:1996xx,PhysRevA.81.033611,Berges:2013eia,PhysRevLett.108.161601,Berges:2008wm} 

The reheat temperature, $T_{\rm reh}$, governs several important phenomena, such as the fate of dangerous relics, phenomena associated with nonequilibrium effects such as out-of-equilibrium production of heavy particles and topological defects, nonthermal phase transitions, and gravitational waves. \cite{PhysRevLett.93.142002}. Most important, the Standard Model degrees of freedom (at least) must attain complete thermalization before the start of big bang nucleosynthesis, which places a lower bound of $T_{\rm reh} \geq 1$ MeV.

\subsection{Conditions for thermalization}
To reach complete thermalization, two conditions must be satisfied: (1) the system attains a nearly constant equation of state, with $w = p / \rho \sim 1/3$, and hence the universe is radiation dominated; and (2) the system reaches Local Thermal Equilibrium (LTE). Criterion (1), sometimes dubbed ``prethermalization," can occur much earlier than (2).\cite{PhysRevLett.93.142002,Mazumdar:2013gya} 

The condition of reaching LTE itself entails two separate requirements: (2a) kinetic equilibrium, which ensures that the momentum distribution of the particles maximizes the entropy, and (2b) chemical equilibrium, which ensures the stability of different species of matter interacting with each other.\cite{Davidson:2000er,Enqvist:1990dp}. For weakly interacting particles, kinetic equilibrium implies that the particles in the system will be close to the Bose-Einstein distribution (for bosons) or a Fermi-Dirac distribution (for fermions).

In order to achieve thermalization, the inflaton must complete its decay. If a massive inflaton remains, then the universe may become matter-dominated before nucleosynthesis. For example, for a coupling of the form $g^2 \phi^2 \chi^2$, scattering becomes inefficient due to the expansion of the universe and the number of inflaton quanta becomes constant. One way to achieve complete decay of the inflaton is to introduce three-leg interactions, such as $\phi \chi^2$ or a Yukawa interaction with fermions, $\phi \bar{\psi} \psi$.\cite{Podolsky:2005bw,Mazumdar:2013gya}. Hence perturbative decays of the inflaton remain critical to the process of reheating, even if they are overshadowed at early stages by nonperturbative resonances.

\subsection{The stages of thermalization}

Thermalization proceeds in stages.\cite{Son:1996uv,Micha:2002ey,Micha:2004bv,Podolsky:2005bw} These stages are associated with different time-scales, in addition to the usual time-scales of Hubble expansion, $H^{-1}$, and the inflaton oscillation time, $m_\phi^{-1}$. For example, for a simple model with $V (\phi, \chi) = {1 \over 2} m_\phi^2 \phi^2 + {1 \over 2} g^2 \phi^2 \chi^2$, one finds four distinct regimes\cite{Podolsky:2005bw}:

\begin{itemize}
\item {\it Preheating}: The duration of this first phase, dominated by parametric resonance, is typically of order $\delta t_1 \sim 100 \> m_{\phi}^{-1}$ in simple models.
\item {\it Nonlinear dynamics and chaos}: At the end of preheating, a short and violent stage occurs in which nonlinear effects evolve in a chaotic way, erasing details of initial conditions. The duration is typically of order $\delta t_2 \sim 10 \> m_{\phi}^{-1}$.
\item {\it Turbulent regime}: The spectrum of fluctuations cascades toward both ultraviolet and infrared modes on a time-scale $\delta t_3$ which is longer than either $\delta t_1$ or $\delta t_2$. 
\item {\it Thermalization}: The last stage is characterized by particle fusion and off-shell processes. The spectrum relaxes to a thermal distribution on a time-scale $\delta t_4$ which is the longest of the four stages.
\end{itemize}

As we found in Section \ref{ParametricResonanceSection}, preheating yields a highly nonthermal spectrum. From Eq. (\ref{Eknk}), we find that the number density of created particles grows as $n_k \sim e^{2 \mu_k t}$ for modes that lie within a resonance band. Parametric resonance is most efficient for long-wavelength modes, and thus the particle-number distribution is strongly peaked in the infrared at this early stage. Next comes a stage marked by strongly nonlinear dynamics, as described in Section \ref{NonlinearSection}. Particle occupation number is ill-defined at this stage, and a nonlinear wave description is more appropriate.\cite{Podolsky:2005bw}  Note that while the stage can be short in the particular model discussed above, in many cases metastable objects such as oscillons and Q-balls can emerge during this stage and significantly extend the time-scale of this stage.

The turbulent regime begins with a phase of driven turbulence, in which the system is driven by the energy in the infrared modes of the inflaton. When the energy stored in the inflaton condensate drops below the energy stored in the created particles, the evolution of the system transitions from driven to free turbulence. During the turbulence regime, the particle-number distributions smooth out and begin evolving toward higher comoving momenta (at a much slower rate, $\delta t_3 \gg \delta t_1, \delta t_2$). The spectra in the infrared approach a saturated power-law state, which then slowly propagates toward the ultraviolet. Although one may observe a greater tendency toward equilibrium distributions among the infrared modes (for which the rescaled spectra are closer to flat), the overall distributions remain typical of the turbulent regime and are far from thermal.\cite{Micha:2002ey,Micha:2004bv,Podolsky:2005bw} 

The turbulence regime has been studied analytically and with lattice simulations in Refs. \refcite{Micha:2002ey,Micha:2004bv}. They find that the occupation number is characterized by self-similar evolution. For example, for a $\lambda \phi^4$ model,  they find
\beq
n_k (\tau) = \tau^{-q} n_0 (k\tau^{-p})
\eeq
where $\tau= \eta / \eta_c$ is the rescaled conformal time and $n_0(k)$ is the distribution function at some late time $\eta_c$, chosen during the self-similar regime. The best numerical fits correspond to $q \sim 3.5p$ and $p \sim 1/5$.\cite{Micha:2002ey,Micha:2004bv}

Finally, during the last stage of free turbulence, the front of the distribution propagates into the ultraviolet until it relaxes to a Bose-Einstein (or Fermi-Dirac) spectrum. At that stage, quantum effects dominate over thermal fluctuations, and the system is commonly taken to have achieved thermal equilibrium. Despite these promising recent studies, however, an understanding of the entire thermalization process remains incomplete, and deserves further study. In particular, given any particular model, it remains an open challenge to trace the evolution of the system through each of the four major stages and compute a robust, equilibrium reheat temperature, $T_{\rm reh}$.

\section{Particle Physics Models}
\label{ParticleModelsSection}

At the very high energies relevant to (p)reheating, the full set of degrees of freedom and interactions remains unknown. The governing particle theory could involve many degrees of freedom beyond the Standard Model. Indeed, the inflaton itself might consist of multiple fields with noncanonical dynamics. In this Section, we provide an overview of some of the various possibilities that have been explored in the literature.

\subsection{Multifield inflation}
\label{MultifieldSubsection}

Let us consider the case of inflation governed by $N$ scalar fields, $\phi^I$, each of which couples to the same daughter field, $\chi$, with the potential
\beq
V (\phi^I , \chi) = U (\phi^I ) + \frac{1}{2} \sum_I g_I^2 (\phi^I )^2 \chi^2 ,
\label{VUmulti}
\eeq
where $U (\phi^I)$ only involves the $N$ inflaton fields, and for now we assume that each coupling $g_I^2 > 0$. One could include higher-dimension operators, but there is a consistent power-counting scheme in which such higher-order terms are subdominant at the end of inflation, when the inflaton field values are small. In particular, at the end of inflation it is often sufficient to approximate the potential $U (\phi^I)$ by Taylor expanding around its minimum to quadratic order,
\beq
U (\phi^I ) = \frac{1}{2} \sum_I m_I^2 (\phi^I )^2 + . . .
\label{Uexpand}
\eeq
Eq. (\ref{Uexpand}) would obviously need to be modified for massless inflaton fields.

For couplings between $\phi^I$ and $\chi$ as in Eq. (\ref{VUmulti}), and with unequal inflaton masses $m_I$, the effective mass of the $\chi$ field will rarely pass through zero. The parametric resonance for fluctuations $\delta \chi_{ \bf k}$ will therefore be less efficient than in the models considered in Section \ref{PreheatingSection}. Indeed, if there are many inflaton fields, as in models like ``N-flation" \cite{Liddle:1998jc,Copeland:1999cs,Jokinen:2004bp,Dimopoulos:2005ac}, then there is considerable reduction in the efficiency of preheating, related to the de-phasing of the pump fields $\varphi^I (t)$\cite{Battefeld:2008bu,2009PhRvD..79l3510B}(however, see Refs. \refcite{Ashoorioon:2009wa,2012JCAP...11..062B}).

On the other hand, preheating in multifield models could become more efficient than in simple models by incorporating other types of couplings. Couplings beyond the simple $g^2 \phi^2 \chi^2$ form --- such as $\phi^n \chi$ or $\phi \chi^n$, with $n = 2, 3$ --- arise in the low-energy effective actions for supersymmetric and supergravity models, and such couplings produce very efficient resonances.\cite{Bassett:1999ta,Dufaux:2006ee,DeMelo:2001nr} Likewise, even for couplings of the form $g_I^2 (\phi^I)^2 \chi^2$, if at least one coupling $g_I^2 < 0$, then the negative-coupling instability can drive efficient, broad-resonance preheating.\cite{Greene:1997ge,Bassett:1999ta} (The potential for such models will be well-behaved for large field values if one includes quartic self-couplings for the fields.)

 \subsection{Higher-spin daughter fields}
 
 It is quite important to consider daughter fields that are not simply scalars, since it is reasonable to assume that the inflaton will couple to various degrees of freedom, including fermions and gauge bosons. 
 
 \subsubsection{Fermions}
 
 We begin by considering the possibility that the daughter species consists of spin-1/2 particles.
 For example, a Yukawa interaction between the inflaton $\phi$ and a fermion $\psi$ of the form
 \beq
 \Delta\mathcal{L} = y\, \phi\, \bar{\psi}\,\psi
 \eeq
 allows the inflaton to decay into fermion/anti-fermion pairs. The usual expectation is that this process is inefficient due to ``Pauli blocking,"
 wherein only one fermion can occupy a single mode. However it has been shown that this can still lead to a form of parametric resonance, since the inflaton can decay into a wide band of wavenumbers \cite{Greene:1998nh,Greene:2000ew,Peloso:2000hy,Tsujikawa:2000ik}. So although each mode cannot grow substantially, nevertheless, many modes can be excited. This still leads to an enhanced decay compared to standard perturbative decays. For reheating into higher-spin fermions see, for example, Refs. \refcite{Maroto:1999ch,Kallosh:1999jj}.

 \subsubsection{Gauge bosons}

Another natural possibility is to couple the inflaton to spin-1 fields. We assume that these are massless spin-1 fields to avoid complications in the ultraviolet, that is, we consider gauge bosons. This leads to two natural options. The first is if the inflaton is a gauge singlet. We can then couple $\phi$ to a gauge field $A_\mu$ through higher dimension operators of the form
\beq
\Delta\mathcal{L} = -W(\phi) F_{\mu\nu}F^{\mu\nu} ,
\eeq
where $W(\phi)$ may be linear or quadratic in $\phi$. Resonances are especially efficient if the fields are coupled with such a conformal factor.\cite{Deskins:2013dwa} 

The second possibility arises if the inflaton is charged \cite{Davis:2000zp,Braden:2010wd}. If we take $\phi$ to be a complex scalar field under an Abelian $U(1)$ symmetry, then the standard kinetic term takes the form
\beq
\Delta\mathcal{L}=-|D_\mu\phi|^2 = -|\partial_\mu\phi|^2+ig(\phi\,\partial_\mu\phi^*-\phi^*\partial_\mu\phi) A^\mu
+g^2|\phi|^2A_\mu A^\mu .
\eeq
The final term is naturally reminiscent of the toy-model interaction $g^2\phi^2\chi^2$ that we examined in Section \ref{PreheatingSection}, but with the scalar field $\chi$ replaced by a gauge field. Since spin-1 fields are bosonic this can lead to significant parametric resonance. Further interesting possibilities naturally emerge if we charge $\phi$ under a non-Abelian symmetry.\cite{GarciaBellido:2003wd,Allahverdi:2011aj}
 
 \subsection{Higher-derivative interactions}
 
In Section \ref{MultifieldSubsection} we considered a standard power-counting scheme in which higher-derivative operators are suppressed by some high cutoff. One may instead consider an alternative power-counting scheme in which higher-derivative terms do not remain so strongly suppressed. A well-known example is so-called ``DBI inflation,"\cite{Alishahiha:2004eh} in which the kinetic term for the inflaton takes the form
\beq
\Delta\mathcal{L} = M^4\sqrt{1-(\partial\phi)^2/\Lambda^4} ,
\eeq
where $\Lambda$ sets the cutoff on the field theory with respect to the Minkowski vacuum. (The effective cutoff will be altered in the inflationary de Sitter background.) Similar to having a nonlinear potential $V(\phi)$, this can lead to self-resonance. It was claimed that this can improve the efficiency of preheating in Ref.~\refcite{Zhang:2013asa}, but contrary results were found in Ref.~\refcite{Karouby:2011xs}.
 
 Another possibility is to couple $\phi$ to a daughter field with derivative interactions. For example, if $\phi$ carries some type of shift symmetry, then the usual $g^2\phi^2\chi^2$ will be forbidden, but a term of the form
 \beq
 \Delta\mathcal{L} = -{1\over 2 F^2}(\partial\phi)^2\chi^2
 \eeq
 can be allowed. Although this is a dimension-6 operator, it can still lead to appreciable parametric resonance. This is obvious when $\phi$ is exhibiting standard harmonic oscillation with frequency $m_\phi$, leading to 
 \beq
 \Delta\mathcal{L} \approx-{m_\phi^2\over 2 F^2}\phi^2\chi^2 .
 \eeq
 Hence such derivative interactions can take on the form of standard couplings due to the coherent behavior of the inflaton.

``Natural inflation" presents another possibility, in which $\phi$ is a pseudo-scalar Goldstone boson arising from the spontaneous breaking of an axial symmetry, i.e., a type of ``axion." Since the axion carries an almost exact shift symmetry, it can only couple to total derivatives. This leads to couplings of the form
\beq
\Delta\mathcal{L} = {1\over F}\phi\, \varepsilon^{\mu\nu\alpha\beta}F_{\mu\nu}F_{\alpha\beta}
\eeq
where $\varepsilon^{\mu\nu\alpha\beta}$ is the totally anti-symmetric tensor. This model of inflation is perhaps one of the most compelling as it is dictated almost entirely by symmetry and symmetry breaking. It is interesting that it also allows for reheating and may even give rise to gravity waves \cite{Barnaby:2011qe}.

 \subsection{Standard Model and beyond}

A complete model would include all the relevant degrees of freedom at the inflationary scale. For simple models of inflation with efficient reheating, the typical reheating temperature is of order $T_{\rm{reh}}\sim 10^{15}$\,GeV.  This suggests that we need to include all degrees of freedom of masses less than or comparable to $m\sim 10^{15}$\,GeV. At such energies, we may be able to ignore GUT-mass particles or stringy states, but almost all other degrees of freedom would need to be accounted for. Colliders have given us significant information regarding the degrees of freedom that exist and their interactions up to $\sim$\,TeV, still many orders of magnitude lower than the characteristic reheating temperature. This leaves a great deal of ambiguity regarding the appropriate dynamics for reheating. 

One simple possibility is to just focus on the Standard Model degrees of freedom. This approach ignores corrections coming from dark matter\cite{2012arXiv1201.3942P,Bernabei:2003za,Peebles:2013hla,Hertzberg:2012zc,Okada:2010jd,Kohri:2009ka,Allahverdi:2007wt,Cardenas:2007xh}, baryogenesis\cite{Murayama:1993xu,Delepine:2006rn,Copeland:2001qw,Hertzberg:2013mba,Hertzberg:2013jba,Harigaya:2014tla}, unification\cite{Georgi:1974sy,Georgi:1974yf,Dimopoulos:1981yj,Carena:1993ag,Hertzberg:2014sza}, and so on, but may nonetheless be informative. For example, one could consider models like ``Higgs inflation"\cite{Bezrukov:2007ep} and the associated preheating dynamics.\cite{Bezrukov:2008ut,GarciaBellido:2008ab,Figueroa:2009jw} These models incorporate a large nonminimal coupling to gravity, whose validity as an effective field theory has been analyzed in various works (see, e.g., Refs. \refcite{Barbon:2009ya,Bezrukov:2010jz,Burgess:2010zq,Hertzberg:2011rc,Bezrukov:2013fka,George:2013iia} and references therein).

Another possibility is to build a fuller model including new physics beyond the Standard Model, such as supersymmetry and/or supergravity. These models have the drawback that we have very poor constraints on many of the parameters in these models. The huge landscape of string theory is indicative of this ambiguity in parameters (though some intriguing models include Refs. \refcite{Barnaby:2004gg,Ashoorioon:2013oha}). However, if a particular model of inflation were confirmed through a combination of observations\cite{Dodelson:2009kq}, including CMB, gravitational waves, and so on, then we may be able to infer the dynamics relevant to reheating. Furthermore, observations can constrain the details of reheating. For example, if collider data forces baryogenesis or dark matter production to be at high scales, then the reheat temperature would have to be correspondingly high.

\section{Observational Consequences}

Reheating is an important and dynamically rich period after inflation, but it is difficult to constrain observationally. There are two main reasons for this. First, since reheating occurs after inflation has ended, the complicated dynamics on subhorizon scales tend to get masked by later, nonlinear evolution of structure on short scales (though see the discussion of modulated reheating and non-Gaussianity from preheating below, in Section \ref{NongaussianitySubsection}). Second, we know that the universe has to be in a thermal state (at least for the Standard Model species) by the time big bang nucleosynthesis occurs. This ``late" thermal state essentially hides information about the early stages of the universe after inflation. Nevertheless, direct and indirect signatures of reheating are possible, some of which we discuss in this Section.

\subsection{Expansion history effects}
\begin{figure}[t] 
   \centering
    \includegraphics[width=4in]{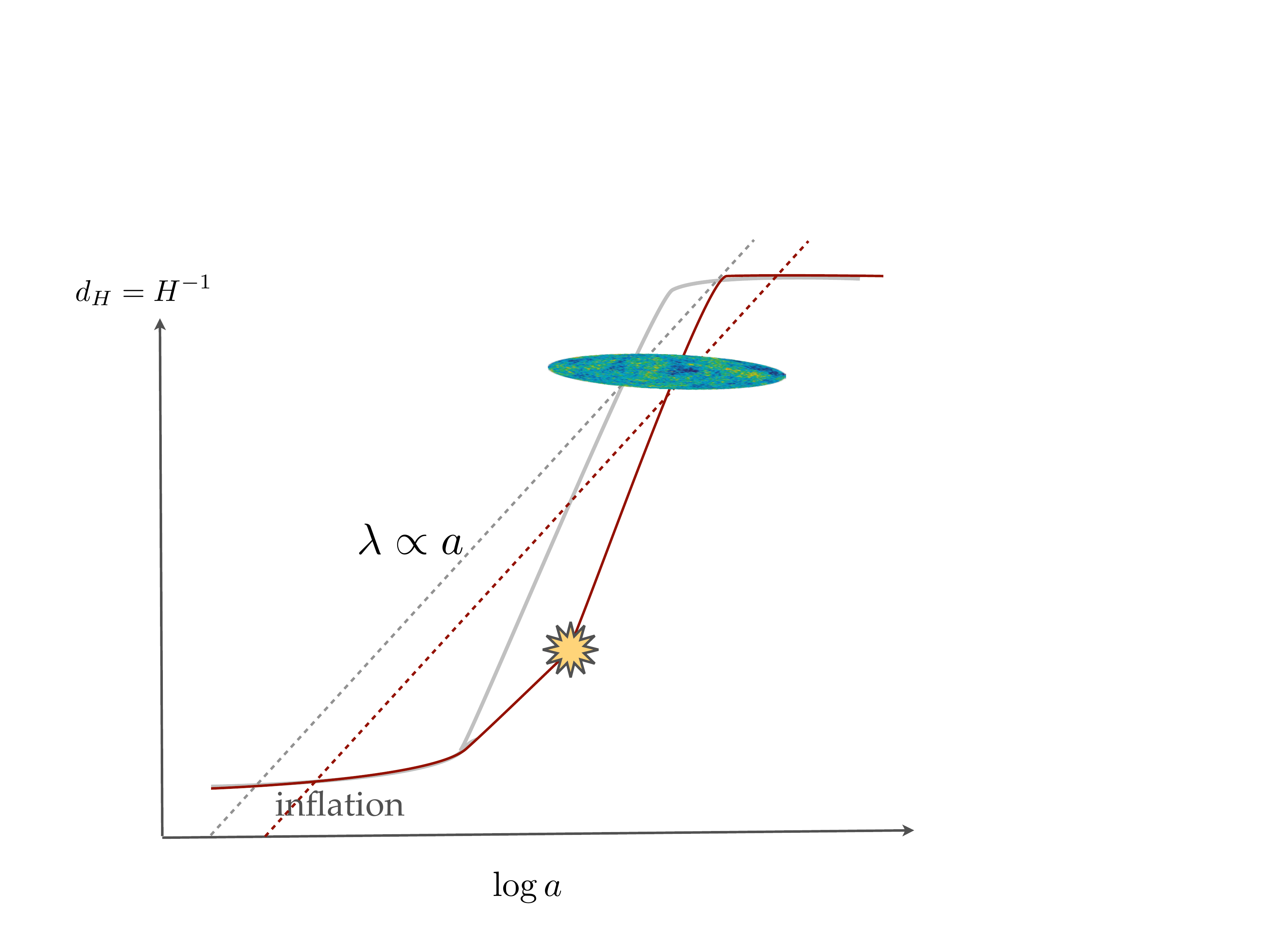}
   \caption{\small Although the thermal history does not impact the evolution of adiabatic curvature perturbations on superhorizon scales, it does change the mapping of observed scales in the CMB to horizon-exit during inflation. The figure show two possible expansion histories: (1) instantaneous thermalization and radiation domination (gray) after the end of inflation; and (2) radiation domination and thermalization preceded by a long matter-dominated phase after end of inflation (red). Uncertainty in the duration between the end of inflation and the thermalization that completes reheating leads to an uncertainty in the number of e-folds after the end of inflation.}
   \label{fig:ExpansionHistory}
\end{figure}

In the standard lore, the universe goes from being inflaton-dominated during inflation, with equation of state $w \equiv p / \rho = -1$, to radiation-dominated with $w=1/3$ at the end of inflation. In reality, however, we have very little observational evidence for the equation of state (and hence for the expansion history) between the end of inflation and the time of big bang nucleosynthesis. 

If reheating is very inefficient, the universe can undergo a prolonged period of matter-dominated expansion following inflation. While not observable directly, such a phase can affect our determination of inflationary parameters \cite{Adshead:2010mc,Dodelson:2014exa,Dai:2014jja,Creminelli:2014fca}. For example, the expansion history affects how one matches perturbations on observable scales today to the time those perturbations first crossed outside the Hubble radius during inflation (see Fig. \ref{fig:ExpansionHistory}). The number of e-folds before the end of inflation, $N_*$, when perturbations of interest first crossed the Hubble radius may be written\cite{Dodelson:2003vq,Liddle:2003as,Tegmark:2004qd}
\begin{equation}
N_* = 63.3  + \frac{1}{4}\ln \frac{V_*}{(10^{16}\>  \rm{GeV})^4} + \frac{1}{4}\ln \frac{V_*}{\rho_{\rm end}} - \frac{1}{12}\ln\frac{\rho_{\rm end}}{\rho_{\rm reh}} ,
\label{Nstar}
\end{equation}
where quantities marked with an asterisk correspond to the time during inflation when $k_* = (a H)_*$, and ``end" and ``reh" refer to the end of inflation and the onset of radiation domination, respectively. Uncertainty in the duration of reheating affects $N_*$ via the last term on the righthand side of Eq. (\ref{Nstar}); and that uncertainty, in turn, translates into uncertainties in predictions for spectral observables such as the spectral tilt, $n_s$, which typically varies as $N_*^{-1}$. Conversely, in Refs. \refcite{MartinRingeval:2010} and \refcite{Martin:2014nya}, the authors provide constraints based on the CMB data on the reheating temperature in single-field models. 

 \begin{figure}[t] 
   \centering
    \includegraphics[width=5in]{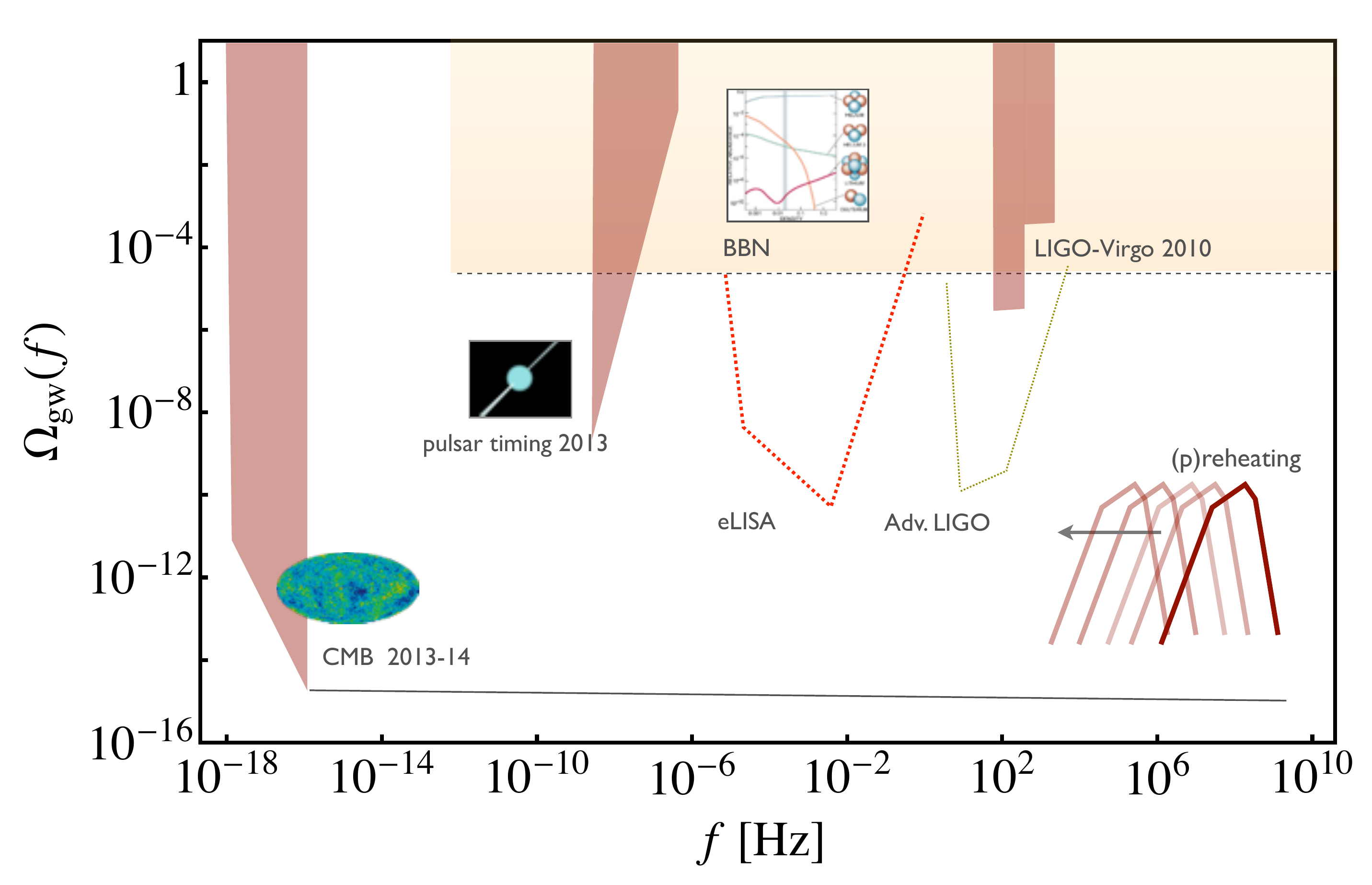}
    \caption{\small The energy density of the current universe in gravitational waves per logarithmic interval in frequency. The thick, peaked curve at the right end of the plot is typical of what is expected from preheating for GUT-scale inflation/reheating. Some low-scale inflation models, however, might be accessible to future experiments since the location of the peak is proportional to the energy scale at the end of inflation.\cite{Easther:2006gt} Shaded regions are ruled out by current observations. Dotted line ``buckets" for eLISA and Advanced LIGO indicate approximate sensitivities for these experiments . The current observational upper limits from the CMB, pulsar timing arrays, BBN, and terrestrial detection are taken from Refs. \cite{Ade:2013zuv,Shannon:2013wma,Aasi:2014zwg} (also see Ref. \cite{Moore:2014lga}). For the approximately constant line (thin grey) we assumed an approximately scale-invariant inflationary gravitational wave spectrum, without considering reheating effects or effects due to transient changes in the expansion history. The ``bend" at low frequencies is due to a transition from radiation to matter dominated era.}
 \label{fig:OmegaGW}
\end{figure}

\subsection{Gravitational waves}

The rapid fragmentation of the inflaton field(s) at the end of inflation and subsequent thermalization often generates gravitational waves \cite{Khlebnikov:1997di,Easther:2006gt,Easther:2006vd,Dufaux:2007pt,GarciaBellido:2007af, GarciaBellido:2007dg,Dufaux:2008dn,Giblin:2010sp,2011JCAP...11..015F,Bethke:2013vca,Figueroa:2013vif,Ashoorioon:2013oha,Figueroa:2014aya}. These waves can in principle be observed today as a stochastic gravitational wave background. This background is in addition to the usual spectrum of gravitational waves generated during inflation and observable via CMB polarization. While both arise in the early universe, their origins are quite different. The inflationary one is distinctly quantum in nature whereas the gravitational wave background from (p)reheating arises mostly from classical motion of inhomogeneities at the end of inflation. In general, the inflationary spectrum is nearly flat, whereas the preheating one is relatively peaked. This peak is roughly determined by the characteristic length scale of inflation fragmentation at the end of inflation,
\begin{equation}
\lambda^*_{\rm end}=\beta H_{\rm end}^{-1} .
\end{equation} 
The prefactor $\beta$  allows for the scale $\lambda_{\rm end}^*$ to be either larger or smaller than the horizon $H_{\rm end}^{-1}$ at that time, and can estimated from a linear analysis of the instabilities at the end of inflation. This peak frequency redshifted to today is then given by \cite{Easther:2006gt}  
\begin{equation}
f_*\sim \left(\frac{a_{\rm end}}{a_0}\right)\beta^{-1}H_{\rm end}\sim \beta^{-1}10^{-4} \left( \frac{ E_{\textrm{end}}}{\textrm{1 TeV} } \right) \> \textrm{Hz},
\end{equation}
where $a_0$ is the scale factor today and $a_{\rm end}$ is the scale factor at the end of inflation. In the final expression we have assumed that the universe becomes radiation dominated soon after the end of inflation. 

The fraction of energy density in gravitational waves per logarithmic interval in frequency, at the peak frequency, can be written as \footnote{M. A. Amin and E. A. Lim, {\it unpublished notes}, 2009. For different parameterizations, see for example Refs. \refcite{Felder:2006cc,Dufaux:2008dn} and \refcite{Giblin:2014gra}.}
\begin{equation}
\Omega_{gw}(f_*)\sim \Omega_{{\rm r} 0}\delta_{\pi}^2\left(H_{\rm end}\lambda_{\rm end}^*\right)^2\sim \Omega_{{\rm r}0}\delta_\pi^2\beta^{2},
\end{equation}
 where $\Omega_{{\textrm{r}} 0}$ is the current energy density in radiation and $\delta_\pi$ is the fraction of the energy density in anisotropic stresses at the time of preheating. The appearance of $\Omega_{\rm r0}$ is due to the fact that after matter domination the gravitational wave energy density dilutes faster than the total energy density in the universe. We caution that this is a very rough estimate provided to highlight the scaling with relevant physical quantities. To improve upon this rough estimate (which assumes very rapid reheating), one may determine the detailed shape and amplitude of the gravitational wave spectrum from simulations. This simple scaling works for a number of (though not all) reheating models.

For simple models one typically finds $\Omega_{gw}(f_*)\sim 10^{-9}-10^{-11}$, which follows from assuming that $\delta_{\pi}\sim 1/3$, that is $1/3$ of the energy is in anisotropic stresses responsible for gravitational wave generation and $\beta \sim 10^{-2}-10^{-3}$. For comparison, the inflationary contribution at high frequencies (assuming no running of the spectral index) is of order $10^{-15}$. Nevertheless, note that for inflation at high energy scales, the peak frequency is very high, $f_*\sim \beta^{-1}10^{9}\,\textrm{Hz}$ (assuming $E_{\rm end}=10^{16}\,\rm GeV$). This makes such a signal highly unlikely to be detected by conventional detectors for two reasons. First, the frequency is too high for such detectors, since their sensitivity deteriorates at high frequencies (Advanced LIGO has its best sensitivity near $10^3-10^4 \,\textrm{Hz}$). Second, the chirp amplitude scales inversely with frequency $h_c(f)\propto \sqrt{\Omega_{gw}(f)} /f$ \cite{Allen:1996vm} (See Fig. \ref{fig:OmegaGW}.) For low enough energy scales at the time of gravitational wave production, the frequency shifts towards the detectable regime.
We also note that there exist proposals for non-interferometric, high-frequency detectors \cite{Arvanitaki:2012cn, Cruise:2006zt}. 

Phase transitions after the end of inflation can lead to defect formation and energetically subdominant scaling seed networks which can generate gravitational waves potentially observable in CMB polarization.\cite{2011PhLB..695...26G}.  Unlike the peaked spectrum from fragmentation during preheating discussed earlier, such gravitational waves can yield a scale invariant spectrum over a wide range of frequencies (see for example Refs. \refcite{2009JCAP...10..005F,Figueroa:2012kw}).

Finally, we note that even without fragmentation, the gravitational wave spectrum generated {\it during} inflation is affected by reheating . The observed spectrum damps close to the reheating scale, and is also sensitive to changes in the expansion history after the end of inflation. If the reheating scale is low enough, this damping could be probed by proposed experiments such as DECIGO \cite{Nakayama:2008wy,Kawamura:2011zz}. In addition, atom interferometry might provide an exciting avenue for direct gravitational wave detection in the future. \cite{Dimopoulos:2008sv}

\subsection{Density perturbations from modulated reheating}
The essential idea of `modulated' reheating is that the value of coupling `constants' that determine the decay of the inflaton are functions of a light field. Such a light field develops fluctuations during inflation, resulting in spatially varying decay rates, which in turn generates spatial perturbations in the densities of the decay products.  \cite{Dvali:2003em,Kofman:2003nx,Bernardeau:2004zz,Battefeld:2007st} Modulated reheating allows the fluctuations in the CMB to be sourced by the decay rates after inflation, rather than the density perturbations of the inflaton itself. Such a modulated reheating scenario can generate adiabatic as well as isocurvature fluctuations. 

\subsection{Non-Gaussianity}
\label{NongaussianitySubsection}

In multifield models, entropy (or isocurvature) perturbations can cause modes of the curvature perturbation to evolve even after they have been stretched beyond the Hubble radius; they stop evolving once the system has achieved the adiabatic limit. Hence reheating --- an epoch marked by enormous entropy generation --- can affect the behavior of observables such as the primordial bispectrum of non-Gaussianities in the curvature perturbation spectrum. (The squeezed, local shape for the primordial bispectrum, in particular, couples modes of very different wavelengths, and hence can depend on dynamics around the end of inflation.\cite{Bartolo:2004if,Chen:2010xka,Byrnes:2010em}) The non-Gaussian parameter $f_{\rm NL}$ can be sensitive to the duration of the reheating epoch, even in the case of perturbative reheating with no strong resonances.\cite{Elliston:2011dr,Leung:2012ve}

\begin{figure}[t] 
   \centering
    \includegraphics[width=5in]{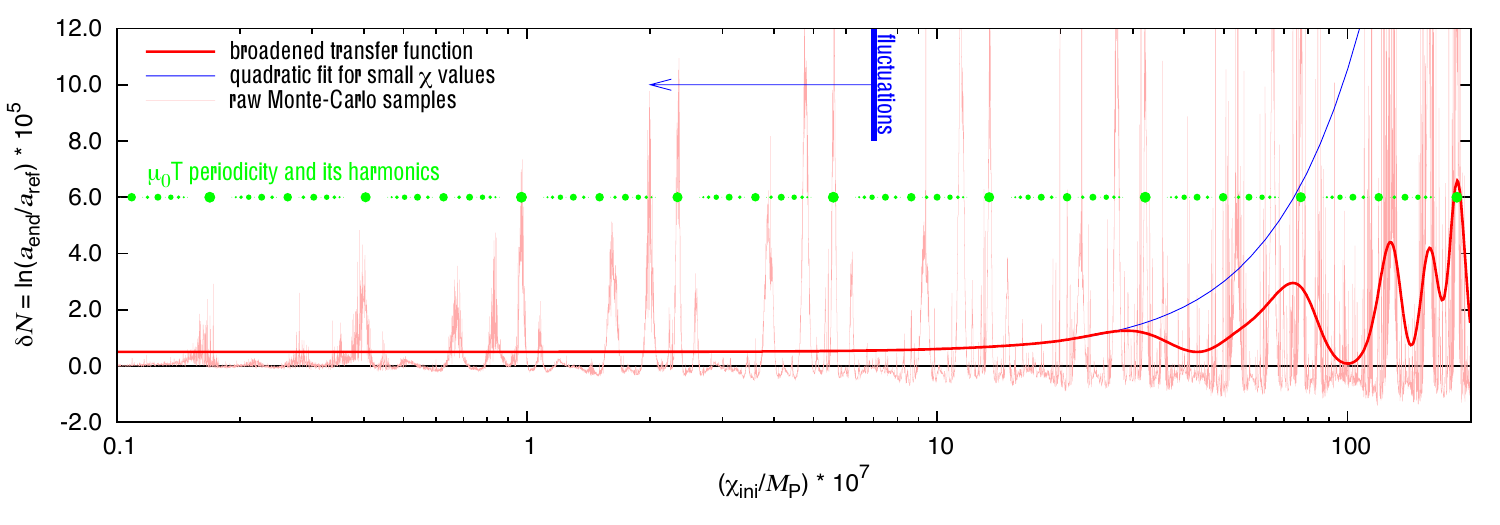}
   \caption{Non-Gaussianity in the curvature perturbation can arise from an extreme sensitivity of the expansion rate to the vacuum expectation value of light fields at the end of inflation. (From Ref. \cite{Bond:2009xx}.) }
   \label{fig:NG}
\end{figure}

For models that do feature strongly nonperturbative, nonlinear, chaotic dynamics of massless fields at the end of inflation, even stronger non-Gaussianities can be generated in the curvature perturbations, and hence could in principle be observable in the CMB.\cite{Chambers:2007se,Chambers:2009ki,Bond:2009xx} The essential idea is that the expansion (or equivalently the curvature perturbation) in different Hubble patches shows extreme sensitivity to the vacuum expectation value of the light scalar field(s) in that patch. (See Fig. \ref{fig:NG}.) Using the same mechanism, others have calculated an anisotropic gravitational wave background from preheating.\cite{Bethke:2013vca,Bethke:2013aba}

Along with the non-perturbative preheating dynamics, significant non-Gaussianity (compared to usual single-field inflation) is also generated in scenarios such as modulated reheating (discussed above). For a recent analysis, see Refs. \refcite{Cicoli:2012cy,Kobayashi:2013nwa} and references therein. 

\subsection{Primordial black holes}

The production of large inhomogeneities at the end of inflation could potentially lead to the formation of primordial black holes \cite{Khlopov:1985,GarciaBellido:1996qt,Green:2000he,Bassett:2000ha,Hidalgo:2011fj,Torres-Lomas:2014bua}. If such black holes decay quickly, they would contribute to reheating the universe. If they survive for a long time, they would eventually dominate the radiation component, leading to a matter-dominated expansion history. Such considerations can put constraints on the mechanisms of black hole production after reheating.\cite{2010PhRvD..81j4019C}

\subsection{Primordial magnetic fields}
\label{MagneticSubsection}

Large-scale (galactic and cluster scale) magnetic fields in the contemporary universe might require a primordial origin (see for example, Refs. \refcite{Kronberg:1993vk,Enquist:1998}). Phase transitions in the early universe, including reheating after inflation, provide possible mechanisms for helical magnetic field generation.\cite{DiazGil:2007dy,DiazGil:2008tf} However, connecting the typical causal length scale (the size of the horizon) at the time of generation in the early universe to galactic and cluster scales in the contemporary universe poses serious challenges. Such primordial magnetic fields can also generate stochastic gravitational waves.\cite{DiazGil:2008tf}

\subsection{Reheating and baryogenesis}
In  the Standard Model, baryogenesis is difficult \cite{1985PhLB..155...36K}. A number of ideas for the generation of baryon asymmetry, with ingredients from beyond the Standard Model, have been put forth (see, for example, Refs. \refcite{Dolgov:1982th,Abbott:1982hn,Dine:2003ax,Cline:2006ts}). It is possible and interesting to try to connect reheating to the observed baryon asymmetry. For example, a variation of the Affleck-Dine mechanism\cite{Affleck:1984fy} using the inflaton as the Affleck-Dine scalar field was proposed in Refs. \refcite{Hertzberg:2013mba,Hertzberg:2013jba}. In these cases, the resonance and nonlinear dynamics at the end of inflation can significantly affect the generated baryon asymmetry.\cite{Hertzberg:2014jza,Lozanov:2014zfa}. Some other interesting connections between reheating and baryogenesis include Refs. \refcite{Dolgov:1994zq,Dolgov:1996qq,GarciaBellido:1999sv,Davidson:2000dw,Megevand:2000da,Tranberg:2003gi,Tranberg:2006ip}.


\section{Conclusions }

With ever increasing evidence for inflation coming from the CMB and large-scale structure, the post-inflationary era becomes one of the major frontiers in our understanding of the early universe. 
The details of inflation are starting to be pinned down with measurement of the spectral index, constraints on non-Gaussianity and isocurvature perturbations, as well as possible evidence of the energy scale of inflation from B-modes in the CMB.
By contrast, there is as yet relatively little known about the precise details of the universe between the inflationary phase and the later thermal phase of big bang nucleosynthesis. The reason for this discrepancy in understanding is due to the fact that inflation stretches modes to huge length scales that are cosmologically observable, while post-inflationary dynamics tends to yield effects on much smaller scales.

This observational gap leads to a wide variety of possibilities regarding the dynamics and degrees of freedom that were relevant at the end of inflation and during the post-inflationary era before radiation domination. In particular, various forms of nonperturbative dynamics can take place, including the formation of solitons and defects. Much of this is being studied carefully with lattice simulations.

In addition to the many works on the post-inflationary era that we surveyed in this review, there are various directions for future work in this field. On the more theoretical side, it is important to develop and compute the details of more realistic post-inflationary models. This includes a full study of models involving not only scalar fields (as has been typical), but also Abelian and non-Abelian gauge fields. Ultimately, this should include $W$ and $Z$ bosons and all the various degrees of freedom of the Standard Model and beyond. 

Other open questions concern the production of relics: in particular the formation of the matter/anti-matter asymmetry (baryogenesis), the production of dark matter, and other relics such as photons and neutrinos. Ultimately these degrees of freedom must arise from the (possibly nonperturbative) decay of the inflaton. Furthermore, if inflation occurred at a high energy scale, moduli fields can play an important role. A detailed model of the particle dynamics in the post-inflation era is currently unknown, but it is important to narrow the range. 

Another outstanding direction for future work is a full understanding of thermalization. In particular, although much work has gone into understanding the fragmentation of the inflaton condensate and its decay, most analyses stop short of seeing the full development into a proper thermal distribution of particles. A number of studies have found long-lived metastable configurations of the fields, which could delay thermalization. A full accounting of thermalization, even without metastable field configurations, would presumably involve a detailed quantum mechanical computation to see evolution towards a state of maximum local entropy, such as a type of Bose-Einstein or Fermi-Dirac distribution.

On the observational side there is also much work to be done, as nonperturbative dynamics after inflation can lead to observational consequences. Of course, progress in this direction is difficult because the relevant modes of interest tend to be on small scales (although there do exist multifield models in which superhorizon perturbations may also be amplified during reheating). 
Indirect clues could come in the CMB by narrowing down to a specific model for inflation. 
More direct information on the post-inflationary era could come from the detection of gravitational waves arising from the nonperturbative fragmentation of the inflaton condensate. If the scale of inflation is very low, then the corresponding frequencies may be obtainable with large interferometers. On the other hand, if the scale of inflation is high, then the frequencies of gravitational waves from reheating would make a direct detection challenging.

Another observational signature would be the detection of non-Gaussianity on small scales. But perhaps the most direct impact on observations from the post-inflationary era is its impact on the expansion history of the universe, which affects how we map perturbations on cosmologically relevant length scales today to the time during inflation when they first crossed outside the Hubble radius.
In the meantime, further constraints on the relevant degrees of freedom in the early universe may come from particle accelerators or cosmic rays, giving clues as to the physics at very high energies. 

Ultimately, a unified theory of physics must provide a consistent theory of inflation, reheating, high energy physics, and late-time observations. 
A full understanding of the nonperturbative dynamics after inflation remains an ongoing challenge in this quest for unification.

\vskip 14pt

{\bf Acknowledgments} We would like to thank Bruce Bassett and Kaloian Lozanov for helpful comments on an earlier draft, and thank Christopher Moore, Yi Wang and Daniel Baumann for useful conversations. We would also like to acknowledge support from the Center for Theoretical Physics at MIT and the Kavli Institute for Cosmology at Cambridge (KICC). This work is supported by the U.S. Department of Energy under grant Contract Number DE-SC00012567. JK is also supported by an NSERC PDF fellowship. MA is supported by a Senior Kavli Fellowship at KICC.


\begin{thebibliography}{100}

\bibitem{Guth:2005zr}
A.~H. Guth and D.~I. Kaiser, {\em Science} {\bf 307}  (2005) 884,
  \href{http://arxiv.org/abs/astro-ph/0502328}{{\ttfamily
  arXiv:astro-ph/0502328 [astro-ph]}}.

\bibitem{Bassett:2005xm}
B.~A. Bassett, S.~Tsujikawa and D.~Wands, {\em {Rev. Mod. Phys.}} {\bf 78}
  (2006) 537, \href{http://arxiv.org/abs/astro-ph/0507632}{{\ttfamily
  arXiv:astro-ph/0507632 [astro-ph]}}.

\bibitem{Lyth:2009zz}
D.~H. Lyth and A.~R. Liddle, {\em {The primordial density perturbation:
  Cosmology, inflation and the origin of structure}} (New York: Cambridge
  University Press, 2009).

\bibitem{Baumann:2009ds}
D.~Baumann  (2009) \href{http://arxiv.org/abs/0907.5424}{{\ttfamily
  arXiv:0907.5424 [hep-th]}}.

\bibitem{Linde:2014nna}
A.~Linde  (2014) \href{http://arxiv.org/abs/1402.0526}{{\ttfamily
  arXiv:1402.0526 [hep-th]}}.

\bibitem{Martin:2014vha}
J.~Martin, C.~Ringeval and V.~Vennin, {\em Phys.Dark Univ.}   (2014)
  \href{http://arxiv.org/abs/1303.3787}{{\ttfamily arXiv:1303.3787
  [astro-ph.CO]}}.

\bibitem{Steigman:2007xt}
G.~Steigman, {\em Ann. Rev. Nucl. Part. Sci.} {\bf 57}  (2007) 463,
  \href{http://arxiv.org/abs/0712.1100}{{\ttfamily arXiv:0712.1100
  [astro-ph]}}.

\bibitem{Taruya:1997iv}
A.~Taruya and Y.~Nambu, {\em Phys.Lett.} {\bf B428}  (1998) 37,
  \href{http://arxiv.org/abs/gr-qc/9709035}{{\ttfamily arXiv:gr-qc/9709035
  [gr-qc]}}.

\bibitem{Bassett:1999cg}
B.~A. Bassett and F.~Viniegra, {\em Phys.Rev.} {\bf D62}  (2000)   043507,
  \href{http://arxiv.org/abs/hep-ph/9909353}{{\ttfamily arXiv:hep-ph/9909353
  [hep-ph]}}.

\bibitem{Finelli:2000ya}
F.~Finelli and R.~H. Brandenberger, {\em Phys.Rev.} {\bf D62}  (2000)   083502,
  \href{http://arxiv.org/abs/hep-ph/0003172}{{\ttfamily arXiv:hep-ph/0003172
  [hep-ph]}}.

\bibitem{Tsujikawa:2002nf}
S.~Tsujikawa and B.~A. Bassett, {\em Phys.Lett.} {\bf B536}  (2002) 9,
  \href{http://arxiv.org/abs/astro-ph/0204031}{{\ttfamily
  arXiv:astro-ph/0204031 [astro-ph]}}.

\bibitem{Chambers:2007se}
A.~Chambers and A.~Rajantie, {\em Phys.Rev.Lett.} {\bf 100}  (2008)   041302,
  \href{http://arxiv.org/abs/0710.4133}{{\ttfamily arXiv:0710.4133
  [astro-ph]}}.

\bibitem{Bond:2009xx}
J.~R. Bond, A.~V. Frolov, Z.~Huang and L.~Kofman, {\em Phys. Rev. Lett.} {\bf
  103}  (2009)   071301, \href{http://arxiv.org/abs/0903.3407}{{\ttfamily
  arXiv:0903.3407 [astro-ph.CO]}}.

\bibitem{Bethke:2013aba}
L.~Bethke, D.~G. Figueroa and A.~Rajantie, {\em Phys.Rev.Lett.} {\bf 111}
  (2013)   011301, \href{http://arxiv.org/abs/1304.2657}{{\ttfamily
  arXiv:1304.2657 [astro-ph.CO]}}.

\bibitem{Moghaddam:2014ksa}
H.~B. Moghaddam, R.~H. Brandenberger, Y.-F. Cai and E.~G.~M. Ferreira  (2014)
  \href{http://arxiv.org/abs/1409.1784}{{\ttfamily arXiv:1409.1784
  [astro-ph.CO]}}.

\bibitem{Dodelson:2003vq}
S.~Dodelson and L.~Hui, {\em Phys. Rev. Lett.} {\bf 91}  (2003)   131301,
  \href{http://arxiv.org/abs/astro-ph/0305113}{{\ttfamily
  arXiv:astro-ph/0305113 [astro-ph]}}.

\bibitem{Liddle:2003as}
A.~Liddle and S.~M. Leach, {\em Phys. Rev.} {\bf D68}  (2003)   103503,
  \href{http://arxiv.org/abs/astro-ph/0305263}{{\ttfamily
  arXiv:astro-ph/0305263 [astro-ph]}}.

\bibitem{Tegmark:2004qd}
M.~Tegmark, {\em JCAP} {\bf 0504}  (2005)   001,
  \href{http://arxiv.org/abs/astro-ph/0410281}{{\ttfamily
  arXiv:astro-ph/0410281 [astro-ph]}}.

\bibitem{Adshead:2010mc}
P.~Adshead, R.~Easther, J.~Pritchard and A.~Loeb, {\em JCAP} {\bf 1102}  (2011)
    021, \href{http://arxiv.org/abs/1007.3748}{{\ttfamily arXiv:1007.3748
  [astro-ph.CO]}}.

\bibitem{Dodelson:2014exa}
S.~Dodelson, {\em Phys. Rev. Lett.} {\bf 112}  (2014)   191301,
  \href{http://arxiv.org/abs/1403.6310}{{\ttfamily arXiv:1403.6310
  [astro-ph.CO]}}.

\bibitem{Dai:2014jja}
L.~Dai, M.~Kamionkowski and J.~Wang  (2014)
  \href{http://arxiv.org/abs/1404.6704}{{\ttfamily arXiv:1404.6704
  [astro-ph.CO]}}.

\bibitem{Creminelli:2014fca}
P.~Creminelli, D.~L. Nacir, M.~Simonovi, G.~Trevisan and M.~Zaldarriaga  (2014)
  \href{http://arxiv.org/abs/1405.6264}{{\ttfamily arXiv:1405.6264
  [astro-ph.CO]}}.

\bibitem{Albrecht:1982mp}
A.~Albrecht, P.~J. Steinhardt, M.~S. Turner and F.~Wilczek, {\em Phys. Rev.
  Lett.} {\bf 48}  (1982)   1437.

\bibitem{Dolgov:1982th}
A.~Dolgov and A.~D. Linde, {\em Phys. Lett.} {\bf B116}  (1982)   329.

\bibitem{Abbott:1982hn}
L.~Abbott, E.~Farhi and M.~B. Wise, {\em Phys. Lett.} {\bf B117}  (1982)  ~29.

\bibitem{Dolgov:1989us}
A.~Dolgov and D.~Kirilova, {\em Sov. J. Nucl. Phys.} {\bf 51}  (1990) 172.

\bibitem{Traschen:1990sw}
J.~H. Traschen and R.~H. Brandenberger, {\em Phys. Rev.} {\bf D42}  (1990)
  2491.

\bibitem{Kofman:1994rk}
L.~Kofman, A.~D. Linde and A.~A. Starobinsky, {\em Phys. Rev. Lett.} {\bf 73}
  (1994) 3195, \href{http://arxiv.org/abs/hep-th/9405187}{{\ttfamily
  arXiv:hep-th/9405187 [hep-th]}}.

\bibitem{Shtanov:1994ce}
Y.~Shtanov, J.~H. Traschen and R.~H. Brandenberger, {\em Phys. Rev.} {\bf D51}
  (1995) 5438, \href{http://arxiv.org/abs/hep-ph/9407247}{{\ttfamily
  arXiv:hep-ph/9407247 [hep-ph]}}.

\bibitem{Boyanovsky:1995ud}
D.~Boyanovsky, M.~D'Attanasio, H.~de~Vega, R.~Holman, D.-S. Lee {\em et~al.}
  (1995) \href{http://arxiv.org/abs/hep-ph/9505220}{{\ttfamily
  arXiv:hep-ph/9505220 [hep-ph]}}.

\bibitem{Yoshimura:1995gc}
M.~Yoshimura, {\em Prog. Theor. Phys.} {\bf 94}  (1995) 873,
  \href{http://arxiv.org/abs/hep-th/9506176}{{\ttfamily arXiv:hep-th/9506176
  [hep-th]}}.

\bibitem{Kaiser:1995fb}
D.~I. Kaiser, {\em Phys. Rev.} {\bf D53}  (1996) 1776,
  \href{http://arxiv.org/abs/astro-ph/9507108}{{\ttfamily
  arXiv:astro-ph/9507108 [astro-ph]}}.

\bibitem{Kofman:1997yn}
L.~Kofman, A.~D. Linde and A.~A. Starobinsky, {\em Phys. Rev.} {\bf D56}
  (1997) 3258, \href{http://arxiv.org/abs/hep-ph/9704452}{{\ttfamily
  arXiv:hep-ph/9704452 [hep-ph]}}.

\bibitem{Boyanovsky:1996sv}
D.~Boyanovsky, H.~de~Vega and R.~Holman  (1996) 183,
  \href{http://arxiv.org/abs/hep-ph/9701304}{{\ttfamily arXiv:hep-ph/9701304
  [hep-ph]}}.

\bibitem{Allahverdi:2010xz}
R.~Allahverdi, R.~Brandenberger, F.-Y. Cyr-Racine and A.~Mazumdar, {\em Ann.
  Rev. Nucl. Part. Sci.} {\bf 60}  (2010) 27,
  \href{http://arxiv.org/abs/1001.2600}{{\ttfamily arXiv:1001.2600 [hep-th]}}.

\bibitem{Frolov:2010sz}
A.~V. Frolov, {\em Class. Quant. Grav.} {\bf 27}  (2010)   124006,
  \href{http://arxiv.org/abs/1004.3559}{{\ttfamily arXiv:1004.3559 [gr-qc]}}.

\bibitem{Mukhanov:2005sc}
V.~Mukhanov, {\em {Physical foundations of cosmology}} (New York: Cambridge
  University Press, 2005).

\bibitem{Kaiser:2010ps}
D.~I. Kaiser, {\em Phys. Rev.} {\bf D81}  (2010)   084044,
  \href{http://arxiv.org/abs/1003.1159}{{\ttfamily arXiv:1003.1159 [gr-qc]}}.

\bibitem{Bardeen:1980kt}
J.~M. Bardeen, {\em Phys.Rev.} {\bf D22}  (1980) 1882.

\bibitem{Sasaki:1995aw}
M.~Sasaki and E.~D. Stewart, {\em Prog.Theor.Phys.} {\bf 95}  (1996) 71,
  \href{http://arxiv.org/abs/astro-ph/9507001}{{\ttfamily
  arXiv:astro-ph/9507001 [astro-ph]}}.

\bibitem{GrootNibbelink:2001qt}
S.~Groot~Nibbelink and B.~van Tent, {\em Class.Quant.Grav.} {\bf 19}  (2002)
  613, \href{http://arxiv.org/abs/hep-ph/0107272}{{\ttfamily
  arXiv:hep-ph/0107272 [hep-ph]}}.

\bibitem{Langlois:2008mn}
D.~Langlois and S.~Renaux-Petel, {\em JCAP} {\bf 0804}  (2008)   017,
  \href{http://arxiv.org/abs/0801.1085}{{\ttfamily arXiv:0801.1085 [hep-th]}}.

\bibitem{Weinberg:2008zzc}
S.~Weinberg, {\em {Cosmology}} (New York: Cambridge University Press, 2008).

\bibitem{Gong:2011uw}
J.-O. Gong and T.~Tanaka, {\em JCAP} {\bf 1103}  (2011)   015,
  \href{http://arxiv.org/abs/1101.4809}{{\ttfamily arXiv:1101.4809
  [astro-ph.CO]}}.

\bibitem{Peterson:2011yt}
C.~M. Peterson and M.~Tegmark, {\em Phys.Rev.} {\bf D87}  (2013)   103507,
  \href{http://arxiv.org/abs/1111.0927}{{\ttfamily arXiv:1111.0927
  [astro-ph.CO]}}.

\bibitem{Elliston:2012ab}
J.~Elliston, D.~Seery and R.~Tavakol, {\em JCAP} {\bf 1211}  (2012)   060,
  \href{http://arxiv.org/abs/1208.6011}{{\ttfamily arXiv:1208.6011
  [astro-ph.CO]}}.

\bibitem{Kaiser:2012ak}
D.~I. Kaiser, E.~A. Mazenc and E.~I. Sfakianakis, {\em Phys. Rev.} {\bf D87}
  (2013)   064004, \href{http://arxiv.org/abs/1210.7487}{{\ttfamily
  arXiv:1210.7487 [astro-ph.CO]}}.

\bibitem{Lozanov:2014zfa}
K.~D. Lozanov and M.~A. Amin  (2014)
  \href{http://arxiv.org/abs/1408.1811}{{\ttfamily arXiv:1408.1811 [hep-ph]}}.

\bibitem{Mukhanov:1985rz}
V.~F. Mukhanov, {\em JETP Lett.} {\bf 41}  (1985) 493.

\bibitem{Sasaki:1986hm}
M.~Sasaki, {\em Prog. Theor. Phys.} {\bf 76}  (1986)   1036.

\bibitem{Bassett:1999mt}
B.~A. Bassett, F.~Tamburini, D.~I. Kaiser and R.~Maartens, {\em Nucl. Phys.}
  {\bf B561}  (1999) 188, \href{http://arxiv.org/abs/hep-ph/9901319}{{\ttfamily
  arXiv:hep-ph/9901319 [hep-ph]}}.

\bibitem{Podolsky:2002qv}
D.~Podolsky and A.~A. Starobinsky, {\em Grav. Cosmol. Suppl.} {\bf 8N1}  (2002)
  13, \href{http://arxiv.org/abs/astro-ph/0204327}{{\ttfamily
  arXiv:astro-ph/0204327 [astro-ph]}}.

\bibitem{Jin:2004bf}
Y.~Jin and S.~Tsujikawa, {\em Class. Quant. Grav.} {\bf 23}  (2006) 353,
  \href{http://arxiv.org/abs/hep-ph/0411164}{{\ttfamily arXiv:hep-ph/0411164
  [hep-ph]}}.

\bibitem{Salopek:1988qh}
D.~Salopek, J.~Bond and J.~M. Bardeen, {\em Phys.Rev.} {\bf D40}  (1989)
  1753.

\bibitem{Price:2014xpa}
L.~C. Price, J.~Frazer, J.~Xu, H.~V. Peiris and R.~Easther  (2014)
  \href{http://arxiv.org/abs/1410.0685}{{\ttfamily arXiv:1410.0685
  [astro-ph.CO]}}.

\bibitem{Camilleri:2008zz}
L.~Camilleri, E.~Lisi and J.~F. Wilkerson, {\em Ann. Rev. Nucl. Part. Sci.}
  {\bf 58}  (2008) 343.

\bibitem{Birrell:1982ix}
N.~Birrell and P.~Davies, {\em {Quantum Fields in Curved Space}} (New York:
  Cambridge University Press, 1982).

\bibitem{Polarski:1995jg}
D.~Polarski and A.~A. Starobinsky, {\em Class. Quant. Grav.} {\bf 13}  (1996)
  377, \href{http://arxiv.org/abs/gr-qc/9504030}{{\ttfamily arXiv:gr-qc/9504030
  [gr-qc]}}.

\bibitem{Mukhanov:2007}
V.~F. Mukhanov and S.~Winitzki, {\em {Introduction to Quantum Effects in
  Gravity}} (New York: Cambridge University Press, 2007).

\bibitem{Khlebnikov:1996mc}
S.~Y. Khlebnikov and I.~Tkachev, {\em Phys. Rev. Lett.} {\bf 77}  (1996) 219,
  \href{http://arxiv.org/abs/hep-ph/9603378}{{\ttfamily arXiv:hep-ph/9603378
  [hep-ph]}}.

\bibitem{Khlebnikov:1996wr}
S.~Y. Khlebnikov and I.~Tkachev, {\em Phys. Lett.} {\bf B390}  (1997) 80,
  \href{http://arxiv.org/abs/hep-ph/9608458}{{\ttfamily arXiv:hep-ph/9608458
  [hep-ph]}}.

\bibitem{Khlebnikov:1996zt}
S.~Y. Khlebnikov and I.~Tkachev, {\em Phys. Rev. Lett.} {\bf 79}  (1997) 1607,
  \href{http://arxiv.org/abs/hep-ph/9610477}{{\ttfamily arXiv:hep-ph/9610477
  [hep-ph]}}.

\bibitem{Prokopec:1996rr}
T.~Prokopec and T.~G. Roos, {\em Phys. Rev.} {\bf D55}  (1997) 3768,
  \href{http://arxiv.org/abs/hep-ph/9610400}{{\ttfamily arXiv:hep-ph/9610400
  [hep-ph]}}.

\bibitem{PeskinSchroeder}
M.~E. Peskin and D.~V. Schroeder, {\em {An Introduction to Quantum Field
  Theory}} (Reading, Mass.: Addison-Wesley, 1995).

\bibitem{Mazumdar:2013gya}
A.~Mazumdar and B.~Zaldivar  (2013)
  \href{http://arxiv.org/abs/1310.5143}{{\ttfamily arXiv:1310.5143 [hep-ph]}}.

\bibitem{Gleiser:1993ea}
M.~Gleiser and R.~O. Ramos, {\em Phys.Rev.} {\bf D50}  (1994) 2441,
  \href{http://arxiv.org/abs/hep-ph/9311278}{{\ttfamily arXiv:hep-ph/9311278
  [hep-ph]}}.

\bibitem{Boyanovsky:1994me}
D.~Boyanovsky, H.~de~Vega, R.~Holman, D.~Lee and A.~Singh, {\em Phys.Rev.} {\bf
  D51}  (1995) 4419, \href{http://arxiv.org/abs/hep-ph/9408214}{{\ttfamily
  arXiv:hep-ph/9408214 [hep-ph]}}.

\bibitem{Boyanovsky:1996xx}
D.~Boyanovsky, I.~Lawrie and D.~Lee, {\em Phys.Rev.} {\bf D54}  (1996) 4013,
  \href{http://arxiv.org/abs/hep-ph/9603217}{{\ttfamily arXiv:hep-ph/9603217
  [hep-ph]}}.

\bibitem{Boyanovsky:1996rw}
D.~Boyanovsky, D.~Cormier, H.~de~Vega and R.~Holman, {\em Phys.Rev.} {\bf D55}
  (1997) 3373, \href{http://arxiv.org/abs/hep-ph/9610396}{{\ttfamily
  arXiv:hep-ph/9610396 [hep-ph]}}.

\bibitem{Berges:2002wr}
J.~Berges, S.~Borsanyi and J.~Serreau, {\em Nucl. Phys.} {\bf B660}  (2003) 51,
  \href{http://arxiv.org/abs/hep-ph/0212404}{{\ttfamily arXiv:hep-ph/0212404
  [hep-ph]}}.

\bibitem{Boyanovsky:1995ema}
D.~Boyanovsky, M.~D'Attanasio, H.~de~Vega, R.~Holman and D.-S. Lee, {\em Phys.
  Rev.} {\bf D52}  (1995) 6805,
  \href{http://arxiv.org/abs/hep-ph/9507414}{{\ttfamily arXiv:hep-ph/9507414
  [hep-ph]}}.

\bibitem{Drewes:2013iaa}
M.~Drewes and J.~U. Kang, {\em Nucl.Phys.} {\bf B875}  (2013) 315,
  \href{http://arxiv.org/abs/1305.0267}{{\ttfamily arXiv:1305.0267 [hep-ph]}}.

\bibitem{Drewes:2014pfa}
M.~Drewes  (2014) \href{http://arxiv.org/abs/1406.6243}{{\ttfamily
  arXiv:1406.6243 [hep-ph]}}.

\bibitem{Podolsky:2005bw}
D.~I. Podolsky, G.~N. Felder, L.~Kofman and M.~Peloso, {\em Phys. Rev.} {\bf
  D73}  (2006)   023501, \href{http://arxiv.org/abs/hep-ph/0507096}{{\ttfamily
  arXiv:hep-ph/0507096 [hep-ph]}}.

\bibitem{Amin:2010xe}
M.~A. Amin  (2010) \href{http://arxiv.org/abs/1006.3075}{{\ttfamily
  arXiv:1006.3075 [astro-ph.CO]}}.

\bibitem{Amin:2010dc}
M.~A. Amin, R.~Easther and H.~Finkel, {\em JCAP} {\bf 1012}  (2010)   001,
  \href{http://arxiv.org/abs/1009.2505}{{\ttfamily arXiv:1009.2505
  [astro-ph.CO]}}.

\bibitem{Amin:2011hj}
M.~A. Amin, R.~Easther, H.~Finkel, R.~Flauger and M.~P. Hertzberg, {\em
  Phys.Rev.Lett.} {\bf 108}  (2012)   241302,
  \href{http://arxiv.org/abs/1106.3335}{{\ttfamily arXiv:1106.3335
  [astro-ph.CO]}}.

\bibitem{Braden:2010wd}
J.~Braden, L.~Kofman and N.~Barnaby, {\em JCAP} {\bf 1007}  (2010)   016,
  \href{http://arxiv.org/abs/1005.2196}{{\ttfamily arXiv:1005.2196 [hep-th]}}.

\bibitem{Zanchin:1997gf}
V.~Zanchin, J.~Maia, A., W.~Craig and R.~H. Brandenberger, {\em Phys.Rev.} {\bf
  D57}  (1998) 4651, \href{http://arxiv.org/abs/hep-ph/9709273}{{\ttfamily
  arXiv:hep-ph/9709273 [hep-ph]}}.

\bibitem{Zanchin:1998fj}
V.~Zanchin, J.~Maia, A., W.~Craig and R.~H. Brandenberger, {\em Phys.Rev.} {\bf
  D60}  (1999)   023505, \href{http://arxiv.org/abs/hep-ph/9901207}{{\ttfamily
  arXiv:hep-ph/9901207 [hep-ph]}}.

\bibitem{Berges:2002cz}
J.~Berges and J.~Serreau, {\em Phys.Rev.Lett.} {\bf 91}  (2003)   111601,
  \href{http://arxiv.org/abs/hep-ph/0208070}{{\ttfamily arXiv:hep-ph/0208070
  [hep-ph]}}.

\bibitem{HillBook}
W.~Magnus and S.~Winkler, {\em {Hill's Equation}} (Dover Publications, 1979).

\bibitem{dif}
G.~Teschl, {\em {Ordinary Differential Equations and Dynamical Systems}}
  (American Mathematical Society, 2012).

\bibitem{Bassett:1999ta}
B.~A. Bassett, C.~Gordon, R.~Maartens and D.~I. Kaiser, {\em Phys. Rev.} {\bf
  D61}  (2000)   061302, \href{http://arxiv.org/abs/hep-ph/9909482}{{\ttfamily
  arXiv:hep-ph/9909482 [hep-ph]}}.

\bibitem{Dufaux:2006ee}
J.~F. Dufaux, G.~N. Felder, L.~Kofman, M.~Peloso and D.~Podolsky, {\em JCAP}
  {\bf 0607}  (2006)   006,
  \href{http://arxiv.org/abs/hep-ph/0602144}{{\ttfamily arXiv:hep-ph/0602144
  [hep-ph]}}.

\bibitem{DeMelo:2001nr}
F.~d. R. V.~B. De~Melo, R.~H. Brandenberger and J.~Maia, Adolfo, {\em
  Int.J.Mod.Phys.} {\bf A17}  (2002) 4413,
  \href{http://arxiv.org/abs/hep-ph/0110003}{{\ttfamily arXiv:hep-ph/0110003
  [hep-ph]}}.

\bibitem{Greene:1997ge}
B.~R. Greene, T.~Prokopec and T.~G. Roos, {\em Phys.Rev.} {\bf D56}  (1997)
  6484, \href{http://arxiv.org/abs/hep-ph/9705357}{{\ttfamily
  arXiv:hep-ph/9705357 [hep-ph]}}.

\bibitem{tach1}
G.~N. Felder, J.~Garcia-Bellido, P.~B. Greene, L.~Kofman, A.~D. Linde {\em
  et~al.}, {\em Phys. Rev. Lett.} {\bf 87}  (2001)   011601,
  \href{http://arxiv.org/abs/hep-ph/0012142}{{\ttfamily arXiv:hep-ph/0012142
  [hep-ph]}}.

\bibitem{Arrizabalaga:2004iw}
A.~Arrizabalaga, J.~Smit and A.~Tranberg, {\em JHEP} {\bf 0410}  (2004)   017,
  \href{http://arxiv.org/abs/hep-ph/0409177}{{\ttfamily arXiv:hep-ph/0409177
  [hep-ph]}}.

\bibitem{Boyanovsky:1996sq}
D.~Boyanovsky, H.~de~Vega, R.~Holman and J.~Salgado, {\em Phys.Rev.} {\bf D54}
  (1996) 7570, \href{http://arxiv.org/abs/hep-ph/9608205}{{\ttfamily
  arXiv:hep-ph/9608205 [hep-ph]}}.

\bibitem{Greene:1997fu}
P.~B. Greene, L.~Kofman, A.~D. Linde and A.~A. Starobinsky, {\em Phys.Rev.}
  {\bf D56}  (1997) 6175, \href{http://arxiv.org/abs/hep-ph/9705347}{{\ttfamily
  arXiv:hep-ph/9705347 [hep-ph]}}.

\bibitem{Kaiser:1997mp}
D.~I. Kaiser, {\em Phys.Rev.} {\bf D56}  (1997) 706,
  \href{http://arxiv.org/abs/hep-ph/9702244}{{\ttfamily arXiv:hep-ph/9702244
  [hep-ph]}}.

\bibitem{Kaiser:1997hg}
D.~I. Kaiser, {\em Phys.Rev.} {\bf D57}  (1998) 702,
  \href{http://arxiv.org/abs/hep-ph/9707516}{{\ttfamily arXiv:hep-ph/9707516
  [hep-ph]}}.

\bibitem{Lachapelle:2008sy}
J.~Lachapelle and R.~H. Brandenberger, {\em JCAP} {\bf 0904}  (2009)   020,
  \href{http://arxiv.org/abs/0808.0936}{{\ttfamily arXiv:0808.0936 [hep-th]}}.

\bibitem{Karouby:2011xs}
J.~Karouby, B.~Underwood and A.~C. Vincent, {\em Phys. Rev.} {\bf D84}  (2011)
   043528, \href{http://arxiv.org/abs/1105.3982}{{\ttfamily arXiv:1105.3982
  [hep-th]}}.

\bibitem{Hertzberg:2014iza}
M.~P. Hertzberg, J.~Karouby, W.~G. Spitzer, J.~C. Becerra and L.~Li  (2014)
  \href{http://arxiv.org/abs/1408.1396}{{\ttfamily arXiv:1408.1396 [hep-th]}}.

\bibitem{Amin:2011hu}
M.~A. Amin, P.~Zukin and E.~Bertschinger, {\em Phys. Rev.} {\bf D85}  (2012)
  103510, \href{http://arxiv.org/abs/1108.1793}{{\ttfamily arXiv:1108.1793
  [astro-ph.CO]}}.

\bibitem{Khlopov:1985}
M.~I. {Khlopov}, B.~A. {Malomed} and I.~B. {Zeldovich}, {\em MNRAS} {\bf 215}
  (August 1985) 575.

\bibitem{Johnson:2008se}
M.~C. Johnson and M.~Kamionkowski, {\em Phys.Rev.} {\bf D78}  (2008)   063010,
  \href{http://arxiv.org/abs/0805.1748}{{\ttfamily arXiv:0805.1748
  [astro-ph]}}.

\bibitem{Silverstein:2008sg}
E.~Silverstein and A.~Westphal, {\em Phys.Rev.} {\bf D78}  (2008)   106003,
  \href{http://arxiv.org/abs/0803.3085}{{\ttfamily arXiv:0803.3085 [hep-th]}}.

\bibitem{McAllister:2014mpa}
L.~McAllister, E.~Silverstein, A.~Westphal and T.~Wrase, {\em JHEP} {\bf 1409}
  (2014)   123, \href{http://arxiv.org/abs/1405.3652}{{\ttfamily
  arXiv:1405.3652 [hep-th]}}.

\bibitem{Greenwood:2012aj}
R.~N. Greenwood, D.~I. Kaiser and E.~I. Sfakianakis, {\em Phys.Rev.} {\bf D87}
  (2013)   064021, \href{http://arxiv.org/abs/1210.8190}{{\ttfamily
  arXiv:1210.8190 [hep-ph]}}.

\bibitem{Gordon:2000hv}
C.~Gordon, D.~Wands, B.~A. Bassett and R.~Maartens, {\em Phys. Rev.} {\bf D63}
  (2001)   023506, \href{http://arxiv.org/abs/astro-ph/0009131}{{\ttfamily
  arXiv:astro-ph/0009131 [astro-ph]}}.

\bibitem{Hertzberg:2014jza}
M.~P. Hertzberg, J.~Karouby, W.~G. Spitzer, J.~C. Becerra and L.~Li  (2014)
  \href{http://arxiv.org/abs/1408.1398}{{\ttfamily arXiv:1408.1398 [hep-th]}}.

\bibitem{Levasseur:2010rk}
L.~Perreault~Levasseur, G.~Laporte and R.~Brandenberger, {\em Phys.Rev.} {\bf
  D82}  (2010)   123524, \href{http://arxiv.org/abs/1004.1425}{{\ttfamily
  arXiv:1004.1425 [hep-th]}}.

\bibitem{Felder:2000hq}
G.~N. Felder and I.~Tkachev, {\em Comput.Phys.Commun.} {\bf 178}  (2008) 929,
  \href{http://arxiv.org/abs/hep-ph/0011159}{{\ttfamily arXiv:hep-ph/0011159
  [hep-ph]}}.

\bibitem{Frolov:2008hy}
A.~V. Frolov, {\em JCAP} {\bf 0811}  (2008)   009,
  \href{http://arxiv.org/abs/0809.4904}{{\ttfamily arXiv:0809.4904 [hep-ph]}}.

\bibitem{Easther:2010qz}
R.~Easther, H.~Finkel and N.~Roth, {\em JCAP} {\bf 1010}  (2010)   025,
  \href{http://arxiv.org/abs/1005.1921}{{\ttfamily arXiv:1005.1921
  [astro-ph.CO]}}.

\bibitem{Huang:2011gf}
Z.~Huang, {\em Phys.Rev.} {\bf D83}  (2011)   123509,
  \href{http://arxiv.org/abs/1102.0227}{{\ttfamily arXiv:1102.0227
  [astro-ph.CO]}}.

\bibitem{Child:2013ria}
H.~L. Child, J.~Giblin, John~T., R.~H. Ribeiro and D.~Seery, {\em
  Phys.Rev.Lett.} {\bf 111}  (2013)   051301,
  \href{http://arxiv.org/abs/1305.0561}{{\ttfamily arXiv:1305.0561
  [astro-ph.CO]}}.

\bibitem{Sainio:2009hm}
J.~Sainio, {\em Comput.Phys.Commun.} {\bf 181}  (2010) 906,
  \href{http://arxiv.org/abs/0911.5692}{{\ttfamily arXiv:0911.5692
  [astro-ph.IM]}}.

\bibitem{Sainio:2012mw}
J.~Sainio, {\em JCAP} {\bf 1204}  (2012)   038,
  \href{http://arxiv.org/abs/1201.5029}{{\ttfamily arXiv:1201.5029
  [astro-ph.IM]}}.

\bibitem{Jedamzik:2010dq}
K.~Jedamzik, M.~Lemoine and J.~Martin, {\em JCAP} {\bf 1009}  (2010)   034,
  \href{http://arxiv.org/abs/1002.3039}{{\ttfamily arXiv:1002.3039
  [astro-ph.CO]}}.

\bibitem{Jedamzik:2010hq}
K.~Jedamzik, M.~Lemoine and J.~Martin, {\em JCAP} {\bf 1004}  (2010)   021,
  \href{http://arxiv.org/abs/1002.3278}{{\ttfamily arXiv:1002.3278
  [astro-ph.CO]}}.

\bibitem{Easther:2010mr}
R.~Easther, R.~Flauger and J.~B. Gilmore, {\em JCAP} {\bf 1104}  (2011)   027,
  \href{http://arxiv.org/abs/1003.3011}{{\ttfamily arXiv:1003.3011
  [astro-ph.CO]}}.

\bibitem{Bogolyubsky:1976yu}
I.~L. Bogolyubsky and V.~G. Makhankov, {\em {Pisma Zh. Eksp. Teor. Fiz.}} {\bf
  24}  (1976)  ~15.

\bibitem{Gleiser:1993pt}
M.~Gleiser, {\em Phys.Rev.} {\bf D49}  (1994) 2978,
  \href{http://arxiv.org/abs/hep-ph/9308279}{{\ttfamily arXiv:hep-ph/9308279
  [hep-ph]}}.

\bibitem{CopGle95}
E.~J. Copeland, M.~Gleiser and H.-R. Muller, {\em Phys.Rev.} {\bf D52}  (1995)
  1920, \href{http://arxiv.org/abs/hep-ph/9503217}{{\ttfamily
  arXiv:hep-ph/9503217 [hep-ph]}}.

\bibitem{Salmi:2012ta}
P.~Salmi and M.~Hindmarsh, {\em Phys.Rev.} {\bf D85}  (2012)   085033,
  \href{http://arxiv.org/abs/1201.1934}{{\ttfamily arXiv:1201.1934 [hep-th]}}.

\bibitem{Hertzberg:2010yz}
M.~P. Hertzberg, {\em Phys.Rev.} {\bf D82}  (2010)   045022,
  \href{http://arxiv.org/abs/1003.3459}{{\ttfamily arXiv:1003.3459 [hep-th]}}.

\bibitem{Coleman:1985ki}
S.~R. Coleman, {\em Nucl.Phys.} {\bf B262}  (1985)   263.

\bibitem{Lee:1991ax}
T.~Lee and Y.~Pang, {\em Phys.Rept.} {\bf 221}  (1992) 251.

\bibitem{Kasuya:2001hg}
S.~Kasuya and M.~Kawasaki, {\em Phys.Rev.} {\bf D64}  (2001)   123515,
  \href{http://arxiv.org/abs/hep-ph/0106119}{{\ttfamily arXiv:hep-ph/0106119
  [hep-ph]}}.

\bibitem{2010PhRvD..81h3503C}
T.~{Chiba}, K.~{Kamada} and M.~{Yamaguchi}, {\em Phys.Rev. D} {\bf 81} (April
  2010)   083503, \href{http://arxiv.org/abs/0912.3585}{{\ttfamily
  arXiv:0912.3585 [astro-ph.CO]}}.

\bibitem{Zhou:2013tsa}
S.-Y. Zhou, E.~J. Copeland, R.~Easther, H.~Finkel, Z.-G. Mou {\em et~al.}, {\em
  JHEP} {\bf 1310}  (2013)   026,
  \href{http://arxiv.org/abs/1304.6094}{{\ttfamily arXiv:1304.6094
  [astro-ph.CO]}}.

\bibitem{Amin:2013ika}
M.~A. Amin, {\em Phys.Rev.} {\bf D87}  (2013)   123505,
  \href{http://arxiv.org/abs/1303.1102}{{\ttfamily arXiv:1303.1102
  [astro-ph.CO]}}.

\bibitem{Felder:2006cc}
G.~N. Felder and L.~Kofman, {\em Phys.Rev.} {\bf D75}  (2007)   043518,
  \href{http://arxiv.org/abs/hep-ph/0606256}{{\ttfamily arXiv:hep-ph/0606256
  [hep-ph]}}.

\bibitem{GarciaBellido:2002aj}
J.~Garcia-Bellido, M.~Garcia~Perez and A.~Gonzalez-Arroyo, {\em Phys.Rev.} {\bf
  D67}  (2003)   103501, \href{http://arxiv.org/abs/hep-ph/0208228}{{\ttfamily
  arXiv:hep-ph/0208228 [hep-ph]}}.

\bibitem{Copeland:2002ku}
E.~J. Copeland, S.~Pascoli and A.~Rajantie, {\em Phys.Rev.} {\bf D65}  (2002)
  103517, \href{http://arxiv.org/abs/hep-ph/0202031}{{\ttfamily
  arXiv:hep-ph/0202031 [hep-ph]}}.

\bibitem{Gleiser:2010}
M.~{Gleiser}, N.~{Graham} and N.~{Stamatopoulos}, {\em Phys.Rev.} {\bf 82}
  (August 2010)   043517, \href{http://arxiv.org/abs/1004.4658}{{\ttfamily
  arXiv:1004.4658 [astro-ph.CO]}}.

\bibitem{Gleiser:2011xj}
M.~Gleiser, N.~Graham and N.~Stamatopoulos, {\em Phys.Rev.} {\bf D83}  (2011)
  096010, \href{http://arxiv.org/abs/1103.1911}{{\ttfamily arXiv:1103.1911
  [hep-th]}}.

\bibitem{Gleiser:2014ipa}
M.~Gleiser and N.~Graham, {\em Phys.Rev.} {\bf D89}  (2014)   083502,
  \href{http://arxiv.org/abs/1401.6225}{{\ttfamily arXiv:1401.6225
  [astro-ph.CO]}}.

\bibitem{Mazumdar:2008up}
A.~Mazumdar and H.~Stoica, {\em Phys.Rev.Lett.} {\bf 102}  (2009)   091601,
  \href{http://arxiv.org/abs/0807.2570}{{\ttfamily arXiv:0807.2570 [hep-th]}}.

\bibitem{Dufaux:2010cf}
J.-F. Dufaux, D.~G. Figueroa and J.~Garcia-Bellido, {\em Phys.Rev.} {\bf D82}
  (2010)   083518, \href{http://arxiv.org/abs/1006.0217}{{\ttfamily
  arXiv:1006.0217 [astro-ph.CO]}}.

\bibitem{Deskins:2013dwa}
J.~T. Deskins, J.~T. Giblin and R.~R. Caldwell, {\em Phys. Rev.} {\bf D88}
  (2013)   063530, \href{http://arxiv.org/abs/1305.7226}{{\ttfamily
  arXiv:1305.7226 [astro-ph.CO]}}.

\bibitem{GarciaBellido:2003wd}
J.~Garcia-Bellido, M.~Garcia-Perez and A.~Gonzalez-Arroyo, {\em Phys.Rev.} {\bf
  D69}  (2004)   023504, \href{http://arxiv.org/abs/hep-ph/0304285}{{\ttfamily
  arXiv:hep-ph/0304285 [hep-ph]}}.

\bibitem{PhysRevA.81.033611}
C.~Scheppach, J.~Berges and T.~Gasenzer, {\em Phys. Rev. A} {\bf 81} (Mar 2010)
    033611.

\bibitem{Berges:2013eia}
J.~Berges, K.~Boguslavski, S.~Schlichting and R.~Venugopalan  (2013)
  \href{http://arxiv.org/abs/1303.5650}{{\ttfamily arXiv:1303.5650 [hep-ph]}}.

\bibitem{PhysRevLett.108.161601}
J.~Berges and D.~Sexty, {\em Phys. Rev. Lett.} {\bf 108} (Apr 2012)   161601.

\bibitem{Berges:2008wm}
J.~Berges, A.~Rothkopf and J.~Schmidt, {\em Phys. Rev. Lett.} {\bf 101}  (2008)
    041603, \href{http://arxiv.org/abs/0803.0131}{{\ttfamily arXiv:0803.0131
  [hep-ph]}}.

\bibitem{PhysRevLett.93.142002}
J.~Berges, S.~Bors\'anyi and C.~Wetterich, {\em Phys. Rev. Lett.} {\bf 93} (Sep
  2004)   142002.

\bibitem{Davidson:2000er}
S.~Davidson and S.~Sarkar, {\em JHEP} {\bf 0011}  (2000)   012,
  \href{http://arxiv.org/abs/hep-ph/0009078}{{\ttfamily arXiv:hep-ph/0009078
  [hep-ph]}}.

\bibitem{Enqvist:1990dp}
K.~Enqvist and K.~Eskola, {\em Mod. Phys. Lett.} {\bf A5}  (1990) 1919.

\bibitem{Son:1996uv}
D.~Son, {\em Phys. Rev.} {\bf D54}  (1996) 3745,
  \href{http://arxiv.org/abs/hep-ph/9604340}{{\ttfamily arXiv:hep-ph/9604340
  [hep-ph]}}.

\bibitem{Micha:2002ey}
R.~Micha and I.~I. Tkachev, {\em Phys.Rev.Lett.} {\bf 90}  (2003)   121301,
  \href{http://arxiv.org/abs/hep-ph/0210202}{{\ttfamily arXiv:hep-ph/0210202
  [hep-ph]}}.

\bibitem{Micha:2004bv}
R.~Micha and I.~I. Tkachev, {\em Phys. Rev.} {\bf D70}  (2004)   043538,
  \href{http://arxiv.org/abs/hep-ph/0403101}{{\ttfamily arXiv:hep-ph/0403101
  [hep-ph]}}.

\bibitem{Liddle:1998jc}
A.~R. Liddle, A.~Mazumdar and F.~E. Schunck, {\em Phys.Rev.} {\bf D58}  (1998)
   061301, \href{http://arxiv.org/abs/astro-ph/9804177}{{\ttfamily
  arXiv:astro-ph/9804177 [astro-ph]}}.

\bibitem{Copeland:1999cs}
E.~J. Copeland, A.~Mazumdar and N.~Nunes, {\em Phys.Rev.} {\bf D60}  (1999)
  083506, \href{http://arxiv.org/abs/astro-ph/9904309}{{\ttfamily
  arXiv:astro-ph/9904309 [astro-ph]}}.

\bibitem{Jokinen:2004bp}
A.~Jokinen and A.~Mazumdar, {\em Phys.Lett.} {\bf B597}  (2004)   222,
  \href{http://arxiv.org/abs/hep-th/0406074}{{\ttfamily arXiv:hep-th/0406074
  [hep-th]}}.

\bibitem{Dimopoulos:2005ac}
S.~Dimopoulos, S.~Kachru, J.~McGreevy and J.~G. Wacker, {\em JCAP} {\bf 0808}
  (2008)   003, \href{http://arxiv.org/abs/hep-th/0507205}{{\ttfamily
  arXiv:hep-th/0507205 [hep-th]}}.

\bibitem{Battefeld:2008bu}
D.~Battefeld and S.~Kawai, {\em Phys. Rev.} {\bf D77}  (2008)   123507,
  \href{http://arxiv.org/abs/0803.0321}{{\ttfamily arXiv:0803.0321
  [astro-ph]}}.

\bibitem{2009PhRvD..79l3510B}
D.~{Battefeld}, T.~{Battefeld} and J.~T. {Giblin}, Jr., {\em Phys.~Rev.~}
  (June 2009)   123510, \href{http://arxiv.org/abs/0904.2778}{{\ttfamily
  arXiv:0904.2778 [astro-ph.CO]}}.

\bibitem{Ashoorioon:2009wa}
A.~Ashoorioon, H.~Firouzjahi and M.~Sheikh-Jabbari, {\em JCAP} {\bf 0906}
  (2009)   018, \href{http://arxiv.org/abs/0903.1481}{{\ttfamily
  arXiv:0903.1481 [hep-th]}}.

\bibitem{2012JCAP...11..062B}
T.~{Battefeld}, A.~{Eggemeier} and J.~T. {Giblin}, Jr., {\em JCAP} {\bf 11}
  (November 2012)  ~62, \href{http://arxiv.org/abs/1209.3301}{{\ttfamily
  arXiv:1209.3301 [astro-ph.CO]}}.

\bibitem{Greene:1998nh}
P.~B. Greene and L.~Kofman, {\em Phys. Lett.} {\bf B448}  (1999) 6,
  \href{http://arxiv.org/abs/hep-ph/9807339}{{\ttfamily arXiv:hep-ph/9807339
  [hep-ph]}}.

\bibitem{Greene:2000ew}
P.~B. Greene and L.~Kofman, {\em Phys. Rev.} {\bf D62}  (2000)   123516,
  \href{http://arxiv.org/abs/hep-ph/0003018}{{\ttfamily arXiv:hep-ph/0003018
  [hep-ph]}}.

\bibitem{Peloso:2000hy}
M.~Peloso and L.~Sorbo, {\em JHEP} {\bf 0005}  (2000)   016,
  \href{http://arxiv.org/abs/hep-ph/0003045}{{\ttfamily arXiv:hep-ph/0003045
  [hep-ph]}}.

\bibitem{Tsujikawa:2000ik}
S.~Tsujikawa, B.~A. Bassett and F.~Viniegra, {\em JHEP} {\bf 0008}  (2000)
  019, \href{http://arxiv.org/abs/hep-ph/0006354}{{\ttfamily
  arXiv:hep-ph/0006354 [hep-ph]}}.

\bibitem{Maroto:1999ch}
A.~L. Maroto and A.~Mazumdar, {\em Phys.Rev.Lett.} {\bf 84}  (2000) 1655,
  \href{http://arxiv.org/abs/hep-ph/9904206}{{\ttfamily arXiv:hep-ph/9904206
  [hep-ph]}}.

\bibitem{Kallosh:1999jj}
R.~Kallosh, L.~Kofman, A.~D. Linde and A.~Van~Proeyen, {\em Phys.Rev.} {\bf
  D61}  (2000)   103503, \href{http://arxiv.org/abs/hep-th/9907124}{{\ttfamily
  arXiv:hep-th/9907124 [hep-th]}}.

\bibitem{Davis:2000zp}
A.-C. Davis, K.~Dimopoulos, T.~Prokopec and O.~Tornkvist, {\em Phys. Lett.}
  {\bf B501}  (2001) 165,
  \href{http://arxiv.org/abs/astro-ph/0007214}{{\ttfamily
  arXiv:astro-ph/0007214 [astro-ph]}}.

\bibitem{Allahverdi:2011aj}
R.~Allahverdi, A.~Ferrantelli, J.~Garcia-Bellido and A.~Mazumdar, {\em
  Phys.Rev.} {\bf D83}  (2011)   123507,
  \href{http://arxiv.org/abs/1103.2123}{{\ttfamily arXiv:1103.2123 [hep-ph]}}.

\bibitem{Alishahiha:2004eh}
M.~Alishahiha, E.~Silverstein and D.~Tong, {\em Phys. Rev.} {\bf D70}  (2004)
  123505, \href{http://arxiv.org/abs/hep-th/0404084}{{\ttfamily
  arXiv:hep-th/0404084 [hep-th]}}.

\bibitem{Zhang:2013asa}
J.~Zhang, Y.~Cai and Y.-S. Piao  (2013)
  \href{http://arxiv.org/abs/1307.6529}{{\ttfamily arXiv:1307.6529 [hep-th]}}.

\bibitem{Barnaby:2011qe}
N.~Barnaby, E.~Pajer and M.~Peloso, {\em Phys. Rev.} {\bf D85}  (2012)
  023525, \href{http://arxiv.org/abs/1110.3327}{{\ttfamily arXiv:1110.3327
  [astro-ph.CO]}}.

\bibitem{2012arXiv1201.3942P}
A.~H.~G. {Peter}, {\em ArXiv e-prints}  (January 2012)
  \href{http://arxiv.org/abs/1201.3942}{{\ttfamily arXiv:1201.3942
  [astro-ph.CO]}}.

\bibitem{Bernabei:2003za}
R.~Bernabei, P.~Belli, F.~Cappella, R.~Cerulli, F.~Montecchia {\em et~al.},
  {\em Riv.Nuovo Cim.} {\bf 26N1}  (2003) 1,
  \href{http://arxiv.org/abs/astro-ph/0307403}{{\ttfamily
  arXiv:astro-ph/0307403 [astro-ph]}}.

\bibitem{Peebles:2013hla}
P.~Peebles  (2013) \href{http://arxiv.org/abs/1305.6859}{{\ttfamily
  arXiv:1305.6859 [astro-ph.CO]}}.

\bibitem{Hertzberg:2012zc}
M.~P. Hertzberg  (2012) \href{http://arxiv.org/abs/1210.3624}{{\ttfamily
  arXiv:1210.3624 [hep-ph]}}.

\bibitem{Okada:2010jd}
N.~Okada and Q.~Shafi, {\em Phys.Rev.} {\bf D84}  (2011)   043533,
  \href{http://arxiv.org/abs/1007.1672}{{\ttfamily arXiv:1007.1672 [hep-ph]}}.

\bibitem{Kohri:2009ka}
K.~Kohri, A.~Mazumdar and N.~Sahu, {\em Phys.Rev.} {\bf D80}  (2009)   103504,
  \href{http://arxiv.org/abs/0905.1625}{{\ttfamily arXiv:0905.1625 [hep-ph]}}.

\bibitem{Allahverdi:2007wt}
R.~Allahverdi, B.~Dutta and A.~Mazumdar, {\em Phys.Rev.Lett.} {\bf 99}  (2007)
   261301, \href{http://arxiv.org/abs/0708.3983}{{\ttfamily arXiv:0708.3983
  [hep-ph]}}.

\bibitem{Cardenas:2007xh}
V.~H. Cardenas, {\em Phys.Rev.} {\bf D75}  (2007)   083512,
  \href{http://arxiv.org/abs/astro-ph/0701624}{{\ttfamily
  arXiv:astro-ph/0701624 [astro-ph]}}.

\bibitem{Murayama:1993xu}
H.~Murayama, H.~Suzuki, T.~Yanagida and J.~Yokoyama, {\em Phys.Rev.} {\bf D50}
  (1994) 2356, \href{http://arxiv.org/abs/hep-ph/9311326}{{\ttfamily
  arXiv:hep-ph/9311326 [hep-ph]}}.

\bibitem{Delepine:2006rn}
D.~Delepine, C.~Martinez and L.~A. Urena-Lopez, {\em Phys.Rev.Lett.} {\bf 98}
  (2007)   161302, \href{http://arxiv.org/abs/hep-ph/0609086}{{\ttfamily
  arXiv:hep-ph/0609086 [hep-ph]}}.

\bibitem{Copeland:2001qw}
E.~J. Copeland, D.~Lyth, A.~Rajantie and M.~Trodden, {\em Phys.Rev.} {\bf D64}
  (2001)   043506, \href{http://arxiv.org/abs/hep-ph/0103231}{{\ttfamily
  arXiv:hep-ph/0103231 [hep-ph]}}.

\bibitem{Hertzberg:2013mba}
M.~P. Hertzberg and J.~Karouby, {\em Phys. Rev.} {\bf D89}  (2014)   063523,
  \href{http://arxiv.org/abs/1309.0010}{{\ttfamily arXiv:1309.0010 [hep-ph]}}.

\bibitem{Hertzberg:2013jba}
M.~P. Hertzberg and J.~Karouby, {\em Phys. Lett.} {\bf B737}  (2014) 34,
  \href{http://arxiv.org/abs/1309.0007}{{\ttfamily arXiv:1309.0007 [hep-ph]}}.

\bibitem{Harigaya:2014tla}
K.~Harigaya, A.~Kamada, M.~Kawasaki, K.~Mukaida and M.~Yamada, {\em Phys.Rev.}
  {\bf D90}  (2014)   043510, \href{http://arxiv.org/abs/1404.3138}{{\ttfamily
  arXiv:1404.3138 [hep-ph]}}.

\bibitem{Georgi:1974sy}
H.~Georgi and S.~Glashow, {\em Phys.Rev.Lett.} {\bf 32}  (1974) 438.

\bibitem{Georgi:1974yf}
H.~Georgi, H.~R. Quinn and S.~Weinberg, {\em Phys.Rev.Lett.} {\bf 33}  (1974)
  451.

\bibitem{Dimopoulos:1981yj}
S.~Dimopoulos, S.~Raby and F.~Wilczek, {\em Phys.Rev.} {\bf D24}  (1981) 1681.

\bibitem{Carena:1993ag}
M.~S. Carena, S.~Pokorski and C.~Wagner, {\em Nucl.Phys.} {\bf B406}  (1993)
  59, \href{http://arxiv.org/abs/hep-ph/9303202}{{\ttfamily
  arXiv:hep-ph/9303202 [hep-ph]}}.

\bibitem{Hertzberg:2014sza}
M.~P. Hertzberg and F.~Wilczek  (2014)
  \href{http://arxiv.org/abs/1407.6010}{{\ttfamily arXiv:1407.6010 [hep-ph]}}.

\bibitem{Bezrukov:2007ep}
F.~L. Bezrukov and M.~Shaposhnikov, {\em Phys. Lett.} {\bf B659}  (2008) 703,
  \href{http://arxiv.org/abs/0710.3755}{{\ttfamily arXiv:0710.3755 [hep-th]}}.

\bibitem{Bezrukov:2008ut}
F.~Bezrukov, D.~Gorbunov and M.~Shaposhnikov, {\em JCAP} {\bf 0906}  (2009)
  029, \href{http://arxiv.org/abs/0812.3622}{{\ttfamily arXiv:0812.3622
  [hep-ph]}}.

\bibitem{GarciaBellido:2008ab}
J.~Garcia-Bellido, D.~G. Figueroa and J.~Rubio, {\em Phys. Rev.} {\bf D79}
  (2009)   063531, \href{http://arxiv.org/abs/0812.4624}{{\ttfamily
  arXiv:0812.4624 [hep-ph]}}.

\bibitem{Figueroa:2009jw}
D.~G. Figueroa, {\em AIP Conf. Proc.} {\bf 1241}  (2010) 578,
  \href{http://arxiv.org/abs/0911.1465}{{\ttfamily arXiv:0911.1465 [hep-ph]}}.

\bibitem{Barbon:2009ya}
J.~Barbon and J.~Espinosa, {\em Phys. Rev.} {\bf D79}  (2009)   081302,
  \href{http://arxiv.org/abs/0903.0355}{{\ttfamily arXiv:0903.0355 [hep-ph]}}.

\bibitem{Bezrukov:2010jz}
F.~Bezrukov, A.~Magnin, M.~Shaposhnikov and S.~Sibiryakov, {\em JHEP} {\bf
  1101}  (2011)   016, \href{http://arxiv.org/abs/1008.5157}{{\ttfamily
  arXiv:1008.5157 [hep-ph]}}.

\bibitem{Burgess:2010zq}
C.~Burgess, H.~M. Lee and M.~Trott, {\em JHEP} {\bf 1007}  (2010)   007,
  \href{http://arxiv.org/abs/1002.2730}{{\ttfamily arXiv:1002.2730 [hep-ph]}}.

\bibitem{Hertzberg:2011rc}
M.~P. Hertzberg, {\em JCAP} {\bf 1208}  (2012)   008,
  \href{http://arxiv.org/abs/1110.5650}{{\ttfamily arXiv:1110.5650 [hep-ph]}}.

\bibitem{Bezrukov:2013fka}
F.~Bezrukov, {\em Class.Quant.Grav.} {\bf 30}  (2013)   214001,
  \href{http://arxiv.org/abs/1307.0708}{{\ttfamily arXiv:1307.0708 [hep-ph]}}.

\bibitem{George:2013iia}
D.~P. George, S.~Mooij and M.~Postma, {\em JCAP} {\bf 1402}  (2014)   024,
  \href{http://arxiv.org/abs/1310.2157}{{\ttfamily arXiv:1310.2157 [hep-th]}}.

\bibitem{Barnaby:2004gg}
N.~Barnaby, C.~Burgess and J.~M. Cline, {\em JCAP} {\bf 0504}  (2005)   007,
  \href{http://arxiv.org/abs/hep-th/0412040}{{\ttfamily arXiv:hep-th/0412040
  [hep-th]}}.

\bibitem{Ashoorioon:2013oha}
A.~Ashoorioon, B.~Fung, R.~B. Mann, M.~Oltean and M.~Sheikh-Jabbari, {\em JCAP}
  {\bf 1403}  (2014)   020, \href{http://arxiv.org/abs/1312.2284}{{\ttfamily
  arXiv:1312.2284 [hep-th]}}.

\bibitem{Dodelson:2009kq}
S.~Dodelson, R.~Easther, S.~Hanany, L.~McAllister, S.~Meyer {\em et~al.}
  (2009) \href{http://arxiv.org/abs/0902.3796}{{\ttfamily arXiv:0902.3796
  [astro-ph.CO]}}.

\bibitem{MartinRingeval:2010}
J.~{Martin} and C.~{Ringeval}, {\em Phys.Rev.} {\bf D82}  (2010)
  \href{http://arxiv.org/abs/1004.5525}{{\ttfamily arXiv:1004.5525
  [astro-ph.CO]}}.

\bibitem{Martin:2014nya}
J.~Martin, C.~Ringeval and V.~Vennin  (2014)
  \href{http://arxiv.org/abs/1410.7958}{{\ttfamily arXiv:1410.7958
  [astro-ph.CO]}}.

\bibitem{Easther:2006gt}
R.~Easther and E.~A. Lim, {\em JCAP} {\bf 0604}  (2006)   010,
  \href{http://arxiv.org/abs/astro-ph/0601617}{{\ttfamily
  arXiv:astro-ph/0601617 [astro-ph]}}.

\bibitem{Ade:2013zuv}
 Planck Collaboration Collaboration (P.~Ade {\em et~al.}), {\em
  Astron.Astrophys.}   (2014) \href{http://arxiv.org/abs/1303.5076}{{\ttfamily
  arXiv:1303.5076 [astro-ph.CO]}}.

\bibitem{Shannon:2013wma}
R.~Shannon, V.~Ravi, W.~Coles, G.~Hobbs, M.~Keith {\em et~al.}, {\em Science}
  {\bf 342}  (2013) 334, \href{http://arxiv.org/abs/1310.4569}{{\ttfamily
  arXiv:1310.4569 [astro-ph.CO]}}.

\bibitem{Aasi:2014zwg}
 LIGO Scientific Collaboration, VIRGO Collaboration Collaboration (J.~Aasi {\em
  et~al.})  (2014) \href{http://arxiv.org/abs/1406.4556}{{\ttfamily
  arXiv:1406.4556 [gr-qc]}}.

\bibitem{Moore:2014lga}
C.~J. Moore, R.~H. Cole and C.~P.~L. Berry  (2014)
  \href{http://arxiv.org/abs/1408.0740}{{\ttfamily arXiv:1408.0740 [gr-qc]}}.

\bibitem{Khlebnikov:1997di}
S.~Khlebnikov and I.~Tkachev, {\em Phys.Rev.} {\bf D56}  (1997) 653,
  \href{http://arxiv.org/abs/hep-ph/9701423}{{\ttfamily arXiv:hep-ph/9701423
  [hep-ph]}}.

\bibitem{Easther:2006vd}
R.~Easther, J.~Giblin, John~T. and E.~A. Lim, {\em Phys.Rev.Lett.} {\bf 99}
  (2007)   221301, \href{http://arxiv.org/abs/astro-ph/0612294}{{\ttfamily
  arXiv:astro-ph/0612294 [astro-ph]}}.

\bibitem{Dufaux:2007pt}
J.~F. Dufaux, A.~Bergman, G.~N. Felder, L.~Kofman and J.-P. Uzan, {\em
  Phys.Rev.} {\bf D76}  (2007)   123517,
  \href{http://arxiv.org/abs/0707.0875}{{\ttfamily arXiv:0707.0875
  [astro-ph]}}.

\bibitem{GarciaBellido:2007af}
J.~Garcia-Bellido, D.~G. Figueroa and A.~Sastre, {\em Phys.Rev.} {\bf D77}
  (2008)   043517, \href{http://arxiv.org/abs/0707.0839}{{\ttfamily
  arXiv:0707.0839 [hep-ph]}}.

\bibitem{GarciaBellido:2007dg}
J.~Garcia-Bellido and D.~G. Figueroa, {\em Phys.Rev.Lett.} {\bf 98}  (2007)
  061302, \href{http://arxiv.org/abs/astro-ph/0701014}{{\ttfamily
  arXiv:astro-ph/0701014 [astro-ph]}}.

\bibitem{Dufaux:2008dn}
J.-F. Dufaux, G.~Felder, L.~Kofman and O.~Navros, {\em JCAP} {\bf 0903}  (2009)
    001, \href{http://arxiv.org/abs/0812.2917}{{\ttfamily arXiv:0812.2917
  [astro-ph]}}.

\bibitem{Giblin:2010sp}
J.~Giblin, John~T., L.~R. Price and X.~Siemens, {\em JCAP} {\bf 1008}  (2010)
  012, \href{http://arxiv.org/abs/1006.0935}{{\ttfamily arXiv:1006.0935
  [astro-ph.CO]}}.

\bibitem{2011JCAP...11..015F}
D.~G. {Figueroa}, J.~{Garc{\'{\i}}a-Bellido} and A.~{Rajantie}, {\em JCAP} {\bf
  11} (November 2011)  ~15, \href{http://arxiv.org/abs/1110.0337}{{\ttfamily
  arXiv:1110.0337 [astro-ph.CO]}}.

\bibitem{Bethke:2013vca}
L.~Bethke, D.~G. Figueroa and A.~Rajantie, {\em JCAP} {\bf 1406}  (2014)   047,
  \href{http://arxiv.org/abs/1309.1148}{{\ttfamily arXiv:1309.1148
  [astro-ph.CO]}}.

\bibitem{Figueroa:2013vif}
D.~G. Figueroa and T.~Meriniemi  (2013)
  \href{http://arxiv.org/abs/1306.6911}{{\ttfamily arXiv:1306.6911
  [astro-ph.CO]}}.

\bibitem{Figueroa:2014aya}
D.~G. Figueroa  (2014) \href{http://arxiv.org/abs/1402.1345}{{\ttfamily
  arXiv:1402.1345 [astro-ph.CO]}}.

\bibitem{Giblin:2014gra}
J.~T. Giblin and E.~Thrane  (2014)
  \href{http://arxiv.org/abs/1410.4779}{{\ttfamily arXiv:1410.4779 [gr-qc]}}.

\bibitem{Allen:1996vm}
B.~Allen  (1996) \href{http://arxiv.org/abs/gr-qc/9604033}{{\ttfamily
  arXiv:gr-qc/9604033 [gr-qc]}}.

\bibitem{Arvanitaki:2012cn}
A.~Arvanitaki and A.~A. Geraci, {\em Phys.Rev.Lett.} {\bf 110}  (2013)
  071105, \href{http://arxiv.org/abs/1207.5320}{{\ttfamily arXiv:1207.5320
  [gr-qc]}}.

\bibitem{Cruise:2006zt}
A.~Cruise and R.~Ingley, {\em Class.Quant.Grav.} {\bf 23}  (2006) 6185.

\bibitem{2011PhLB..695...26G}
J.~{Garc{\'{\i}}a-Bellido}, R.~{Durrer}, E.~{Fenu}, D.~G. {Figueroa} and
  M.~{Kunz}, {\em Physics Letters B} {\bf 695} (January 2011) 26,
  \href{http://arxiv.org/abs/1003.0299}{{\ttfamily arXiv:1003.0299
  [astro-ph.CO]}}.

\bibitem{2009JCAP...10..005F}
E.~{Fenu}, D.~G. {Figueroa}, R.~{Durrer} and J.~{Garc{\'{\i}}a-Bellido}, {\em
  JCAP} {\bf 10} (October 2009)  ~5,
  \href{http://arxiv.org/abs/0908.0425}{{\ttfamily arXiv:0908.0425
  [astro-ph.CO]}}.

\bibitem{Figueroa:2012kw}
D.~G. Figueroa, M.~Hindmarsh and J.~Urrestilla, {\em Phys.Rev.Lett.} {\bf 110}
  (2013)   101302, \href{http://arxiv.org/abs/1212.5458}{{\ttfamily
  arXiv:1212.5458 [astro-ph.CO]}}.

\bibitem{Nakayama:2008wy}
K.~Nakayama, S.~Saito, Y.~Suwa and J.~Yokoyama, {\em JCAP} {\bf 0806}  (2008)
  020, \href{http://arxiv.org/abs/0804.1827}{{\ttfamily arXiv:0804.1827
  [astro-ph]}}.

\bibitem{Kawamura:2011zz}
S.~Kawamura, M.~Ando, N.~Seto, S.~Sato, T.~Nakamura {\em et~al.}, {\em
  Class.Quant.Grav.} {\bf 28}  (2011)   094011.

\bibitem{Dimopoulos:2008sv}
S.~Dimopoulos, P.~W. Graham, J.~M. Hogan, M.~A. Kasevich and S.~Rajendran, {\em
  Phys.Rev.} {\bf D78}  (2008)   122002,
  \href{http://arxiv.org/abs/0806.2125}{{\ttfamily arXiv:0806.2125 [gr-qc]}}.

\bibitem{Dvali:2003em}
G.~Dvali, A.~Gruzinov and M.~Zaldarriaga, {\em Phys.Rev.} {\bf D69}  (2004)
  023505, \href{http://arxiv.org/abs/astro-ph/0303591}{{\ttfamily
  arXiv:astro-ph/0303591 [astro-ph]}}.

\bibitem{Kofman:2003nx}
L.~Kofman  (2003) \href{http://arxiv.org/abs/astro-ph/0303614}{{\ttfamily
  arXiv:astro-ph/0303614 [astro-ph]}}.

\bibitem{Bernardeau:2004zz}
F.~Bernardeau, L.~Kofman and J.-P. Uzan, {\em Phys.Rev.} {\bf D70}  (2004)
  083004, \href{http://arxiv.org/abs/astro-ph/0403315}{{\ttfamily
  arXiv:astro-ph/0403315 [astro-ph]}}.

\bibitem{Battefeld:2007st}
T.~Battefeld, {\em Phys.Rev.} {\bf D77}  (2008)   063503,
  \href{http://arxiv.org/abs/0710.2540}{{\ttfamily arXiv:0710.2540 [hep-th]}}.

\bibitem{Bartolo:2004if}
N.~Bartolo, E.~Komatsu, S.~Matarrese and A.~Riotto, {\em Phys.Rept.} {\bf 402}
  (2004) 103, \href{http://arxiv.org/abs/astro-ph/0406398}{{\ttfamily
  arXiv:astro-ph/0406398 [astro-ph]}}.

\bibitem{Chen:2010xka}
X.~Chen, {\em Adv.Astron.} {\bf 2010}  (2010)   638979,
  \href{http://arxiv.org/abs/1002.1416}{{\ttfamily arXiv:1002.1416
  [astro-ph.CO]}}.

\bibitem{Byrnes:2010em}
C.~T. Byrnes and K.-Y. Choi, {\em Adv.Astron.} {\bf 2010}  (2010)   724525,
  \href{http://arxiv.org/abs/1002.3110}{{\ttfamily arXiv:1002.3110
  [astro-ph.CO]}}.

\bibitem{Elliston:2011dr}
J.~Elliston, D.~J. Mulryne, D.~Seery and R.~Tavakol, {\em JCAP} {\bf 1111}
  (2011)   005, \href{http://arxiv.org/abs/1106.2153}{{\ttfamily
  arXiv:1106.2153 [astro-ph.CO]}}.

\bibitem{Leung:2012ve}
G.~Leung, E.~R. Tarrant, C.~T. Byrnes and E.~J. Copeland, {\em JCAP} {\bf 1209}
   (2012)   008, \href{http://arxiv.org/abs/1206.5196}{{\ttfamily
  arXiv:1206.5196 [astro-ph.CO]}}.

\bibitem{Chambers:2009ki}
A.~Chambers, S.~Nurmi and A.~Rajantie, {\em JCAP} {\bf 1001}  (2010)   012,
  \href{http://arxiv.org/abs/0909.4535}{{\ttfamily arXiv:0909.4535
  [astro-ph.CO]}}.

\bibitem{Cicoli:2012cy}
M.~Cicoli, G.~Tasinato, I.~Zavala, C.~Burgess and F.~Quevedo, {\em JCAP} {\bf
  1205}  (2012)   039, \href{http://arxiv.org/abs/1202.4580}{{\ttfamily
  arXiv:1202.4580 [hep-th]}}.

\bibitem{Kobayashi:2013nwa}
N.~Kobayashi, T.~Kobayashi and A.~L. Erickcek  (2013)
  \href{http://arxiv.org/abs/1308.4154}{{\ttfamily arXiv:1308.4154
  [astro-ph.CO]}}.

\bibitem{GarciaBellido:1996qt}
J.~Garcia-Bellido, A.~D. Linde and D.~Wands, {\em Phys.Rev.} {\bf D54}  (1996)
  6040, \href{http://arxiv.org/abs/astro-ph/9605094}{{\ttfamily
  arXiv:astro-ph/9605094 [astro-ph]}}.

\bibitem{Green:2000he}
A.~M. Green and K.~A. Malik, {\em Phys.Rev.} {\bf D64}  (2001)   021301,
  \href{http://arxiv.org/abs/hep-ph/0008113}{{\ttfamily arXiv:hep-ph/0008113
  [hep-ph]}}.

\bibitem{Bassett:2000ha}
B.~Bassett and S.~Tsujikawa, {\em Phys.Rev.} {\bf D63}  (2001)   123503,
  \href{http://arxiv.org/abs/hep-ph/0008328}{{\ttfamily arXiv:hep-ph/0008328
  [hep-ph]}}.

\bibitem{Hidalgo:2011fj}
J.~Hidalgo, L.~A. Urena-Lopez and A.~R. Liddle, {\em Phys.Rev.} {\bf D85}
  (2012)   044055, \href{http://arxiv.org/abs/1107.5669}{{\ttfamily
  arXiv:1107.5669 [astro-ph.CO]}}.

\bibitem{Torres-Lomas:2014bua}
E.~Torres-Lomas, J.~C. Hidalgo, K.~A. Malik and L.~A. Ureña-López, {\em
  Phys.Rev.} {\bf D89}  (2014)   083008,
  \href{http://arxiv.org/abs/1401.6960}{{\ttfamily arXiv:1401.6960
  [astro-ph.CO]}}.

\bibitem{2010PhRvD..81j4019C}
B.~J. {Carr}, K.~{Kohri}, Y.~{Sendouda} and J.~{Yokoyama}, {\em Phys.Rev.D}
  {\bf 81} (May 2010)   104019,
  \href{http://arxiv.org/abs/0912.5297}{{\ttfamily arXiv:0912.5297
  [astro-ph.CO]}}.

\bibitem{Kronberg:1993vk}
P.~P. Kronberg, {\em Rept.Prog.Phys.} {\bf 57}  (1994) 325.

\bibitem{Enquist:1998}
K.~ENQVIST, {\em International Journal of Modern Physics D} {\bf 07}  (1998)
  331,
  \href{http://arxiv.org/abs/http://www.worldscientific.com/doi/pdf/10.1142/S0218271898000243}{{\ttfamily
  http://www.worldscientific.com/doi/pdf/10.1142/S0218271898000243}}.

\bibitem{DiazGil:2007dy}
A.~Diaz-Gil, J.~Garcia-Bellido, M.~Garcia~Perez and A.~Gonzalez-Arroyo, {\em
  Phys.Rev.Lett.} {\bf 100}  (2008)   241301,
  \href{http://arxiv.org/abs/0712.4263}{{\ttfamily arXiv:0712.4263 [hep-ph]}}.

\bibitem{DiazGil:2008tf}
A.~Diaz-Gil, J.~Garcia-Bellido, M.~G. Perez and A.~Gonzalez-Arroyo, {\em JHEP}
  {\bf 0807}  (2008)   043, \href{http://arxiv.org/abs/0805.4159}{{\ttfamily
  arXiv:0805.4159 [hep-ph]}}.

\bibitem{1985PhLB..155...36K}
V.~A. {Kuzmin}, V.~A. {Rubakov} and M.~E. {Shaposhnikov}, {\em Physics Letters
  B} {\bf 155} (May 1985) 36.

\bibitem{Dine:2003ax}
M.~Dine and A.~Kusenko, {\em Rev. Mod. Phys.} {\bf 76}  (2004)  ~1,
  \href{http://arxiv.org/abs/hep-ph/0303065}{{\ttfamily arXiv:hep-ph/0303065}}.

\bibitem{Cline:2006ts}
J.~M. Cline  (2006) \href{http://arxiv.org/abs/hep-ph/0609145}{{\ttfamily
  arXiv:hep-ph/0609145}}, Les Houches Summer School, Session 86: Particle
  Physics and Cosmology: the Fabric of Spacetime, 7-11 Aug. 2006.

\bibitem{Affleck:1984fy}
I.~Affleck and M.~Dine, {\em Nucl.Phys.} {\bf B249}  (1985)   361.

\bibitem{Dolgov:1994zq}
A.~Dolgov and K.~Freese, {\em Phys.Rev.} {\bf D51}  (1995) 2693,
  \href{http://arxiv.org/abs/hep-ph/9410346}{{\ttfamily arXiv:hep-ph/9410346
  [hep-ph]}}.

\bibitem{Dolgov:1996qq}
A.~Dolgov, K.~Freese, R.~Rangarajan and M.~Srednicki, {\em Phys.Rev.} {\bf D56}
   (1997) 6155, \href{http://arxiv.org/abs/hep-ph/9610405}{{\ttfamily
  arXiv:hep-ph/9610405 [hep-ph]}}.

\bibitem{GarciaBellido:1999sv}
J.~Garcia-Bellido, D.~Y. Grigoriev, A.~Kusenko and M.~E. Shaposhnikov, {\em
  Phys.Rev.} {\bf D60}  (1999)   123504,
  \href{http://arxiv.org/abs/hep-ph/9902449}{{\ttfamily arXiv:hep-ph/9902449
  [hep-ph]}}.

\bibitem{Davidson:2000dw}
S.~Davidson, M.~Losada and A.~Riotto, {\em Phys.Rev.Lett.} {\bf 84}  (2000)
  4284, \href{http://arxiv.org/abs/hep-ph/0001301}{{\ttfamily
  arXiv:hep-ph/0001301 [hep-ph]}}.

\bibitem{Megevand:2000da}
A.~Megevand, {\em Phys.Rev.} {\bf D64}  (2001)   027303,
  \href{http://arxiv.org/abs/hep-ph/0011019}{{\ttfamily arXiv:hep-ph/0011019
  [hep-ph]}}.

\bibitem{Tranberg:2003gi}
A.~Tranberg and J.~Smit, {\em JHEP} {\bf 0311}  (2003)   016,
  \href{http://arxiv.org/abs/hep-ph/0310342}{{\ttfamily arXiv:hep-ph/0310342
  [hep-ph]}}.

\bibitem{Tranberg:2006ip}
A.~Tranberg and J.~Smit, {\em JHEP} {\bf 0608}  (2006)   012,
  \href{http://arxiv.org/abs/hep-ph/0604263}{{\ttfamily arXiv:hep-ph/0604263
  [hep-ph]}}.

\end{thebibliography}
\bibliographystyle{ws-ijmpd}

\end{document}